\documentclass[pdflatex, referee, sn-mathphys-num, iicol, hypersetup={colorlinks=true,allcolors=magenta}]{sn-jnl}

\hypersetup{
    colorlinks=true,
    allcolors=magenta,
    pdftitle={PBDW Paper},
    pdfpagemode=FullScreen,
    }

\usepackage{parskip}
\usepackage{graphicx}%
\usepackage{multirow}%
\usepackage{amsmath,amssymb,amsfonts}%
\usepackage{amsthm}%
\usepackage{mathrsfs}%
\usepackage[title]{appendix}%
\usepackage{xcolor}%
\usepackage{textcomp}%
\usepackage{manyfoot}%
\usepackage{booktabs}%
\usepackage{algorithm}%
\usepackage{algcompatible}
\usepackage{algpseudocode}%
\usepackage{algorithmicx}%
\usepackage{listings}%
\usepackage{caption}
\usepackage{subcaption}

\usepackage{siunitx}
\definecolor{myblue}{RGB}{30, 90, 200} 
\usepackage{bm} 
\usepackage{mathrsfs} 
\usepackage{wrapfig}
\usepackage{cuted}
\usepackage[nameinlink]{cleveref} 
\usepackage{xfrac}

\usepackage[abbreviations]{glossaries-extra}
\makeglossaries
\glssetcategoryattribute{acronym}{indexonlyfirst}{true}
\newabbreviation{lv}{LV}{Left Ventricle}
\newabbreviation{bk}{BK}{best-knowledge}
\newabbreviation{rb}{RB}{Reduced Basis}
\newabbreviation{rbs}{RBs}{Reduced Bases}
\newabbreviation{ss}{SSEL}{Sensor Selection}
\newabbreviation{pod}{POD}{Proper Orthogonal Decomposition}
\newabbreviation{fem}{FEM}{Finite Element Method}
\newabbreviation{fe}{FE}{Finite Element}
\newabbreviation{dofs}{DoFs}{Degrees of Freedom}
\newabbreviation{rom}{ROM}{Reduced Order Model}
\newabbreviation{roms}{ROMs}{Reduced Order Models}
\newabbreviation{fom}{FOM}{Full Order Model}
\newabbreviation{wg}{WG}{Weak-Greedy}
\newabbreviation{pbdw}{PBDW}{Parametrized-Background Data-Weak}
\newabbreviation{mri}{MRI}{Magnetic Resonance Image}
\newabbreviation{mris}{MRIs}{Magnetic Resonance Images}
\newabbreviation{mr}{MR}{Magnetic Resonance}
\newabbreviation{womp}{wOMP}{worst-case Orthogonal Matching Pursuit}
\newabbreviation{comp}{cOMP}{collective Orthogonal Matching Pursuit}
\newabbreviation{svd}{SVD}{Singular Value Decomposition}
\newabbreviation{snr}{SNR}{Signal to Noise Ratio}
\newabbreviation{lhs}{LHS}{Latin Hypercube Sampling}
\newabbreviation{vsc5}{VSC5}{Vienna Scientific Cluster 5}
\newabbreviation{vsc}{VSC}{Vienna Scientific Cluster}
\newabbreviation{3d}{3D}{three-dimensional}
\newabbreviation{ppde}{pPDE}{Parametrized Partial Differential Equation}
\newabbreviation{pde}{PDE}{Partial Differential Equation}
\newabbreviation{pmor}{pMOR}{parametric Model Order Reduction}
\newabbreviation{da}{DA}{Data Assimilation}
\newabbreviation{ssm}{SSM}{Statistical Shape Models}
\newabbreviation{ml}{ML}{Machine Learning}
\newabbreviation{sp}{SPL}{Sensor Placement}

\renewcommand{\vec}{\underline}

\theoremstyle{thmstyleone}%
\theoremstyle{thmstyletwo}%
\newtheorem{remark}{Remark}%

\theoremstyle{thmstylethree}%

\renewcommand{\vec}{\bm}

\begin{document}

\title[PBDW Paper]{Non-Intrusive Parametrized-Background Data-Weak Reconstruction of Cardiac Displacement Fields from Sparse MRI-like Observations}


\author[1,2]{\fnm{Francesco C.} \sur{Mantegazza}}\email{francesco.mantegazza@uni-graz.at}
\author[1,2,3]{\fnm{Federica} \sur{Caforio}}\email{federica.caforio@uni-graz.at}
\author[2,3]{\fnm{Christoph} \sur{Augustin}}\email{christoph.augustin@medunigraz.at}
\author[2,3]{\fnm{Matthias A.F.} \sur{Gsell}}\email{matthias.gsell@medunigraz.at}
\author[1,2]{\fnm{Gundolf} \sur{Haase}}\email{gundolf.haase@uni-graz.at}
\author*[1,2]{\fnm{Elias} \sur{Karabelas}}\email{elias.karabelas@uni-graz.at}

\affil[1]{\orgdiv{Department of Mathematics and Scientific Computing}, \orgname{University of Graz}, \orgaddress{\street{Heinrichstraße 36}, \city{Graz}, \postcode{8010}, \country{Austria}}}
\affil[2]{\orgdiv{BioTechMed-Graz}, \orgaddress{\city{Graz}, \postcode{8010}, \country{Austria}}}

\affil[3]{\orgdiv{Gottfried Schatz Research Center: Division of Biophysics}, \orgname{Medical University of Graz}, \orgaddress{\street{Neue Stiftingtalstraße 6}, \city{Graz}, \postcode{8010}, \country{Austria}}}


\abstract{Personalized cardiac diagnostics requires accurate reconstruction of myocardial displacement fields from sparse clinical imaging data. In this work, we apply the \acrlong{pbdw} approach to \acrlong{3d} cardiac displacement field reconstruction from limited \acrlong{mri}-like observations.
We introduce two methodological enhancements: \textcolor{myblue}{(i)} an $H$-size minibatch \acrlong{womp} algorithm that improves \acrlong{ss} computational efficiency while maintaining reconstruction accuracy, and \textcolor{myblue}{(ii)} memory optimisation techniques exploiting block matrix structures in vectorial problems.
We demonstrate the effectiveness of the method through validation on a \acrlong{3d} left ventricular model with simulated scar tissue.
Starting with noise-free reconstruction, we systematically incorporate Gaussian noise and spatial sparsity mimicking realistic \acrlong{mri} acquisition protocols. Results show exceptional accuracy in noise-free conditions with relative $L_2$ error of \textcolor{myblue}{$\num{1e-5}$}, robust performance with \qty{10}{\percent} noise achieving relative $L_2$ error of \textcolor{myblue}{$\num{1e-2}$}, and effective reconstruction from sparse measurements with relative $L_2$ error of \textcolor{myblue}{$\num{1e-2}$}. The online reconstruction achieves four-order-of-magnitude computational speed-up compared to full \acrlong{fe} simulations, with reconstruction times under one tenth of a second for sparse scenarios, demonstrating significant potential for integration into clinical cardiac modelling workflows.
}




\maketitle

\section{Introduction}
\label{sec:intro}
Cardiovascular diseases remain the leading cause of mortality worldwide, accounting for approximately 17.9 million deaths annually~\textcolor{myblue}{\cite{Roth2020}}. Understanding cardiac mechanics is fundamental for diagnosing, treating, and preventing these diseases, driving the development of computational models that can provide insight into pathophysiology and guide therapeutic interventions. These models require an accurate characterisation of the mechanical behaviour of the myocardium, particularly the displacement field that describes heart deformation during the cardiac cycle~\textcolor{myblue}{\cite{CorralAcero2020,Strocchi2023}}.

From a computational perspective, cardiac mechanics presents several challenges that make it an excellent benchmark for reduced order methods: highly \textcolor{myblue}{non-linear} material behaviour, heterogeneous properties from healthy and scarred tissue, large deformations requiring finite strain formulations, and the need for real-time reconstruction from sparse data. These characteristics test the limits of \gls{rom} methods while addressing a genuine clinical need.

Medical imaging, especially \gls{mri}, has become the gold standard for non-invasive assessment of cardiac structure and function \cite{sarvazyan2011overview}. Cardiac \gls{mri} provides high-resolution anatomical images and enables quantification of various functional parameters using techniques such as cine-\gls{mri} and tagged-\gls{mri}. 
However, extracting complete displacement fields from clinical \gls{mri} data presents significant challenges: standard clinical protocols typically acquire sparse 2D multi-slice data rather than full 3D volumes due to time constraints~\cite{Kramer2020}; measurements contain noise; and through-plane motion is often lost in 2D acquisitions \cite{Gudbjartsson1995,CardenasBlanco2008}. While modern \gls{mri} hardware capabilities have advanced significantly~\cite{Aletras1999}, clinical protocols must balance spatial coverage against practical constraints such as acquisition time, duration of breath hold and patient comfort~\cite{Kocaoglu2020}.



Data assimilation techniques offer a promising approach by integrating sparse experimental observations with computational models with the aim of finding an optimal state that balances fidelity to both the theoretical model and the experimental measurements. Within cardiac biomechanics, several methodological approaches have emerged to address the inverse problem of state and parameter estimation from medical imaging data.

Traditional variational \gls{da} methods have shown considerable success in cardiac applications. Recent comprehensive work~\cite{Chabiniok2016} established a framework for multiphysics cardiac modelling with data-model fusion, demonstrating the potential for clinical translation while highlighting the computational challenges involved. The development of sequential \gls{da} methods \cite{Imperiale2021} specifically addressed tagged-\gls{mri} data processing, showing how complex image data can be incorporated into joint state-parameter estimation frameworks. The comprehensive parameter estimation methodology using \gls{3d} tagged-\gls{mri} was demonstrated with unique identification analysis~\cite{Asner2016}, while the estimation of heterogeneous elastic properties in infarcted human hearts was achieved using high-dimensional adjoint-based \gls{da}~\cite{Balaban2017,Balaban2018}.

However, these traditional approaches typically require numerous evaluations of expensive forward models, compromising overall computational efficiency. The exponentially growing field of \gls{ml} has enabled alternative strategies, yet pure data-driven approaches may not be appropriate for biomedical applications where data acquisition costs can be prohibitive~\cite{Caforio2024}. Recent advances~\cite{Barone2020} addressed some of these challenges through experimental validation of variational \gls{da} procedures for cardiac conductivities, demonstrating robustness against parameter non-identifiability issues.

Within the \gls{da} framework, \gls{roms} have gained attention for their ability to decrease computational complexity while preserving accuracy~\cite{Quarteroni2016}. 
\gls{roms} construct low-dimensional approximation spaces that capture the essential dynamics of high-dimensional systems, making them particularly valuable for real-time applications and parameter estimation problems~\cite{Rozza2007}. Recent advances in cardiac \gls{roms} have demonstrated remarkable computational speed-ups: the work of Cicci et al.~\cite{Cicci2024} achieved orders of magnitude acceleration through deep learning-based operator approximation, while the development by Salvador et al.~\cite{Salvador2024} introduced compact neural networks capable of capturing 43-parameter cardiac models for digital twin applications.

Traditional \gls{roms} have also shown promise for cardiac applications. The \gls{pod} based parametric model order reduction demonstrated by Pfaller et al.~\cite{Pfaller2020} integrated gradient-based optimisation for large \textcolor{myblue}{non-linear} cardiac problems, while matrix-based discrete empirical interpolation methods developed by Bonomi et al.~\cite{Bonomi2017} outperform classical approaches with rigorous \textit{a posteriori} error estimates. These developments highlight the potential for \gls{roms} to enable patient-specific modelling at computationally tractable costs.

The \textcolor{myblue}{\gls{pbdw}} approach represents a state-of-the-art \gls{da} method that combines model reduction with measurement integration. Originally introduced by Maday et al.~\cite{Maday2014}, \gls{pbdw} provides a flexible framework for state estimation that incorporates both a parametrised background model and experimental observations. The method has been extended to handle noisy observations and user-defined update spaces~\cite{Maday2017}, with non-linear formulations developed by Gong et al.~\cite{Gong2019} that demonstrate robustness against measurement noise. \textcolor{myblue}{Related theoretical advances on non-linear approximation spaces for inverse problems~\cite{Cohen2023} provide a broader mathematical foundation for understanding the approximation properties of such data-driven reconstruction methods.}

\textcolor{myblue}{A \gls{pbdw} implementation is termed \emph{non-intrusive} when both the \gls{rom} construction and the \gls{ss} rely solely on solution snapshots, without requiring access to the assembly routines of the forward solver or explicit knowledge of the governing equations (a precise definition is given in Remark~\ref{rem:non_intrusive}). In practice, this means that the method operates as a black-box approach: given a collection of precomputed snapshots, it constructs the \gls{rb} and performs reconstruction without any information about how those snapshots were generated.}
\textcolor{myblue}{Non-intrusive \gls{pbdw} implementations, which construct the \gls{rom} purely from snapshots without requiring residual evaluations or problem-specific error estimators, were pioneered by Hammond et al.~\cite{Hammond2019} for urban dispersion modelling, achieving approximation errors below \qty{10}{\percent}. Extensions to time-dependent problems with sequential processing capabilities were developed by Benaceur~\cite{Benaceur2021}. Such snapshot-only approaches are particularly attractive when the forward solver is available only as a black box, as it is common with commercial softwares, e.g., in clinical settings.}

Furthermore, the selection of optimal observation locations (sensors) remains computationally challenging, especially for complex \gls{3d} problems like cardiac mechanics. Several \gls{ss} algorithms have been proposed, including SGreedy~\cite{Maday2014}, \gls{womp}, and \gls{comp}~\cite{Binev2018}.
Although these methods provide theoretical guarantees for reconstruction accuracy, their computational efficiency remains a concern for large-scale problems. Recent advances~\cite{Farazmand2024} introduced sparse discrete empirical interpolation methods leveraging kernel vectors, while strategic \gls{sp} approaches~\cite{Bidar2024} demonstrated \qty{60}{\percent} error reduction compared to uniform placement strategies.

Parallel developments in medical imaging reconstruction have advanced the state-of-the-art in cardiac motion estimation. Deep learning frameworks for mesh-based cardiac motion tracking were developed by Meng et al.~\cite{Meng2024}, while motion-compensated reconstruction with unrolled optimisation networks was introduced by Pan et al.~\cite{Pan2024}. Regularisation frameworks using spatio-temporal B-splines with physiological constraints were established by Wiputra et al.~\cite{Wiputra2020}, and high-resolution \gls{3d} tissue dynamics measurements using displacement-encoded \gls{mri} techniques were demonstrated by Gomez and Merchant~\cite{Gomez2014}.

These advances in medical imaging reconstruction complement \gls{da} approaches, yet integration between the two domains remains limited. Most existing methods focus either on pure image processing or on model-based approaches without leveraging the synergies between advanced imaging techniques and physics-informed \gls{da}.

Previous applications of \gls{pbdw} in biomedical contexts have shown promising results for problems such as arterial flow~\cite{Galarce2020} and soft tissue deformation of the human brain~\cite{Galarce2023}. However, fully \gls{3d} cardiac mechanics - with its \textcolor{myblue}{non-linear} constitutive laws, heterogeneous properties, and large deformations - presents significantly greater computational challenges that have not yet been addressed.

The integration of \gls{pbdw} methods with cardiac displacement field reconstruction from medical imaging represents a significant research gap, particularly for methods that can operate in non-intrusive settings compatible with diverse computational frameworks.

In this work, we apply the non-intrusive \gls{pbdw} methodology to cardiac mechanics, introducing computational enhancements that make it practical for high-dimensional problems. Following the non-intrusive formulation established by Hammond et al.~\cite{Hammond2019}, we contribute with: (i) an efficient minibatch \gls{ss} algorithm that substantially reduces computational time while maintaining reconstruction accuracy, and (ii) memory-efficient implementations exploiting block structure in vectorial displacement fields, reducing memory requirements by a factor of nine for \gls{3d} problems.
\textcolor{myblue}{The only assumption on the forward model is that its solutions possess $[H^1(\Omega)]^3$ regularity.}
This non-intrusive nature is particularly valuable in cardiac biomechanics where clinical teams use diverse commercial solvers (ANSYS, Abaqus, FEBio) with proprietary implementations, making intrusive methods impractical for widespread adoption.

\textcolor{myblue}{When working with \gls{mri} data, we interpret each image voxel as a sensor measurement acquired within a fixed observational volume that is stationary in both time and space. Consequently, unlike applications in electrophysiology or hemodynamics, where sensor locations can be optimized, \gls{mri} acquisition protocols predetermine the available measurement locations. Our focus is therefore on \gls{ss}—identifying the most informative subset among the available voxel measurements—rather than on \gls{sp} (see Remark~\ref{rem:ss_vs_sp} for a precise distinction).}



\textcolor{myblue}{We equip \gls{pbdw} with an efficient \gls{ss} algorithm that identifies the most informative measurements from the available set. This selection strategy provides two key benefits: first, it reduces computational cost by constructing a smaller update space; second, it improves the numerical conditioning of the reconstruction problem by excluding redundant measurements that provide little additional information. Including all available sensors without selection can lead to an ill-conditioned system where the update space becomes nearly collinear with the background space, degrading reconstruction accuracy.}


\textcolor{myblue}{\gls{ss} and \gls{rom} construction constitute the offline phase of \gls{pbdw}, which is performed once for a given geometry and can then be reused for any number of reconstructions from new measurement data. The efficient linear algebra implementation proposed in this work accelerates both the offline \gls{ss} stage and the online reconstruction phase by exploiting the block structure inherent to vectorial problems.}

We demonstrate the effectiveness of our method through a series of increasingly challenging test cases involving a \gls{3d} left ventricular model with a simulated scar region. Starting with noise-free reconstruction, we gradually incorporate Gaussian noise and spatial sparsity to mimic realistic \gls{mri} data acquisition conditions. 
Our results show that the proposed framework can accurately reconstruct displacement fields even from sparse and noisy observations, with computational efficiency suitable for potential clinical applications. 
The paper is organised as follows: Section~\ref{sec:methods} describes the \gls{pbdw} methodology, including the non-intrusive formulation, the reduced-order modelling approach, and the \gls{ss} algorithm. 
Section~\ref{sec:results} presents numerical results for the test cases of increasing complexity. 
Section~\ref{sec:disc} discusses the implications of these results and potential applications. Finally, Section~\ref{sec:concl} provides concluding remarks and directions for future work.



\section{Methods}
\label{sec:methods}

\subsection{\textcolor{myblue}{Terminology and Scope}}
\label{subsec:terminology}

\textcolor{myblue}{Before presenting the technical details, we clarify two key concepts that are central to our approach and have distinct meanings in literature.
\begin{remark}[Non-intrusive implementations]
    \label{rem:non_intrusive}
    In the context of \gls{rom}-based methods, we call an implementation \emph{non-intrusive} if it requires only solution snapshots $\{\bm{u}^\mathrm{bk}(\bm{\theta}_i)\}_{i=1}^{N_s}$ and the mesh on which they are defined. In particular, non-intrusive methods do not require: i) access to residual evaluations or Jacobian matrices from the forward solver; ii) problem-specific \textit{a posteriori} error estimators; iii) knowledge of the weak form, constitutive model, or discretisation scheme.
    This distinction is practically significant: non-intrusive methods treat the forward solver as a black box, enabling their use with commercial software where source code access is unavailable. Both our \gls{pod}-based background space construction and \gls{womp}-based \gls{ss} satisfy this criterion, as they operate entirely on precomputed snapshots.
\end{remark}
\begin{remark}[\gls{ss} versus \gls{sp}]
    \label{rem:ss_vs_sp}
    We distinguish between two related but fundamentally different problems:
    \begin{description}
        \item[\gls{sp}] concerns the optimisation of sensor \emph{locations} in physical space, typically relevant when sensors can be freely positioned (e.g., strain gauges, electrode arrays).
        \item[\gls{ss}] concerns the identification of the most informative \emph{subset} from a fixed, predetermined set of measurement locations.
    \end{description}
    In the context of \gls{mri} data, the voxel grid is determined by the imaging protocol and hardware, leaving no freedom in sensor positioning. Our problem falls therefore in the category of \gls{ss}: given $K$ available voxel measurements, identify the subset of $M \ll K$ that maximises reconstruction accuracy. This distinction is important because \gls{ss} algorithms (such as \gls{womp}) operate on a discrete candidate set, whereas \gls{sp} methods optimise over continuous spatial domains.
\end{remark}
}

\subsection{\gls{pbdw} for Cardiac Displacement Field Reconstruction}
\label{subsec:pbdw_framework}
\begin{figure*}
    \centering
    \includegraphics[width=0.7\textwidth]{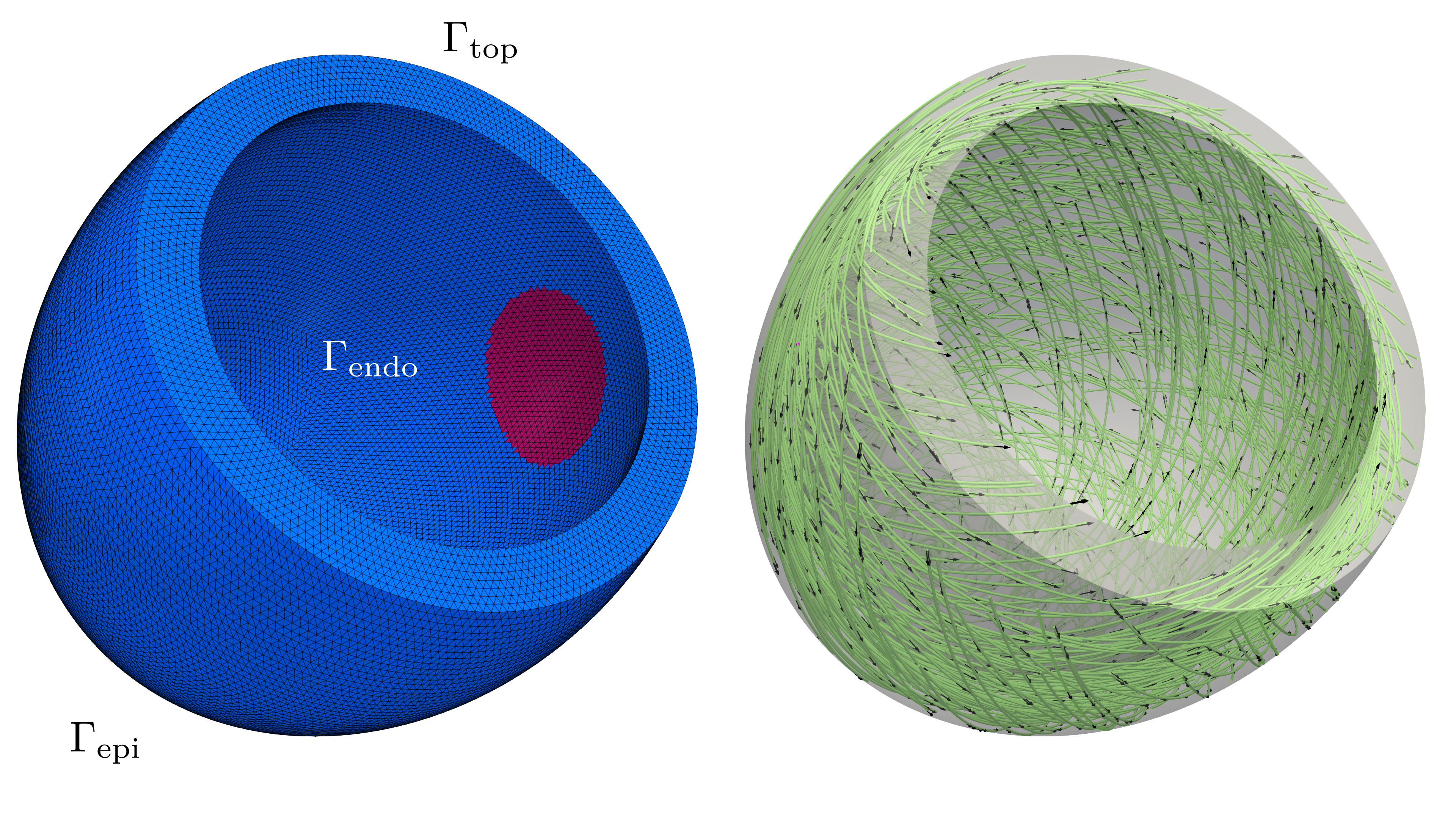}
    \caption{Computational domain representing an idealised \gls{lv}. On the left we indicate the boundaries of the computational domain and different colours refer to an idealised scar region with different material parameters; on the right we indicate the rule-based fiber directions $\bm{f}_0$.}
    \label{fig:ellipsoid}
\end{figure*}

Let us consider the Hilbert space $\mathcal X = [H^1(\Omega)]^3$.
In this work we endow $\mathcal X$ with the weighted inner product~\cite{Benaceur2021}
$$(\bm{u},\bm{v})_{\mathcal{X}} = (\bm{u},\bm{v})_{L^2} + L_g^2(\nabla\bm{u},\nabla\bm{v})_{L^2},$$
with characteristic length scale $L_g$ inducing a norm $\lVert \cdot \rVert_\mathcal{X} = \sqrt{(\cdot, \cdot)_{\mathcal X}}$.
For more details on the notations, and some additional details on functions spaces and discretisation used throughout the paper we refer the interested reader to Section~\ref{app:notations}.
We consider the reconstruction of parametric cardiac displacement fields $\bm{u}^\mathrm{true} \in \mathcal X$ from $K$ \gls{mri}-like observations $\{\vec{y}_m^\mathrm{obs}\}_{m=1}^K$, possibly noisy and sparse, during ventricular inflation.
Here, $K$ refers to the number of voxels in a given \gls{mri}.
We assume a parametrised \gls{bk} mathematical model of our physics for a set of model parameters 
\textcolor{myblue}{$\vec \theta \in \mathcal P^\mathrm{bk}$:}
Find $\bm{u}^\mathrm{bk}_{\vec \theta}$ s. t.:
\begin{equation}
\label{eq:cardiac_pde}
\begin{aligned}
- \mathrm{Div} \, \mathbf{T}(\bm{u}^\mathrm{bk}_{\vec \theta}) &= \bm{0} && \mathrm{in}~\Omega, \\
\mathbf{T}(\bm{u}^\mathrm{bk}_{\vec \theta}) \bm{N} &= -p J(\bm{u}^\mathrm{bk}_{\vec \theta}) \mathbf{F}^{-\top}(\bm{u}^\mathrm{bk}_{\vec \theta}) \bm{N} && \mathrm{on}~\Gamma_{\mathrm{endo}}, \\
\mathbf{T}(\bm{u}^\mathrm{bk}_{\vec \theta}) \bm{N} &= \bm{0} && \mathrm{on}~\Gamma_{\mathrm{epi}}, \\
u^\mathrm{bk}_{{\vec \theta},z} &= 0 && \mathrm{on}~\Gamma_{\mathrm{top}},
\end{aligned}
\end{equation}
where $\Omega \subset \mathbb{R}^3$ is given by a \gls{3d} ellipsoid depicted in Fig.~\ref{fig:ellipsoid}, representing a simplified \gls{lv}, $\bm{N}$ is the outward normal vector with respect to the boundaries, and \textcolor{myblue}{$p$} describes the inflation pressure in kPa.
The inner surface $\Gamma_\mathrm{endo}$ of the ellipsoid represents the endocardium and is endowed with inhomogeneous Neumann boundary conditions. The outer surface $\Gamma_\mathrm{epi}$ represents the epicardium and is endowed with homogeneous Neumann boundary conditions. 
The upper surface $\Gamma_\mathrm{top}$ is endowed with homogeneous Dirichlet boundary conditions in the $z-$direction.
$\mathbf{T}$ is the first Piola-Kirchhoff stress tensor following the almost incompressible Guccione constitutive model~\cite{Guccione1991Passive} based on the Flory split~\cite{Flory1961}, 
\begin{equation}
    \label{eq:Guccione}
    \begin{aligned}
    \mathbf{T}(\bm{u}) &= \frac{\partial W(\bm{u})}{\partial \mathbf{F}},\\
    W&= \frac{\alpha}{2}(e^{\bar{Q}}-1) + \frac{\kappa}{2}(\log J)^2
    \end{aligned}
\end{equation}
with
\begin{equation*}
    \begin{aligned}
    \bar{Q}&=b_f(\bm{f}_0\bar{\mathbf{E}}\bm{f}_0)^2+b_t[(\bm{s}_0\bar{\mathbf{E}}\bm{s}_0)^2+(\bm{n}_0\bar{\mathbf{E}}\bm{n}_0)^2\\&+2(\bm{s}_0\bar{\mathbf{E}}\bm{n}_0)^2]+2b_{fs}[(\bm{f}_0\bar{\mathbf{E}}\bm{s}_0)^2+(\bm{f}_0\bar{\mathbf{E}}\bm{n}_0)^2],      
    \end{aligned}
\end{equation*}
with: deformation gradient $\mathbf{F}:= \mathbf{I} + \nabla\bm{u}$; Jacobian $J=\mathrm{det}(\mathbf{F})$; right Cauchy-Green tensor $\mathbf{C}:=\mathbf{F}^\top\mathbf{F}$; isochoric Cauchy-Green tensor $\bar{\mathbf{C}}=J^{-\frac{2}{3}}\mathbf{C}$; isochoric Green-Lagrange strain tensor $\bar{\mathbf{E}}:=\frac{1}{2}(\bar{\mathbf{C}}-\mathbf{I})$; and fiber, normal, and sheet normal directions $\bm f_0$, $\bm s_0$, $\bm n_0$, generated with a rule-based approach~\cite{Bayer2012}, respectively.
\textcolor{myblue}{In the most general setting, the model parameters $\vec{\theta}$ would comprise all material parameters appearing in equations
  \eqref{eq:cardiac_pde}--\eqref{eq:Guccione}, i.e., $\vec{\theta} = (p, \alpha, \kappa, b_f, b_t, b_{fs})$. For this work we chose to use default values
  of $b_f = 18.48$, $b_t = 3.58$, and $b_{fs} = 1.627$ in accordance with previous work~\cite{Caforio2024}.
  Since the parameter $\kappa$ is related to incompressibility and does not vary in pathological tissue~\cite{sarvazyan2011overview}, we additionally fix
  $\kappa= \qty{650}{\kilo\pascal}$. In Section~\ref{sec:bkerr} we investigate this choice with respect to model error.
  Henceforth we will vary inflation pressure and passive stiffness, defining $\vec{\theta} = (p, \alpha) \in \mathcal P^\mathrm{bk} := [5, 16] \times
  [0.7884, 0.9635]~\mathrm{kPa}$.}
These parameter ranges correspond to physiological variations from healthy to mildly impaired cardiac function, providing a realistic parameter space for testing \gls{rom} performance under practical variability.
With this we can define the \gls{bk} manifold \textcolor{myblue}{$$\mathcal M^\mathrm{bk}:=\left\{\bm{u}^\mathrm{bk}_{\vec \theta} : \vec \theta \in\mathcal{P}^\mathrm{bk}\right\}.$$}
To model measurements obtained from \gls{mri} we assume that we can express the action of measurement as an component-wise averaging process over each voxel defined as the following linear functionals:

\begin{equation}
    \label{eq:mri_observations}
    \vec y_k^\mathrm{obs} := \vec \ell_{k}(\bm{u}_\mathrm{true}) + \vec \varepsilon_k,
\end{equation}

with

\begin{equation}
    \label{eq:mri_functionals}
    \vec \ell_{k,i}(\bm{u}) := \frac{1}{\lvert \Omega_k \rvert}\int_{\Omega_k} u_i \, \mathrm{dx},
\end{equation}

for $k=1,\ldots,K$, and $i=1,2,3$, where $\Omega_k \subset \Omega$ represents the $k$-th voxel and $ \vec \varepsilon_{k}$ denotes unknown disturbances caused by either systematic error in the data acquisition system or experimental random noise.
The collection of all functionals \textcolor{myblue}{$\{\vec \ell_k\}_{k=1,i=1}^{K,3}$} is denoted \emph{sensor library} $\mathcal L_K$.
The key idea of \gls{pbdw} is to find an approximation $$\bm{u}^* = \bm{z}^* + \bm{\eta}^*$$ to $\bm{u}_\mathrm{true}$ using \emph{projection-by-data}.
First, $\bm{z}^* \in \mathcal{Z}_N$ denotes the deduced background estimate.
The $N$-dimensional linear space $\mathcal{Z}_N \subset \mathcal X$ is constructed as a model-informed space utilising snapshots from $\mathcal{M}^\mathrm{bk}$, thus encoding the modeller's available prior knowledge about the physics.
Second, $\bm{\eta}^* \in \mathcal U_M \subset \mathcal X$ is the update estimate.
This $M$-dimensional linear space is spanned by a \textcolor{myblue}{$M<3K$} selection of $\mathcal X$-orthonormal Riesz representers~\cite{Rudin1990} from $\mathcal L_K$.
The update estimate acts as a correction term for modelling errors.

In practice, one has to find suitable methods for constructing $\mathcal Z_N$ and selecting an optimal number of sensors to form $\mathcal U_M$.
Here, the background space $\mathcal{Z}_N = \mathrm{span}\{\bm{\zeta}_n\}_{n=1}^N$ is constructed from $N_s$ solution snapshots via \gls{pod}~\cite{Carlberg2008}, capturing dominant deformation modes (Summary in Sec.~\ref{app:pod_implementation}) while being overall non-intrusive and not requiring information about the underlying \gls{ppde}.
\textcolor{myblue}{Several alternative approaches exist for constructing the background space $\mathcal Z_N$. These include the weak greedy algorithm~\cite{Hesthaven2016,Hammond2019}, greedy reduced basis methods with \textit{a posteriori} error estimators~\cite{Binev2017}, randomised decompositions for large-scale systems~\cite{Levitt2024}, hybrid approaches combining model reduction with machine learning~\cite{Romor2025}, higher-order dynamic mode decomposition variants~\cite{Corrochano2023}, and nonlinear manifold techniques~\cite{Mazzilli2022}; comprehensive reviews can be found in~\cite{Benner2017,Do2025}. We employ \gls{pod} due to its non-intrusive nature (see Remark~\ref{rem:non_intrusive}), computational simplicity, and well-established effectiveness for problems exhibiting rapid singular value decay.}

\textcolor{myblue}{The update space $\mathcal{U}_M=\mathrm{span}\{\tau_m\}_{m=1}^M$ is constructed by employing \gls{womp} (that we describe in detail in Sec.~\ref{subsubsec:minibatch}).}



Following~\cite{Taddei2017}, in the case of possibly noisy measurements, the optimal reconstruction $\bm{u}_\xi^* = \bm z_\xi^* + \bm \eta_\xi^*$ to the unknown field $\bm{u}^\mathrm{true}$ minimises the functional
\begin{equation}
    \label{eq:pbdw_minimization_cardiac}
    J_\xi(\bm{z}, \bm{\eta}) = \xi\|\bm{\eta}\|_{\mathcal{X}}^2 + \frac{1}{M} \sum_{m=1}^M \lVert\vec{\ell}_m(\bm{z}+\bm{\eta})-\vec{y}_m^\mathrm{obs}\rVert_2^2
\end{equation}
where the first term penalizes deviations from the solution manifold and the second one enforces data fidelity.
The term $\xi > 0$ denotes a tunable hyper-parameter encoding the modeler's trust in the \gls{bk} model.
Assuming finite-dimensional bases for $\mathcal Z_N$ and $\mathcal U_M$ we can deduce that the optimal solution to Eq.~\eqref{eq:pbdw_minimization_cardiac} solves the saddle point system

\begin{equation}
\label{eq:pbdw_system_cardiac}
    \begin{bmatrix}
    M\xi\mathbf{I} + \mathbf{T}^\top \mathbf{T} & \quad\mathbf{T}^\top \mathbf{T}\mathbf{P}^\top \\
    \mathbf{P} & \mathbf{0}
    \end{bmatrix}
    \begin{bmatrix}
    \vec{\eta}_\xi^* \\ \vec{z}_\xi^*
    \end{bmatrix}
    =
    \begin{bmatrix}
    \mathbf{T}^\top \vec{y}^{\mathrm{obs}} \\ \mathbf{0}
    \end{bmatrix}
\end{equation}

with matrices $T_{ij} = \vec\ell_i(\bm{\tau}_j)$ and $P_{nm} = (\bm{\zeta}_n, \bm{\tau}_m)_{\mathcal{X}}$ (detailed derivation in Appendix~\ref{app:pbdw_derivation}). 
The reconstructed field is then given as:
\begin{equation}
\label{eq:reconstruction_cardiac}
\bm{u}_\xi^* = \sum_{n=1}^N \underline{z}_{\xi,n}^* \bm{\zeta}_n + \sum_{m=1}^M \underline{\eta}_{\xi,m}^* \bm{\tau}_m
\end{equation}

\textcolor{myblue}{Note that the reconstruction with $\eta^*_{\xi,m}=0\quad\forall m=1,..,M$ is just a static least-squares approach that fits the data through a point in $\mathcal{Z}_N$.}
\subsubsection{Stability of PBDW}
\label{subsubsec:pbdw_stability}
The well-posedness of the \gls{pbdw} system~\eqref{eq:pbdw_system_cardiac} depends on the angle between $\mathcal Z_N$ and $\mathcal U_M$, which is reflected through the inf-sup stability constant 
\begin{equation}
    \beta_{N,M}:= \inf_{\bm w \in \mathcal Z_N} \sup_{\bm q \in \mathcal U_M} \frac{(\bm w, \bm q)_{\mathcal X}}{\lVert \bm w \rVert_{\mathcal X} \lVert \bm q \rVert_{\mathcal X}}.
\end{equation}

\textcolor{myblue}{$\beta_{N,M}$ is, at the discrete level, the condition number of the least-squares normal matrix $C^\top C$, with $C$ the cross-Gramian matrix for the space $\mathcal{U}_M $ and $\mathcal{Z}_N$.}
It is known that Sys.~\eqref{eq:pbdw_system_cardiac} is uniquely solvable provided $\beta_{N,M} >0$, see~\cite{Taddei_phdthesis,Maday2014,Binev2017}, and that $\beta_{N,M} = 0$ for $M < N$.
\textcolor{myblue}{An \textit{a priori} error theory~\cite[Prop.~2, p.~941]{Maday2014} proves that the \gls{pbdw} error estimates depend on $\beta_{N,M}$. However, this classical analysis applies to the noise-free case ($\xi = 0$) only. For noisy measurements with regularization parameter $\xi > 0$, the stability analysis is considerably more involved: Gong et al.~\cite{Gong2019} introduce three distinct stability constants that capture sensitivity to measurement noise, model mismatch, and algorithmic bias, respectively. Crucially, these constants trade off against each other---increasing the background dimension $N$ improves model approximation but can amplify noise sensitivity---and the simple relationship between $\beta_{N,M}$ and the stability constant holds only for $\xi = 0$. Following~\cite{Gong2019,Bui2023}, rather than computing the full analytical stability constants, we focus on demonstrating reconstruction accuracy empirically under various noise levels and sparsity conditions in Section~\ref{sec:results}.}

\subsection{Minibatch \gls{ss}}
\label{subsubsec:minibatch}

\textcolor{myblue}{As clarified in Remark~\ref{rem:ss_vs_sp}, our problem is one of \gls{ss} rather than \gls{sp}.}

\gls{ss} identifies the most informative subset of $M$ measurements from $K$ available sensors, thus providing a compression of the possibly big space $\mathcal U_M$.
Several algorithms exist for \gls{ss}, most notably SGreedy~\cite{Maday2015} and \gls{womp}~\cite{Binev2018}.
In this work we use \gls{womp}.
The basic idea is to first identify in each iteration the state in $\mathcal Z_N$ that realises the minimum observability given by $\beta_{N,M}$, and then to select the sensor from $\mathcal U_M$ that improves its observation the most. The two steps are repeated until a predefined threshold for $\beta_{N,M}$ or the desired sensor number $K_\text{desired}$ has been reached.
This is summarized in Algorithm~\ref{alg:womp} and we provide more details in Appendix~\ref{subsec:ss}.
Most notably, \gls{womp} is inherently sequential, so we propose an acceleration by selecting a minibatch of $H$ sensors per iteration, summarised in Algorithm~\ref{alg:minibatchwomp}.
\begin{algorithm}[!htbp]
    \caption{$H$-size minibatch \gls{womp}}
    \label{alg:minibatchwomp}
    \begin{algorithmic}
        \Require $\mathcal{Z}_N,\mathcal L_K, \beta_\odot \in (0,1), K_\text{desired}$
        \Ensure $\mathcal{U}_M,\mathbf{Q}$ s.t. $\beta_{N,M} \geq \beta_\odot$
        \State $\bm{q}_\beta^\perp  \gets \frac{\bm{\zeta}_1}{||\bm{\zeta}_{1}||_\mathcal{X}}$ 
        \State $M=0$
        \While{$\beta_{N,M}<\beta_\odot$ or $M < K_\text{desired}$}
        \State $\vec\ell_{M+1},\ldots,\vec\ell_{M+H}$
        \Statex \hspace{\algorithmicindent}$= 
               \underset{\vec\ell \in \mathcal{L}_K}{\text{argsort}}
               \frac{1}{\|\vec\ell\|_{\mathcal{X}'}} \,
               |\vec\ell(\bm{q}_\beta^\perp)|[1,\ldots,H]$   
        \State $\bm{\tau}_{M+i}= \frac{\Pi_{\mathcal{U}_M^\perp} \mathcal{R}_{\vec\ell_{M+i}}}{\lVert\Pi_{\mathcal{U}_M^\perp} \mathcal{R}_{\vec{\ell}_{K+i}}\rVert_\mathcal{X}}, \quad i=1,\ldots,H$
        \State $\mathcal{U}_M=\mathrm{span}\{\bm{\tau}_1,...,\bm{\tau}_{M+1},...,\bm{\tau}_{M+H}\}$
        \State $\mathbf{Q}=[\vec{\tau}_1,...,\vec{\tau}_{M+1},...,\vec{\tau}_{M+H}]^\top$
        \State $\mathbf{P}\mathbf{P}^\top \vec{v}_\beta = \lambda_\beta \vec{v}_\beta$
        \State $\beta_{N,M}=\sqrt{\lambda_{\beta,\mathrm{min}}}$
        \State $\bm{\phi}_\beta=\sum_{n=1}^N(\vec{v}_\beta)_n \bm{\zeta}_n$
        \State $\bm{q}_\beta^\perp=\bm{\phi}_\beta-\Pi_{\mathcal{U}_M}\bm{\phi}_\beta$
        \State $\bm{q}_\beta^\perp=\frac{\bm{q}_\beta^\perp}{\lVert\bm{q}_\beta^\perp\rVert_{\mathcal{X}}}$
        \State $M=M+H$
        \EndWhile
    \end{algorithmic}
\end{algorithm}

This reduces selection time from $\mathcal{O}(K_\text{desired} \cdot (K_\text{desired}+N^3+N+M\mathcal{N}))$ to $\mathcal{O}(\frac{K_\text{desired}}{H} \cdot (K_\text{desired}+N^3+N+M\mathcal{N}))$, with optimal $H \in [50,100]$ balancing evaluation reduction against orthonormalisation overhead. We note, that our optimal value of $H$ was estimated empirically related to our considered test problems.

\subsection{Memory Optimisation for Vectorial Fields}
\label{subsubsec:memory}
For vector-valued \gls{fe} functions defined on a mesh with $\mathcal N$ nodes, a naive storage of all \gls{fe} vectors to vector-valued Riesz representers, would result in a dense matrix $\mathbf{R}_{3d} \in \mathbb{R}^{3K \times 3\mathcal{N}}$.
However, we can exploit the block structure of $\mathbf{R}_{3d}$ storing only $\mathbf{R}_{1d} \in \mathbb{R}^{K \times \mathcal{N}}$ and using index mapping:
$$
\mathbf{R}_{3d} = \mathbf{R}_{1d} \otimes \mathbf{I}_3,
$$
with access via 
\begin{equation}
    \label{eq:index}
    \quad (q,r) = (\lfloor i/3 \rfloor, i \bmod 3).
\end{equation}
Strategic permutation matrices $\mathbf{P}_{\mathrm{ord}}$ preserve sparsity in matrix products, reducing memory nine-fold.
This optimisation reduces the storage requirement from $\mathcal{O}(9K\mathcal{N})$ to $\mathcal{O}(K \mathcal{N})$ and transforms dense matrix products into block-sparse operations. 
For our cardiac problem with $\sim$~150000 nodes, this reduces memory usage from $\sim$~\qty{4.5}{\giga\byte} to $\sim$~\qty{0.5}{\giga\byte} per sensor library, enabling practical computation on standard workstations. 
This optimisation strategy is applicable to any vector-valued \gls{rom} problem with component-wise measurements, extending beyond the cardiac application presented here.
The complete derivation is provided in Appendix~\ref{sec:vec_efficient}.

\subsection{Implementation Workflow}
\label{subsec:workflow}
Here we give a high-level overview of our \gls{pbdw} workflow and refer to Appendix~\ref{sec:impl} for details.
\paragraph{Offline Phase}
The offline phase executes once per geometry and parameter range:
\begin{enumerate}
    \item \textbf{Snapshot generation}: Sample $N_s$ parameter pairs $(\alpha_j, p_j)$ via \gls{lhs}, solve Eq.~\eqref{eq:cardiac_pde} using our in-house solver \texttt{carpentry}~\cite{Karabelas2022}.
    \item \textbf{\gls{pod} basis}: Construct $\mathcal{Z}_N$ from snapshots using \gls{pod}, retaining modes capturing \qty{99.9}{\percent} energy, see Sec.~\ref{app:pod_implementation}.
    \item \textbf{Riesz computation}: \textcolor{myblue}{Solve $3K$ linear systems for Riesz representers, derived from the variational formulations:
    \begin{equation}
        \label{eq:var_riesz}
        (\mathcal{R}_{\ell_{k,i}},v)=\ell_{k,i}(v) \quad \forall v\in\mathcal{X}
    \end{equation}
    where the right-hand side is computed as in Eq.~\eqref{eq:mri_functionals}.}
    \item \textbf{\gls{ss}}: Apply Algorithm~\ref{alg:minibatchwomp} until $\beta_{N,M} \geq \beta_\odot$ with a user-defined lower-bound $\beta_\odot \in (0,1)$.
    \item \textbf{Matrix assembly}: Precompute $\mathbf{T}$, $\mathbf{P}$ with memory optimizations.
\end{enumerate}

\paragraph{Online Phase}
For each reconstruction:
\begin{enumerate}
    \item \textcolor{myblue}{Ideally, acquire observations $\vec{y}^{\mathrm{obs}}$ from \gls{mri} data. However, in this paper, we employ synthetic data by computing the observations on \gls{fe} solutions and enriching it with noise, according to Eq.~\eqref{eq:mri_observations}.}
    \item \textcolor{myblue}{In case of noisy input data, choose a suitable value for $\xi > 0$.}
    \item Solve Sys.~\eqref{eq:pbdw_system_cardiac}: $\mathcal{O}((M+N)^3)$ operations.
    \item Reconstruct via Eq.~\eqref{eq:reconstruction_cardiac}: $\mathcal{O}((M+N))$ operations.
\end{enumerate}



\section{Results}
\label{sec:results}

We validate the proposed \gls{pbdw} framework through progressive test cases on the cardiac mechanics problem defined in Section~\ref{sec:methods}. 
All simulations employ an idealised \gls{lv}. The scar region is created by intersecting the geometry with a sphere (radius $r = \qty{10}{\milli\metre}$ centred at $\bm{x}_0 = (\qty{25}{\milli\metre}, \qty{25}{\milli\metre}, \qty{0}{\milli\metre})$ ) and characterised by tenfold increased stiffness $\alpha_{\mathrm{scar}} = 10\alpha$ to simulate infarcted tissue.

\subsection{Experimental Setup}
For the offline phase, we generate $N_s = 150$ solution snapshots using \gls{lhs} over the parameter space $(p, \alpha) \in [5, 16] \times [0.7884, 0.9636]$ kPa, solving the \gls{ppde} \eqref{eq:cardiac_pde} with our internal \gls{fe} solver \texttt{carpentry} on the \gls{vsc5}. 
\textcolor{myblue}{The non-linear finite elasticity equations are discretised using P1–P0 tetrahedral finite elements with element-by-element static condensation to enforce near-incompressibility via a penalty formulation. The resulting non-linear system is solved using Newton's method, where each linearised system is solved iteratively using GMRES preconditioned with an algebraic multigrid (AMG) preconditioner~\cite{Augustin2016, Karabelas2022}.}
We partition the dataset into $N_{\mathrm{train}} = 100$ snapshots for \gls{rom} construction and $N_{\mathrm{test}} = 50$ for validation.
We assume that the sensors operate in the reference (unloaded) configuration, which corresponds to the end-diastolic state when the ventricle is relaxed and minimally loaded. This represents standard practice in cardiac imaging and biomechanics modelling, where the end-diastolic configuration serves as a practical and reproducible reference state~\cite{Rausch2017}.
The size of $\mathcal L_K$ depends on the dimension of the voxels, and we present results for different configurations.
Out of $\mathcal{L}_K$ Alg.~\ref{alg:minibatchwomp} selects $M$ optimal sensors. We construct the background space using \gls{pod} capturing \qty{99.9}{\percent} of the snapshot energy.
For \gls{womp} we heuristically chose $\beta_\odot = 0.1$ to balance computational time and accuracy.

We examine three scenarios of increasing realism.
\begin{enumerate}
    \item \textbf{Noise-free} (Section~\ref{subsec:noisefree}): Full observability, $\vec{\varepsilon}_{k} = \vec{0}$, volumetric voxels of dimensions $\SI{2}{\milli\metre} \times \SI{2}{\milli\metre} \times \SI{8}{\milli\metre}$, resulting in $|\mathcal L_K| = 3315$.
    \item \textbf{Noisy} (Section~\ref{subsec:noisy}): Gaussian noise with different noise levels, volumetric voxels of dimensions $\SI{2}{\milli\metre} \times \SI{2}{\milli\metre} \times \SI{8}{\milli\metre}$, resulting in $|\mathcal L_K| = 3315$.
    \item \textbf{Sparse and noisy} (Section~\ref{subsec:sparse}): \gls{mri}-like slice sampling with noise.
\end{enumerate}
\subsection{Noise-free reconstructions}
\label{subsec:noisefree}
\begin{figure*}[!htbp]
    \centering
    \includegraphics[width=\linewidth]{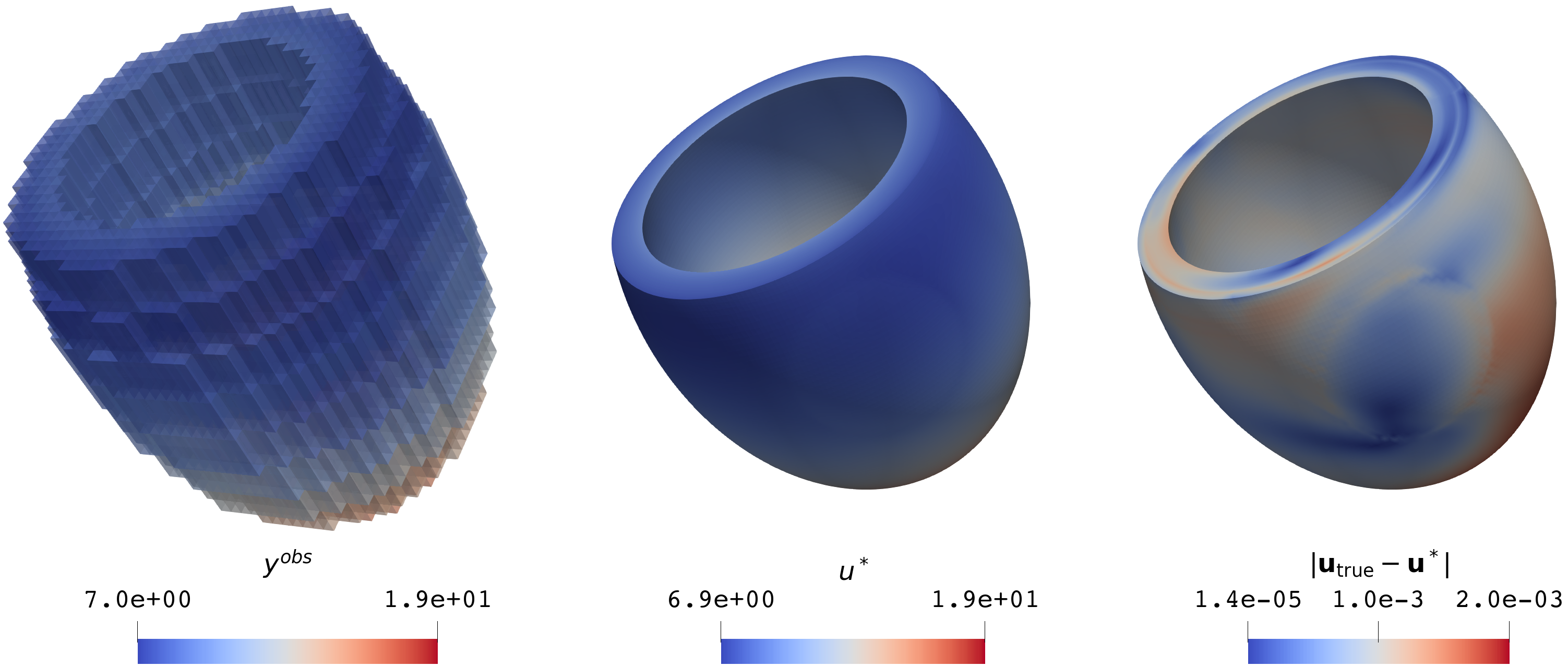}
    \caption{\textcolor{myblue}{Performance of \gls{pbdw} in noiseless test case. Left: norm of the observation vector $\bm{y}^\mathrm{obs}$ in mm; middle: norm of the \gls{pbdw} state reconstruction $\bm{u}^*$ in mm; right: absolute value of the difference between the reconstruction and the ground truth in mm.}}
    \label{fig:noiseless_mag}
\end{figure*}

The first test case is given by a noiseless scenario, i.e. $\vec{\epsilon}_k=0, \quad\forall k=1,\ldots,K$ in Eq.~\eqref{eq:mri_observations}.
\textcolor{myblue}{In Fig.~\ref{fig:noiseless_mag} we show the norm of the observation vector $\bm{y}^\mathrm{obs}$, the \gls{pbdw} state reconstruction $\bm{u}^*=\bm{z}^*+\bm{\eta}^*$ and the absolute value of the difference between reconstruction and ground truth in the reference configuration.}
%
In Tab.~\ref{tab:noiselesstable} we report the relative reconstruction errors and the computational time, highlighting the speed-up achieved with respect to a \gls{fe} solution. Indeed, applying the \gls{rom} by means of \gls{pbdw} almost saves four orders of magnitude in computational time. The timings are consistent, as the architecture is the same. 
Indeed, they both refer to simulations that have been run on the \gls{vsc5} with 128 MPI threads. \textcolor{myblue}{The computational time of the \gls{pbdw} offline stage (snapshot generation, \gls{pod} basis construction, Riesz representer computation, and \gls{ss}) is omitted from this comparison, as it is performed once and subsequently reused for arbitrarily many reconstructions. This offline--online decomposition is the standard paradigm for \gls{rom}-based methods: the offline cost, while non-negligible, is amortized over the many-query scenario that motivates our approach. Once the \gls{rbs} are constructed for a given geometry and parameter range, \gls{pbdw} can be applied to extract high-resolution displacement fields from any compatible \gls{mri} data.}
The reconstruction accuracy is very satisfactory, as even the $L^\infty(\Omega)$-norm returns a relative error $\num{1e-05}$, which means that the absolute error is of order $\num{1e-04}$. 
\begin{table}[ht!]
 \caption{\textcolor{myblue}{From left to right, noiseless relative reconstruction errors computed in the $L^2(\Omega), H^1(\Omega), L^\infty(\Omega)$ norms, respectively, computational time of the \gls{pbdw} online stage and of a full \gls{fe} simulation.}}
    \label{tab:noiselesstable}
    \centering
    \begin{tabular}{ccccc} 
    \toprule
    $\mathrm{err}_{L^2(\Omega)}$ & $\mathrm{err}_{H^1(\Omega)}$ & $\mathrm{err}_{L^\infty(\Omega)}$ & $t_\mathrm{online}$ & $t_\mathrm{FOM}$ \\
    \midrule
    $\num{7.52e-05}$ & $\num{7.75e-05}$ & $\num{9.21e-05}$ & $\qty{2.83e-02}{\second}$ & $\sim\qty{15}{\minute}$\\
    \botrule
    \end{tabular}
\end{table}
On the other hand, one has to account for the \gls{rom} construction.
In Tab.~\ref{tab:wOMPH} we report the computational time of the \gls{womp} \gls{ss} with varying minibatch size $H$. The minibatch approach offers modest speedups, with optimal performance typically achieved for intermediate values of $H$ (between 50-100). Beyond this range, the overhead of orthonormalizing larger sets of vectors offsets the benefits of batch processing.
\begin{figure*}[!htbp]
    \centering
    \includegraphics[width=\linewidth]{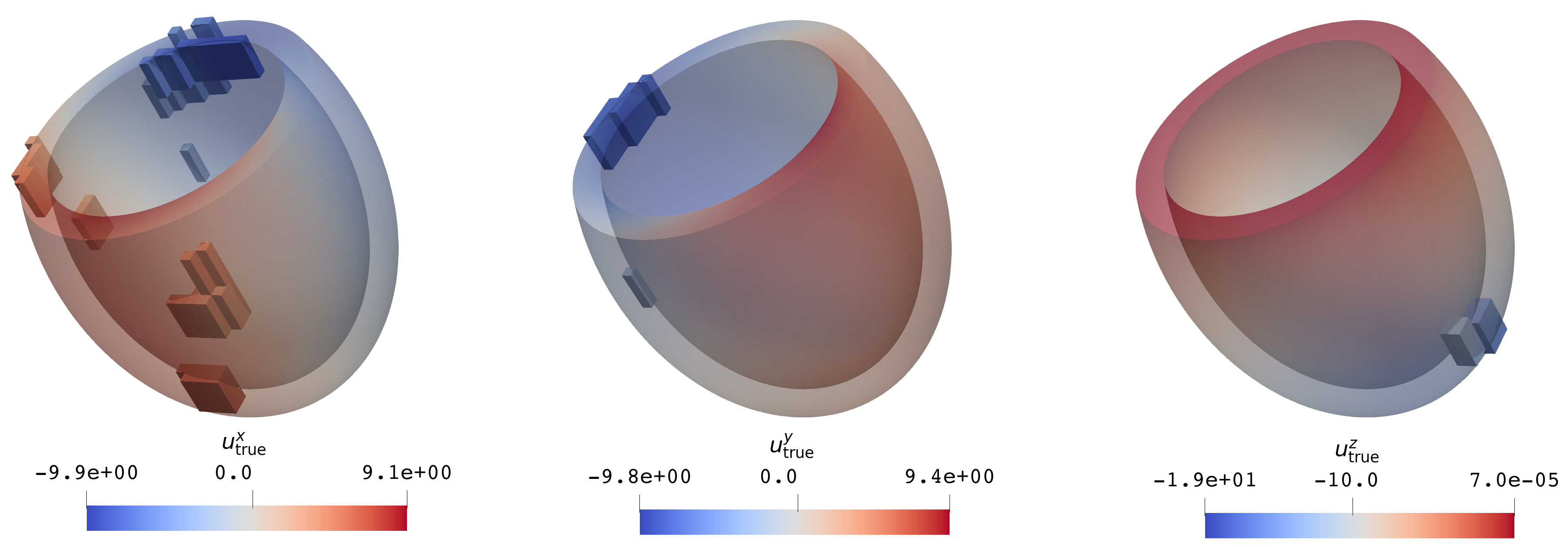}
    \caption{This visualization illustrates the sensors selected by the \gls{womp} algorithm. Each sensor measures the average displacement field within a voxel of fixed dimension. The images display the chosen voxels along with their measured values. For context, the background shows the entire displacement field. Left: $x$-component of the ground truth $\bm{u}_\mathrm{true}$ in mm; middle: $y$-component of the ground truth $\bm{u}_\mathrm{true}$ in mm; right: $z$-component of the ground truth $\bm{u}_\mathrm{true}$ in mm.}
    \label{fig:sensor_selection}
\end{figure*}
Figure~\ref{fig:sensor_selection} visualizes the spatial distribution of sensors selected by \gls{womp} for the noise-free case. Notably, the algorithm preferentially selects sensors on the lateral and septal walls, opposite to the scar region. This counterintuitive pattern likely reflects that the \gls{pod} basis $\mathcal{Z}_N$ already captures the scar's dominant deformation modes, making measurements in complementary regions more informative for reconstruction. The selected voxels show the averaged displacement values within each sensor's observation volume.



%
\begin{table}[ht!]
\caption{Computational time for different minibatch sizes $H$ of \gls{womp}. 
\textcolor{myblue}{From left to right, size of the minibatch, number of degrees of freedom for a \gls{fe} scalar field, number of sensors per component, dimension of $\mathcal{U}_M$, computational time of \gls{womp}.}}
    \label{tab:wOMPH}
    \centering
    \begin{tabular}{ccccc} 
    \toprule
    $H$ & $\sfrac{\mathcal{N}}{3}$ & $\sfrac{N_\mathrm{sens}}{3}$ & $M$ & $t$\\
    \midrule
    \num{1} & \num{148050} & \num{3315} & \num{16} & $\qty{1.17e+02}{\second}$\\
    \num{10} & \num{148050} & \num{3315} & \num{50} & $\qty{3.67e+01}{\second}$\\
    \num{20} & \num{148050} & \num{3315} & \num{100} & $\qty{3.88e+01}{\second}$\\
    \num{50} & \num{148050} & \num{3315} & \num{150} & $\qty{2.59e+01}{\second}$\\
    \num{100} & \num{148050} & \num{3315} & \num{200} & $\qty{2.17e+01}{\second}$\\
    \num{200} & \num{148050} & \num{3315} & \num{400} & $\qty{4.50e+01}{\second}$\\
    \num{500} & \num{148050} & \num{3315} & \num{1000} & $\qty{1.77e+02}{\second}$\\
    \num{1000} & \num{148050} & \num{3315} & \num{1000} & $\qty{2.73e+02}{\second}$\\
    \botrule
    \end{tabular}    
\end{table}
\subsection{Noisy Reconstructions}
\label{subsec:noisy}
When dealing with noise, one quantifies its influence using the \gls{snr}. In general, and also for \gls{mri} applications, the choice of the noise distribution is mainly determined by the \gls{snr} regime, which can be low \gls{snr} (Weak Signal) or high \gls{snr} (Strong Signal).
In the former scenario, the noise follows a Rician distribution \cite{Gudbjartsson1995}. 
This occurs because \gls{mris} are reconstructed from complex-valued data, each containing Gaussian noise, which leads to a Rician-distributed magnitude signal. 
When the true signal is near zero, the Rician distribution appears Rayleigh-distributed.
In the latter case, instead, the Rician distribution converges to a Gaussian distribution \cite{Gudbjartsson1995,CardenasBlanco2008}. 
This happens because the Rician distribution becomes nearly symmetric and resembles a normal distribution as the signal dominates the noise.
Here, we will assume homoscedastic random noise for $\vec \varepsilon_k$ in Eq.~\eqref{eq:mri_observations}, meaning
\begin{equation*}
    \vec \varepsilon_k \overset{iid}{\sim} N(\vec 0, \sigma \mathbf{I}), \quad \forall k=1,\ldots,K.
\end{equation*}
For the value $\sigma$ we will consider two types of variance modelling and show that both yield similar results, demonstrating consistency of the method:
\begin{enumerate}
    \item \emph{Limiting Dispersion (LD)}: Following~\cite{Caforio2024} for a signal $s$ we introduce the LD metric
    \begin{equation*}
        \mathrm{LD} = \frac{3\sigma}{\lVert s\rVert_{\infty}}.
    \end{equation*}
    In our case this translates to
    \begin{equation}
    \label{eq:sigma_ld}
        \mathrm{LD} = \frac{3\sigma}{\underset{\substack{m = 1, \ldots, K \\ i =0,1,2}}{\mathrm{max}}\lvert \vec\ell_{m, i}(\bm u_{\mathrm{true}})\rvert}.
    \end{equation}
    So for a given value of $\mathrm{LD} \in [0,1]$ we can solve for $\sigma_\mathrm{LD}$.
    \item \emph{Sensor standard deviation} (SSTD): Following~\cite{Taddei_phdthesis,Maday2014} for a given signal we can also define the metric
    \begin{equation}
    \label{eq:sigma_snr}
        \widetilde{\mathrm{SNR}} = \underset{\substack{m=1,\ldots K\\i=0,1,2}}{\mathrm{std}}(\vec \ell_{m,i}(\bm u_\mathrm{true})) \frac{1}{\sigma},
    \end{equation}
    yielding a value of $\sigma_{\widetilde{\mathrm{SNR}}}$ for each true solution $\bm u_\mathrm{true}$.
\end{enumerate}

%
    %
    %
    %
    %
    %
    %
%
Note that $\mathrm{LD}, \widetilde{\mathrm{SNR}}\in[0,1]$. 
\gls{pbdw} can cope with noise by tuning the parameter $\xi$ in Sys.~\eqref{eq:pbdw_system_cardiac}. 
In order to find the optimal value for $\xi$, which is problem dependent, a statistical study is carried out by means of K-fold cross validation in Sec.~\ref{sec:kfold}.
The reconstruction errors for the different \gls{snr} definitions are depicted in Tab.~\ref{tab:LDvsSNR}, showing that both definitions deliver comparable results.

\begin{table*}
    \caption{Comparison of \gls{pbdw} performance using different \gls{snr} definitions ($\mathrm{LD}=0.1$ and $\widetilde{\mathrm{SNR}}=0.1$). \textcolor{myblue}{From left to right, \gls{snr} Def.~\eqref{eq:sigma_snr} or \eqref{eq:sigma_ld}, number of degrees of freedom for a \gls{fe} scalar field, number of sensors per component, dimension of $\mathcal{Z}_N$, dimension of $\mathcal{U}_M$, relative reconstruction errors in the $L^2(\Omega), H^1(\Omega), L^\infty(\Omega)$ norms, respectively.}}
    \label{tab:LDvsSNR}
    \centering
    \begin{tabular}{cccccccc} 
    \toprule
    \gls{snr} & $\sfrac{\mathcal{N}}{3}$ & $\sfrac{N_\mathrm{sens}}{3}$ & $N$ & $M$ & $\mathrm{err}_{L^2(\Omega)}$ & $\mathrm{err}_{H^1(\Omega)}$ & $\mathrm{err}_{L^\infty(\Omega)}$\\
    \midrule
    $\mathrm{LD}$ & $\num{148050}$ & $\num{3315}$ & \num{5} & \num{70} & $\num{6.94e-03}$ & $\num{7.11e-03}$ & $\num{8.68e-03}$\\
    \hline
    $\widetilde{\mathrm{SNR}}$ & $\num{148050}$ & $\num{3315}$ & \num{5} & \num{70} & $\num{2.27e-02}$ & $\num{2.32e-02}$ & $\num{2.84e-02}$\\
    \botrule
    \end{tabular}
\end{table*}


\begin{figure*}[!htbp]
    \centering
    \includegraphics[width=\linewidth]{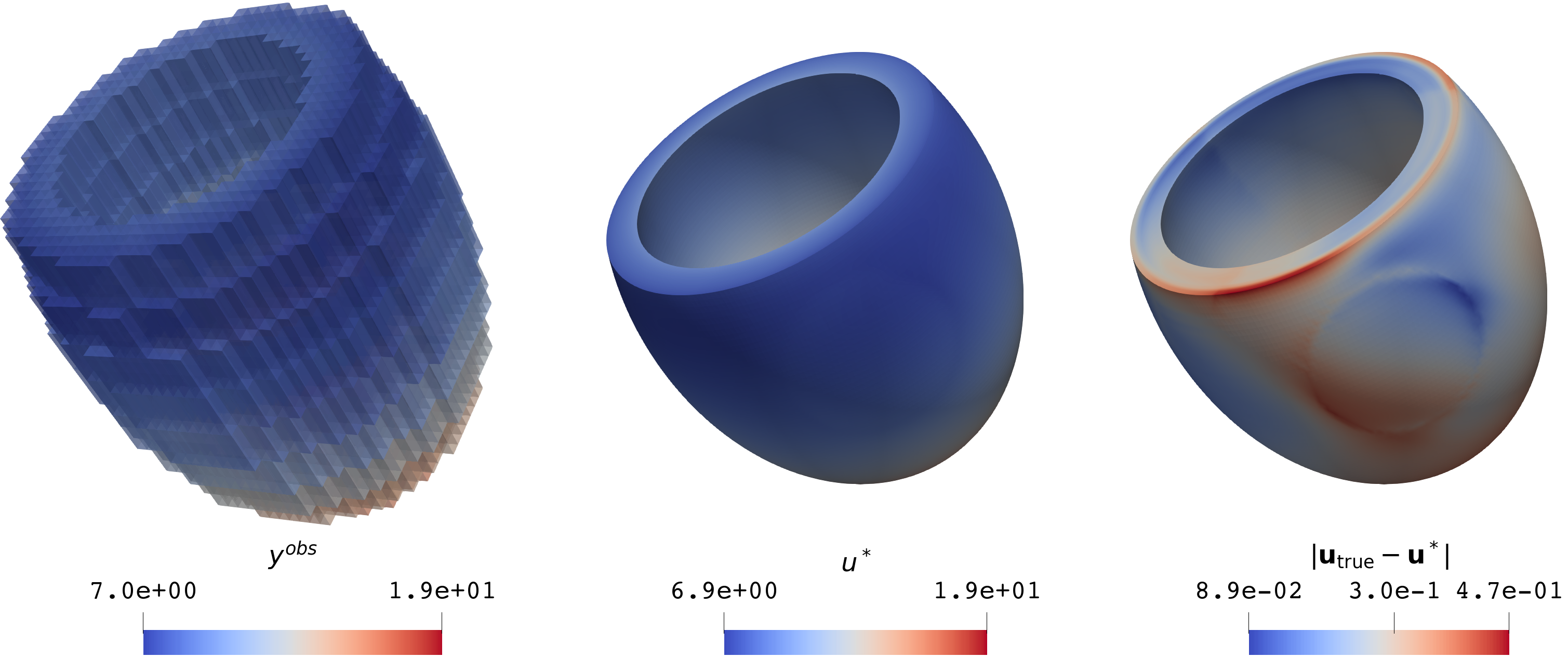}
    \caption{\textcolor{myblue}{Performance of \gls{pbdw} in noisy test case. Left: norm of the observation vector $\bm{y}^\mathrm{obs}$ in mm; middle: norm of the \gls{pbdw} state reconstruction $\bm{u}^*$ in mm; right: absolute value of the difference between the reconstruction and the ground truth in mm.}}
    \label{fig:noisy_mag}
\end{figure*}

We now focus on the second introduced \gls{snr} Def.~\eqref{eq:sigma_snr}.
In particular, we consider $\widetilde{\mathrm{SNR}}=\num{0.1}$ which means that the noise amounts to $\qty{10}{\percent}$ of the signal we observe. 
This is a typical scenario in applications and in particular for \gls{mris}. 
We show multiple plots confirming the quality of reconstruction of \gls{pbdw}: \textcolor{myblue}{Fig.~\ref{fig:noisy_mag} depicts the magnitude of the observation vector $\bm{y}^\mathrm{obs}$ on the left, the \gls{pbdw} reconstruction $\bm{u}^*=\bm{z}^*+\bm{\eta}^*$ in the middle, showing also on the right the absolute value of the difference between the reconstruction and the ground truth $\bm{u}_\mathrm{true}$.}

\begin{figure*}[!htbp]
    \centering
    \includegraphics[width=\linewidth]{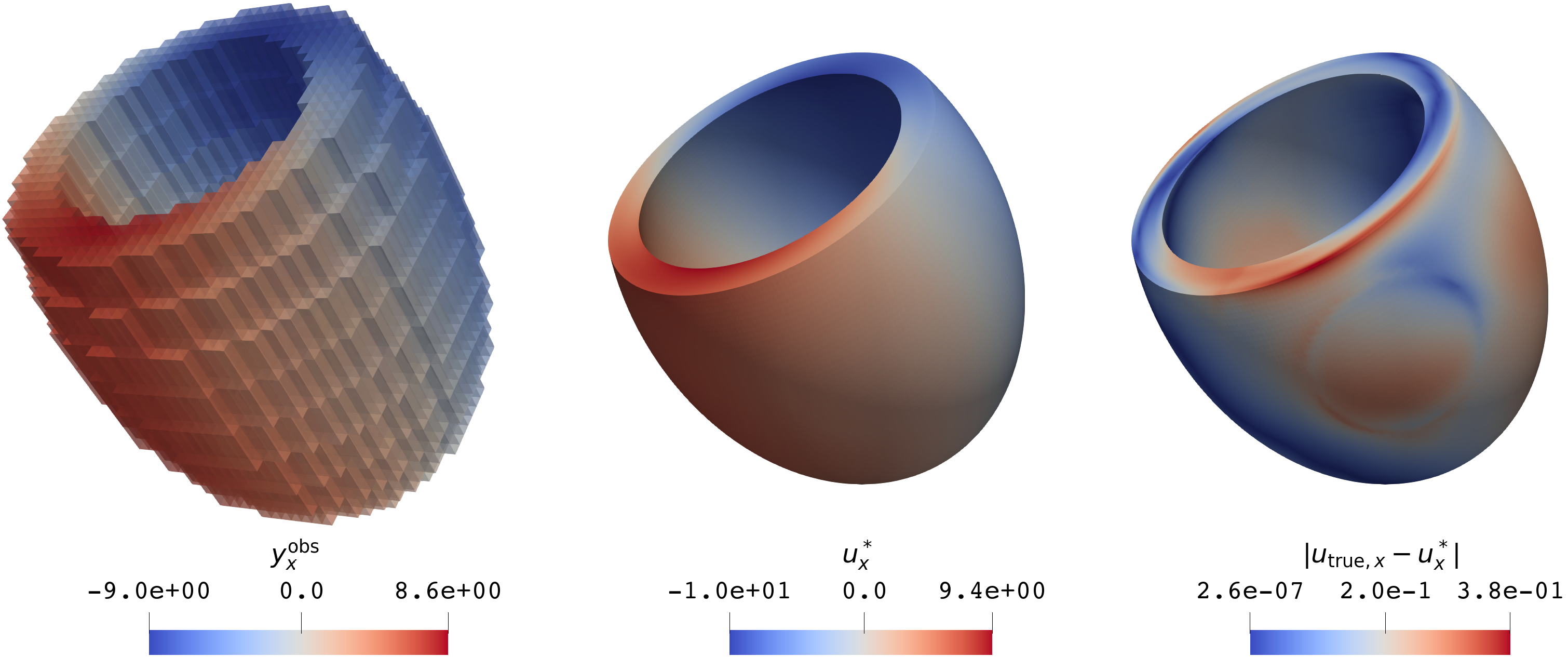}
    \caption{\textcolor{myblue}{Performance of \gls{pbdw} in noisy test case. Left: observation vector $\bm{y}^\mathrm{obs}$, restricted to the $x$-component, in mm; middle: $x$-component of the \gls{pbdw} state reconstruction $\bm{u}^*$ in mm; right: absolute value of the difference between the reconstruction and the ground truth in mm.}}
    \label{fig:noisy_x}
\end{figure*}

\begin{figure*}[!htbp]
    \centering
    \includegraphics[width=\linewidth]{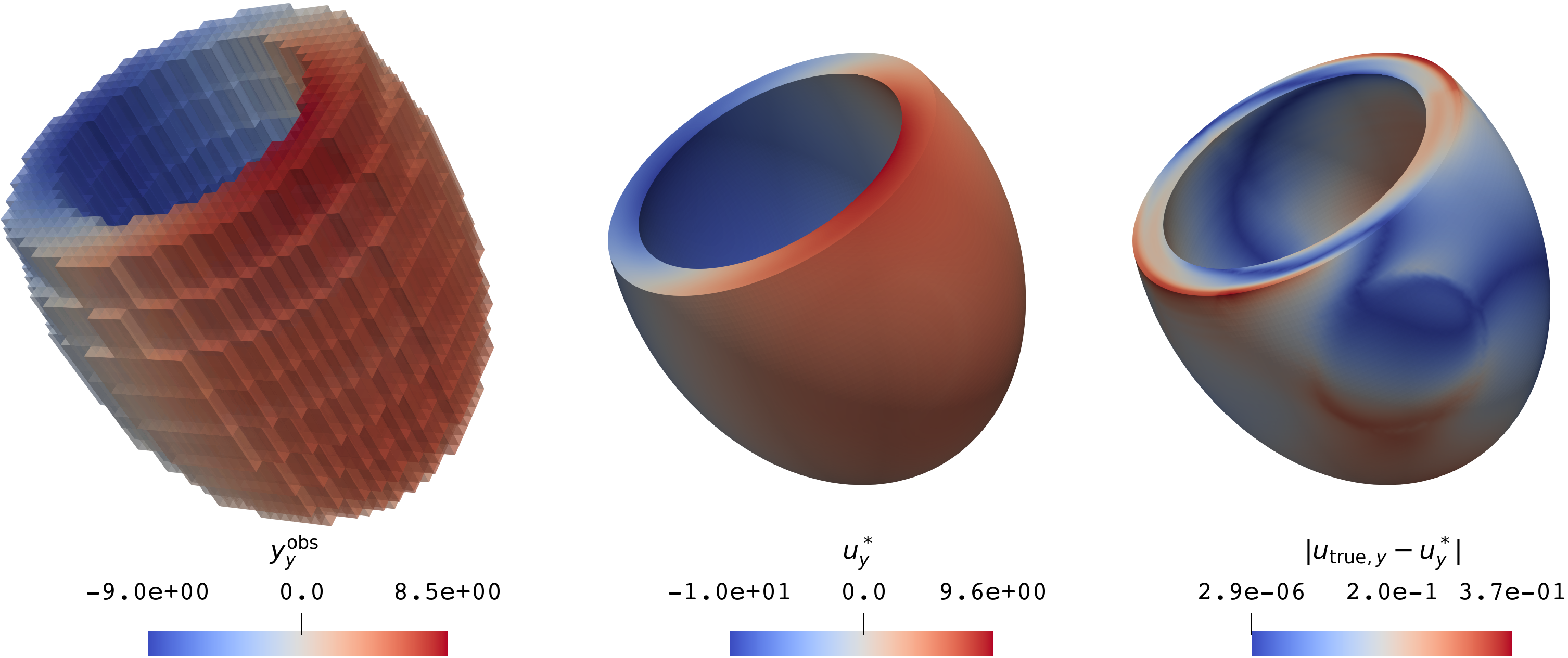}
    \caption{\textcolor{myblue}{Performance of \gls{pbdw} in noisy test case. Left: observation vector $\bm{y}^\mathrm{obs}$, restricted to the $y$-component, in mm; middle: $y$-component of the \gls{pbdw} state reconstruction $\bm{u}^*$ in mm; right: absolute value of the difference between the reconstruction and the ground truth in mm.}}
    \label{fig:noise_y}
\end{figure*}

\begin{figure*}[!htbp]
    \centering
    \includegraphics[width=\linewidth]{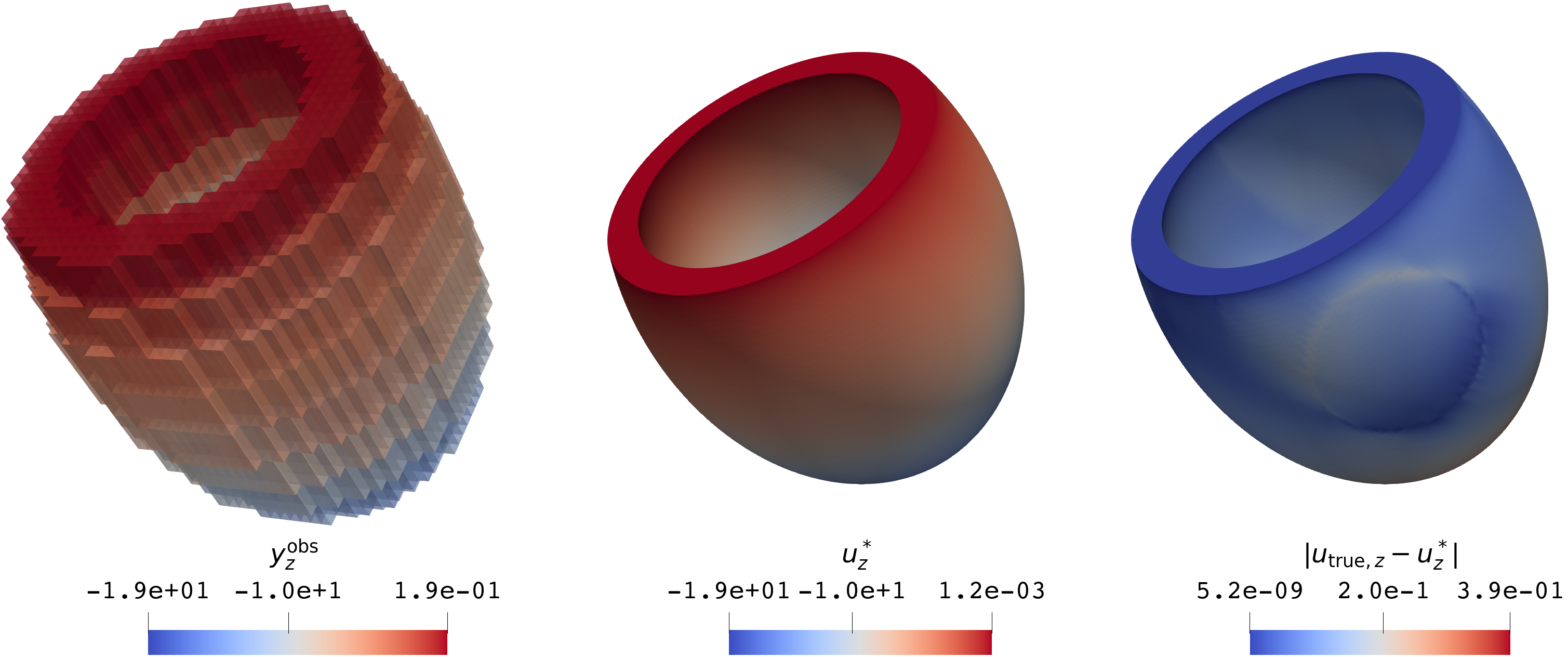}
    \caption{\textcolor{myblue}{Performance of \gls{pbdw} in noisy test case. Left: observation vector $\bm{y}^\mathrm{obs}$, restricted to the $z$-component, in mm; middle: $z$-component of the \gls{pbdw} state reconstruction $\bm{u}^*$ in mm; right: absolute value of the difference between the reconstruction and the ground truth in mm.}}
    \label{fig:noise_z}
\end{figure*}

Figure~\ref{fig:noisy_mag} reveals that the reconstruction error concentrates along the endocardial surface, particularly near the scar region where the tissue stiffness discontinuity creates complex deformation patterns. 
The component-wise analysis (Figures~\ref{fig:noisy_x}-\ref{fig:noise_z}) shows that reconstruction errors are largest for the $x$-component, with maximum errors reaching approximately $\qty{0.4}{\milli\metre}$ in localised regions. The $z$-component exhibits lower errors, consistent with the primarily radial and circumferential deformation during inflation. 
Tab.~\ref{tab:noisytable} shows the relative errors of the \gls{pbdw} reconstructions in different norms across the whole geometry. With the maximum displacement magnitude of approximately $\qty{20}{\milli\metre}$ (Fig.~\ref{fig:noisy_mag}) and relative $L^\infty(\Omega)$ error of $10^{-2}$, the maximum reconstruction error remains below $\qty{0.3}{\milli\metre}$. This demonstrates effective noise filtering, as the reconstruction error is significantly smaller than the $\qty{2}{\milli\metre}$ noise level imposed by $\widetilde{\mathrm{SNR}}=0.1$.

\begin{table}[htbp!]
    \caption{\textcolor{myblue}{From left to right, noisy ($\widetilde{\mathrm{SNR}}=0.1$) relative reconstruction errors computed in the $L^2(\Omega), H^1(\Omega), L^\infty(\Omega)$ norms, respectively, computational time of the \gls{pbdw} online stage and of a full \gls{fe} simulation.}}
    \label{tab:noisytable}
    \begin{tabular}{ccccc} 
    \toprule
    $\mathrm{err}_{L^2(\Omega)}$ & $\mathrm{err}_{H^1(\Omega)}$ & $\mathrm{err}_{L^\infty(\Omega)}$ & $t_\mathrm{online}$ & $t_\mathrm{FOM}$ \\
    \midrule
    $\num{2.27e-02}$ & $\num{2.32e-02}$ & $\num{2.83e-02}$ & $\qty{2.21e-02}{\second}$ & $\sim\qty{15}{\minute}$\\
    \botrule
    \end{tabular}
\end{table}

\subsection{Sparse and Noisy Reconstructions}
\label{subsec:sparse}

\begin{figure*}[!htbp]
    \centering
    \includegraphics[width=\linewidth]{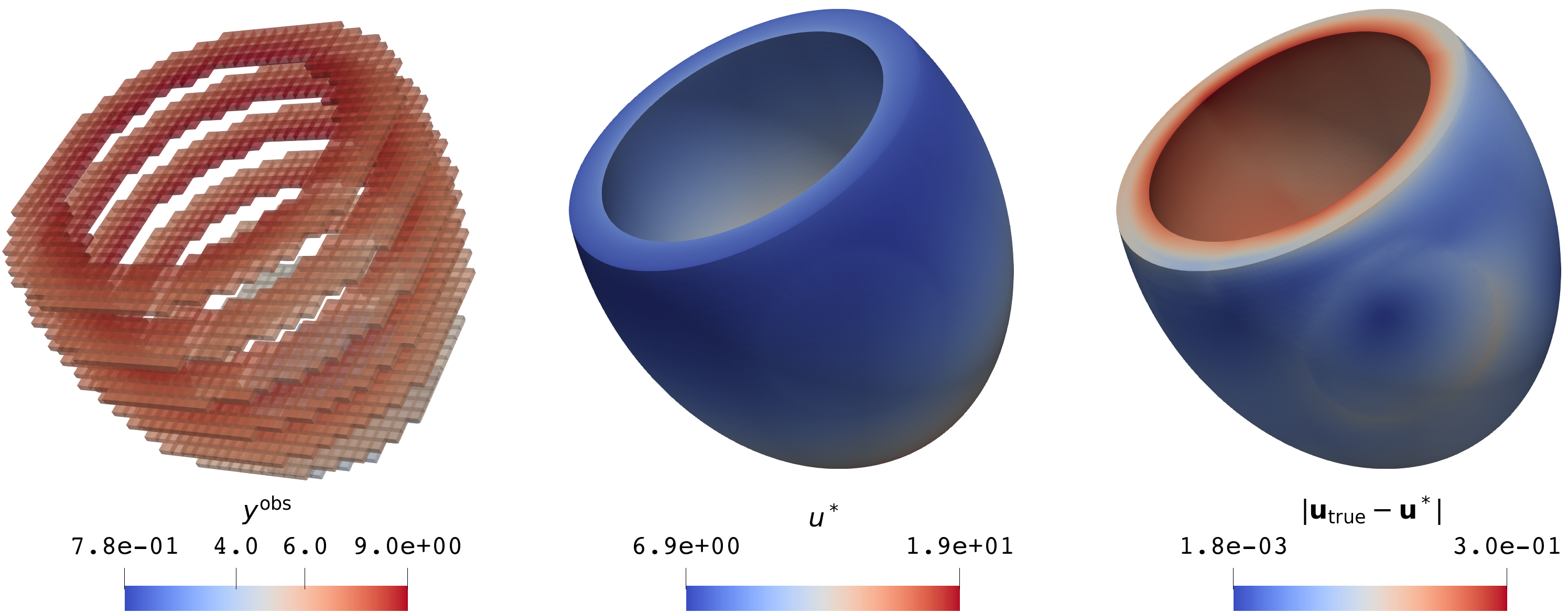}
    \caption{\textcolor{myblue}{Performance of \gls{pbdw} in noisy test case with sparsity. Left: norm of the observation vector $\bm{y}^\mathrm{obs}$ in mm; middle: norm of the \gls{pbdw} state reconstruction $\bm{u}^*$ in mm; right: absolute value of the difference between the reconstruction and the ground truth in mm.}}
    \label{fig:sparsenoisy_mag}
\end{figure*}

In this section we consider sparsity patterns that replicate standard clinical cardiac \gls{mri} protocols.
While modern cardiac \gls{mri} systems can theoretically provide full volumetric coverage with techniques, like DENSE, achieving pixel-by-pixel displacement mapping at \qtyrange[range-phrase = --, range-units = single]{1.5}{2.5}{\milli\metre} resolution~\cite{Aletras1999}, clinical protocols typically employ 2D multi-slice acquisitions due to practical constraints including breath-hold duration, patient comfort, and acquisition time limitations~\cite{Kramer2020,Kramer2013}.
Standard clinical practice involves acquiring \numrange[range-phrase = --]{10}{15} short-axis slices perpendicular to the apicobasal direction (taken as the $z$-direction), with gaps between slices to balance spatial coverage against acquisition time~\cite{Chan2010,Kramer2020}.
We model the $z$-sparsity according to a pattern that tries to replicate the common resolution of an \gls{mr} device\footnote{While techniques like 3D DENSE can provide isotropic resolution throughout the myocardium, such protocols require significantly longer acquisition times and are primarily used in research settings rather than routine clinical practice.}. 
We consider two options:
\begin{itemize}
    \item A pattern 1:8, where we observe a slice of height $h_z=\qty{1}{\milli\metre}$ every $\qty{8}{\milli\metre}$, corresponding to 8 slices;
    \item A pattern 1:4 where we observe a slice of height $h_z=\qty{2}{\milli\metre}$, every $\qty{8}{\milli\metre}$, corresponding to 4 slices.
\end{itemize}
Additionally, each \gls{mr} measurement, performed on the volume $\Omega_k$ within the slice, loses the information about the vectorial field in the $z$-direction. The conjecture is that either the vector is projected onto the $(x,y)$-plane or it is averaged over $h_z$. In order to take into account this additional constraint, we completely discard the sensors acting on the $z$-component of the displacement field $\bm{u}$. In other words, we remove the linear functionals in Eq.~\eqref{eq:mri_functionals} corresponding to $i=3$.
These patterns mimic common clinical acquisition strategies where:
\begin{itemize}
    \item Clinical protocols typically acquire slices with \qtyrange[range-phrase=--, range-units = single]{8}{10}{\milli\metre} thickness and \qtyrange[range-phrase=--, range-units = single]{0}{2}{\milli\metre} gaps~\cite{Kramer2013,Chan2010}.
    \item The trade-off between spatial coverage and patient factors necessitates sparse sampling.
\end{itemize}
%
In what follows, we present the final results by integrating the sparsity pattern and making \gls{pbdw} agnostic of the $z$-component of the vectorial field. 

\begin{figure*}[!htbp]
    \centering
    \includegraphics[width=\linewidth]{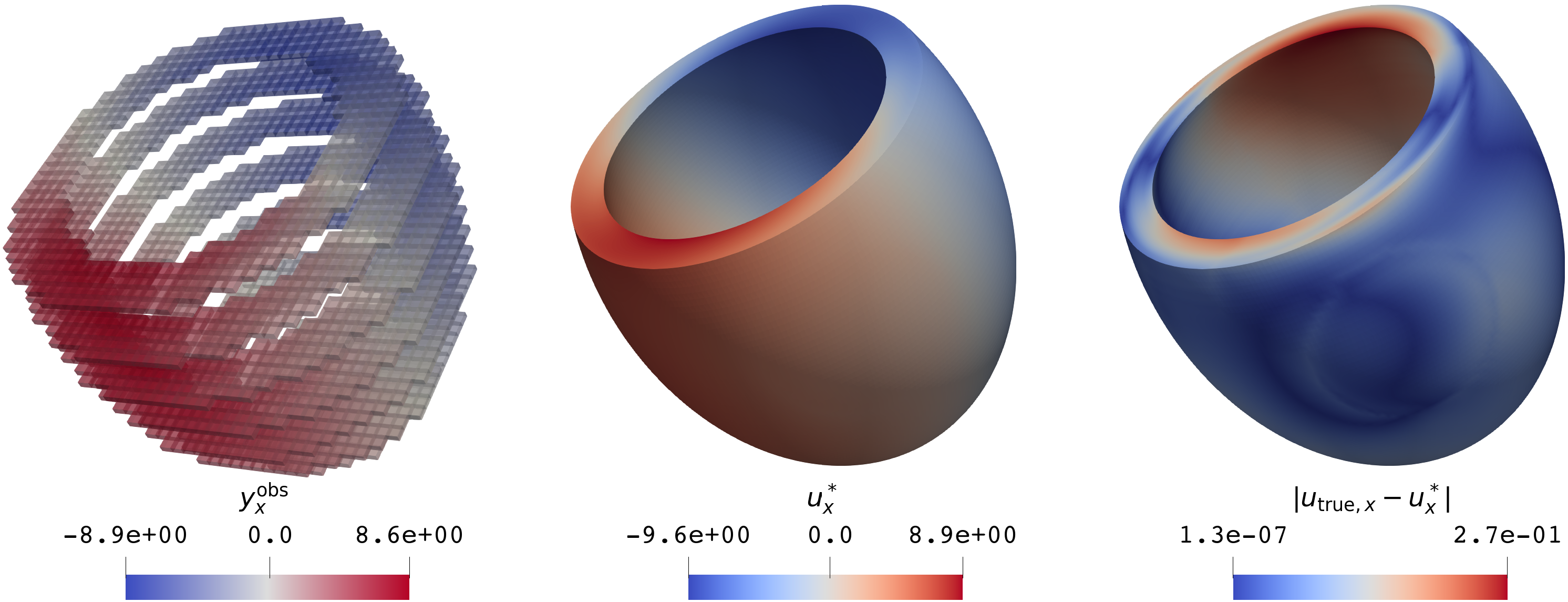}
    \caption{\textcolor{myblue}{Performance of \gls{pbdw} in noisy test case with sparsity. Left: observation vector $\bm{y}^\mathrm{obs}$, restricted to the $x$-component, in mm; middle: $x$-component of the \gls{pbdw} state reconstruction $\bm{u}^*$ in mm; right: absolute value of the difference between the reconstruction and the ground truth in mm.}}
    \label{fig:sparsenoisy_x}
\end{figure*}
\begin{figure*}[!htbp]
    \centering
    \includegraphics[width=\linewidth]{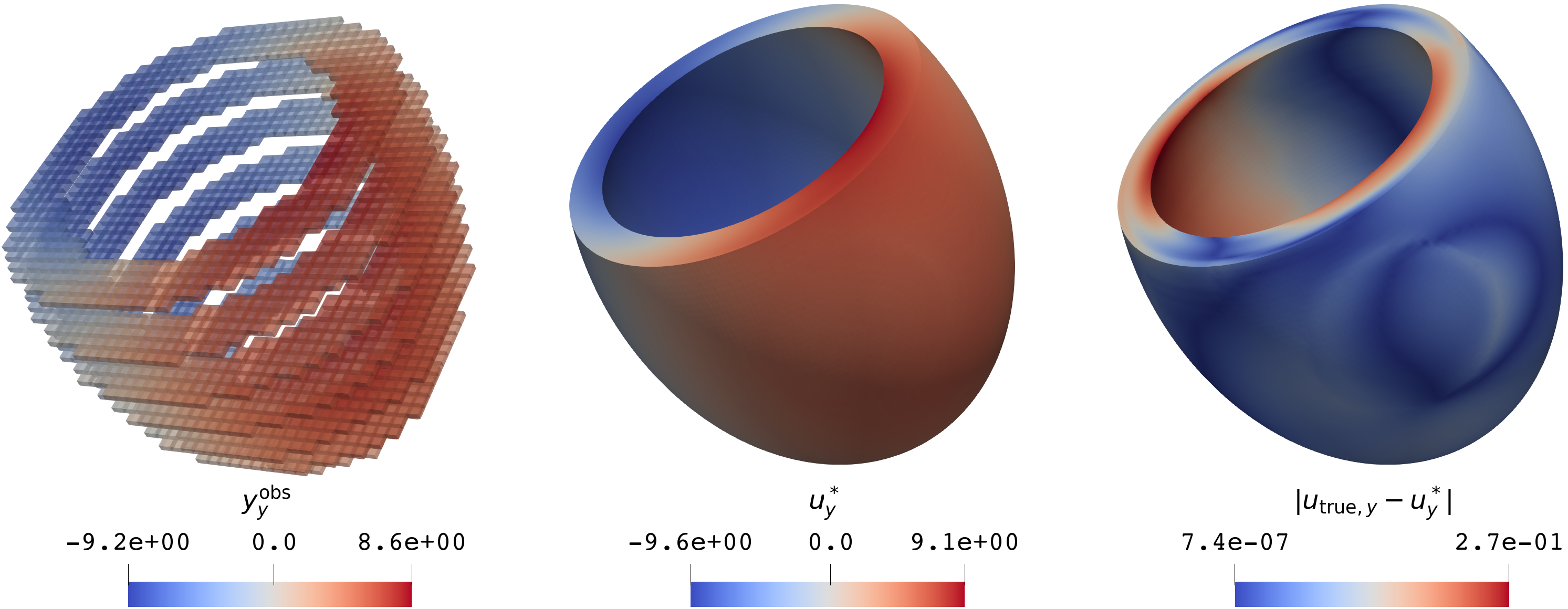}
    \caption{\textcolor{myblue}{Performance of \gls{pbdw} in noisy test case with sparsity. Left: observation vector $\bm{y}^\mathrm{obs}$, restricted to the $y$-component, in mm; middle: $y$-component of the \gls{pbdw} state reconstruction $\bm{u}^*$ in mm; right: absolute value of the difference between the reconstruction and the ground truth in mm.}}
    \label{fig:sparsenoisy_y}
\end{figure*}
\begin{figure*}[!htbp]
    \centering
    \includegraphics[width=0.7\textwidth]{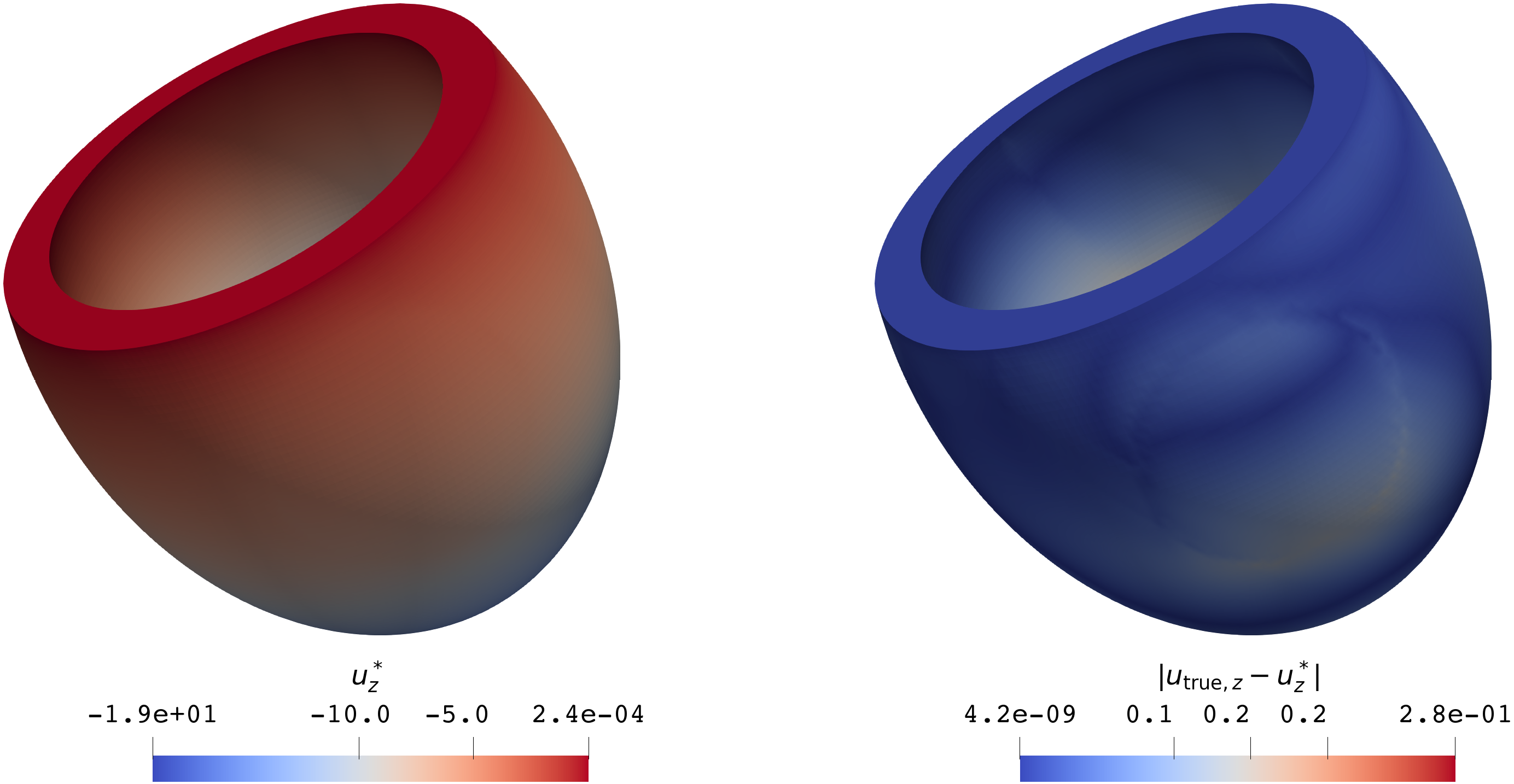}
    \caption{\textcolor{myblue}{Performance of \gls{pbdw} in noisy test case with sparsity. Left: $z$-component of the \gls{pbdw} state reconstruction $\bm{u}^*$ in mm; right: absolute value of the difference between the reconstruction and the ground truth in mm.}}
    \label{fig:sparsenoisy_z}
\end{figure*}

\begin{table}[ht!]
    \caption{\textcolor{myblue}{From left to right, relative reconstruction errors, in presence of noise ($\widetilde{\mathrm{SNR}}=0.1$) and sparsity (sparsity pattern 1:8), computed in the $L^2(\Omega), H^1(\Omega), L^\infty(\Omega)$ norms, respectively, computational time of the \gls{pbdw} online stage and of a full \gls{fe} simulation.}}
    \label{tab:sparsenoisytable}
    \begin{tabular}{ccccc}
    \toprule
    $\mathrm{err}_{L^2(\Omega)}$ & $\mathrm{err}_{H^1(\Omega)}$ & $\mathrm{err}_{L^\infty(\Omega)}$ & $t_\mathrm{online}$ & $t_\mathrm{FOM}$ \\
    \midrule
    $\num{1.74e-02}$ & $\num{1.77e-02}$ & $\num{2.16e-02}$ & $\qty{3.67e-02}{\second}$ & $\sim \qty{15}{\minute}$\\
    \botrule
    \end{tabular}
\end{table}

A comparison of the accuracy of the reconstructions is reported in Tab.~\ref{tab:fullVSsparse} for the two options that we have considered. 
\gls{womp} with minibatch $H=10$ is adopted as \gls{ss} algorithm. 
Moreover, sparsity in the $z$-direction is introduced, according to the 1:8 pattern or the 1:4 pattern. 
For each pattern, the first line refers to a reconstruction that employs sensors acting on all components of the displacement $\bm{u}$. 
In the second line, instead, we report the results where further sparsity is achieved by providing \gls{pbdw} with linear functionals acting on the $x,y$-components only.  
The accuracy, given in terms of relative reconstruction error for a single test case, is the same, or even higher, when  sparser data are given.
\textcolor{myblue}{Figures~\ref{fig:sparsenoisy_mag}-\ref{fig:sparsenoisy_x}-\ref{fig:sparsenoisy_y}-\ref{fig:sparsenoisy_z} show the results in presence of sparsity (sparsity pattern 1:8) and noise ($\widetilde{\mathrm{SNR}}=0.1$) for the magnitude and each component. On the left the observation vector $\bm{y}^\mathrm{obs}$, in the middle the \gls{pbdw} reconstruction $\bm{u}^*$ and on the right the absolute difference between the reconstruction and the ground truth $\bm{u}_\mathrm{true}$. Note that Fig.~\ref{fig:sparsenoisy_z} displays only the $z$-component of \gls{pbdw} reconstruction  and the absolute value of the difference between reconstruction and ground truth. Indeed, we assume to have no sensors acting along the $z$-component, so the observation vector visualization is absent. In Tab.~\ref{tab:sparsenoisytable} the relative reconstruction errors, in multiple norms, are reported, together with the computational time.}

\begin{table*}[ht!]
    \caption{\textcolor{myblue}{Relative reconstruction errors with different sparsity patterns: $h_z=\qty{1}{\milli\metre}, h_z=\qty{2}{\milli\metre}$. 
    From left to right, sparsity pattern, minibatch size, number of degrees of freedom for a \gls{fe} scalar field, total number of sensors, dimension of $\mathcal{Z}_N$, dimension of $\mathcal{U}_M$, relative reconstruction errors in the $L^2(\Omega), H^1(\Omega), L^\infty(\Omega)$ norms, respectively, computational time of the \gls{pbdw} online stage.}
    All runs refer to a sparse coverage in the $z$-direction achieved by considering either a ratio of 1:8 voxels or of 1:4. 
    For each scenario \gls{pbdw} exploits sensors acting on all components (1st row) or only $x,y$ (2nd rows).
    These errors refer to the reconstruction of one displacement only, the same for all, referring to data in presence of noise according to $\widetilde{\mathrm{SNR}}=0.1$.}
    \label{tab:fullVSsparse}
    \centering
    \begin{tabular}{ccccccccc} 
    \toprule
    $z$-sparsity & $H$ & \textcolor{myblue}{$\sfrac{\mathcal{N}}{3}$} & $N_\mathrm{sens}$ & $M$ & $\mathrm{err}_{L^2(\Omega)}$ & $\mathrm{err}_{H^1(\Omega)}$ & $\mathrm{err}_{L^\infty(\Omega)}$ & $t_\mathrm{online}$ \\
    \midrule
    1:8 & 10 & \textcolor{myblue}{$\num{148050}$} & $\num{2202}\cdot3$ & $\num{70}$ & $\num{2.41e-02}$ & $\num{2.47e-02}$ & $\num{3.17e-02}$ & $\qty{2.07e-02}{\second}$\\
    1:8 & 10 & \textcolor{myblue}{$\num{148050}$} & $\num{2202}\cdot2$ & $\num{80}$ & $\num{1.74e-02}$ & $\num{1.77e-02}$ & $\num{2.16e-02}$ & $\qty{3.67e-02}{\second}$\\
    1:4 & 10 & \textcolor{myblue}{$\num{148050}$} & $\num{2269}\cdot3$ & $\num{70}$ & $\num{2.41e-02}$ & $\num{2.46e-02}$ & $\num{3.04e-02}$ & $\qty{2.28e-02}{\second}$\\
    1:4 & 10 & \textcolor{myblue}{$\num{148050}$} & $\num{2269}\cdot2$ & $\num{70}$ & $\num{1.99e-02}$ & $\num{2.03e-02}$ & $\num{2.33e-02}$ & $\qty{3.94e-02}{\second}$\\   
    \botrule
    \end{tabular}
\end{table*}

\subsection{K-Fold Cross Validation}
\label{sec:kfold}

\begin{figure*}[htbp]
    \centering
    \begin{subfigure}[t]{0.48\textwidth}
        \centering
        \includegraphics[width=\textwidth]{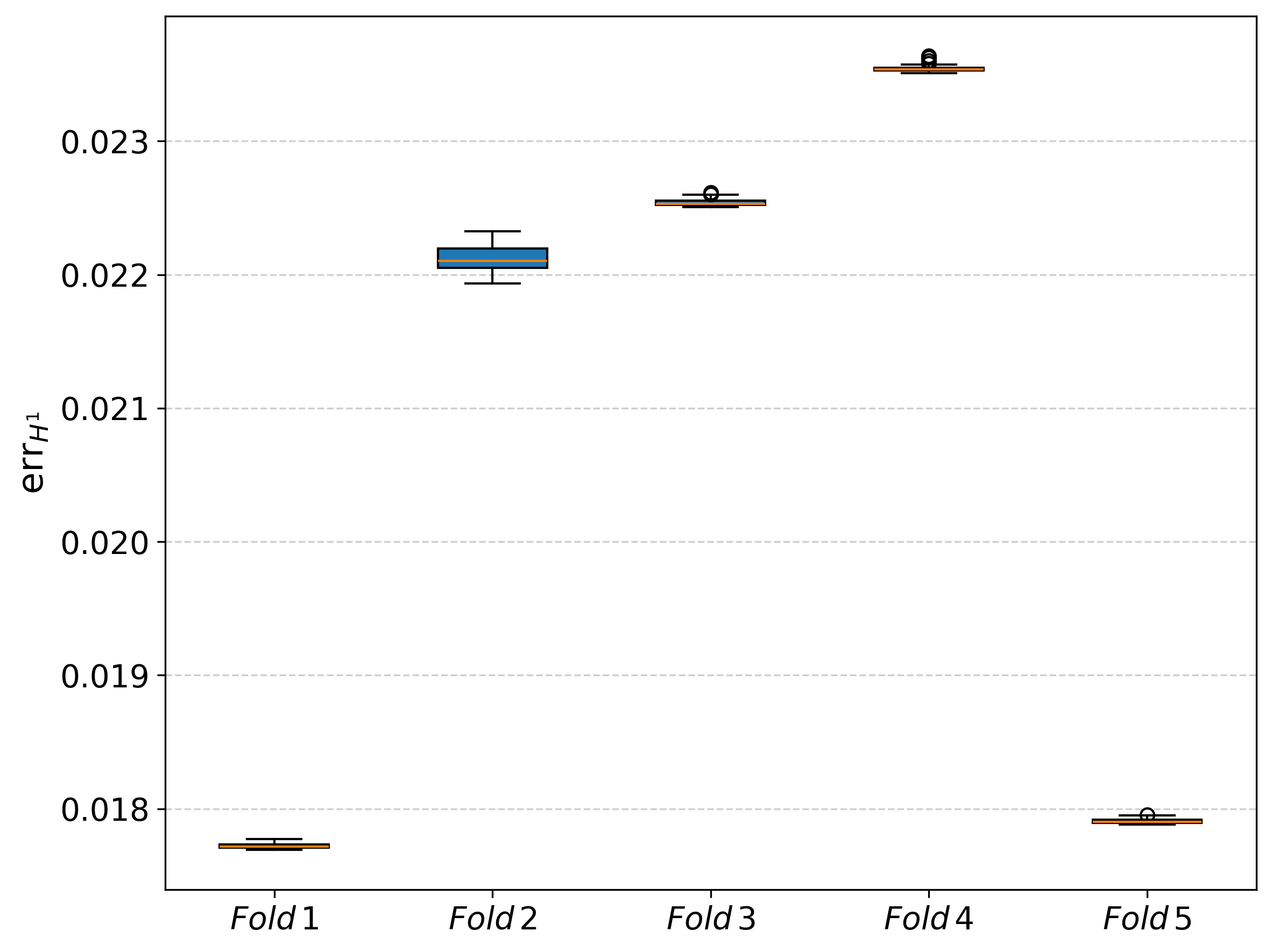}
        \caption{Distribution of the relative reconstruction errors in $H^1(\Omega)$ referring to different splits of the 150 available snapshots. The ratio (100 for the \gls{rb} construction and 50 for testing polluted with noise) is kept fixed for all different folds.}
        \label{fig:5fold}
    \end{subfigure}
    \hfill
    \begin{subfigure}[t]{0.48\textwidth}
        \centering
        \includegraphics[width=\textwidth]{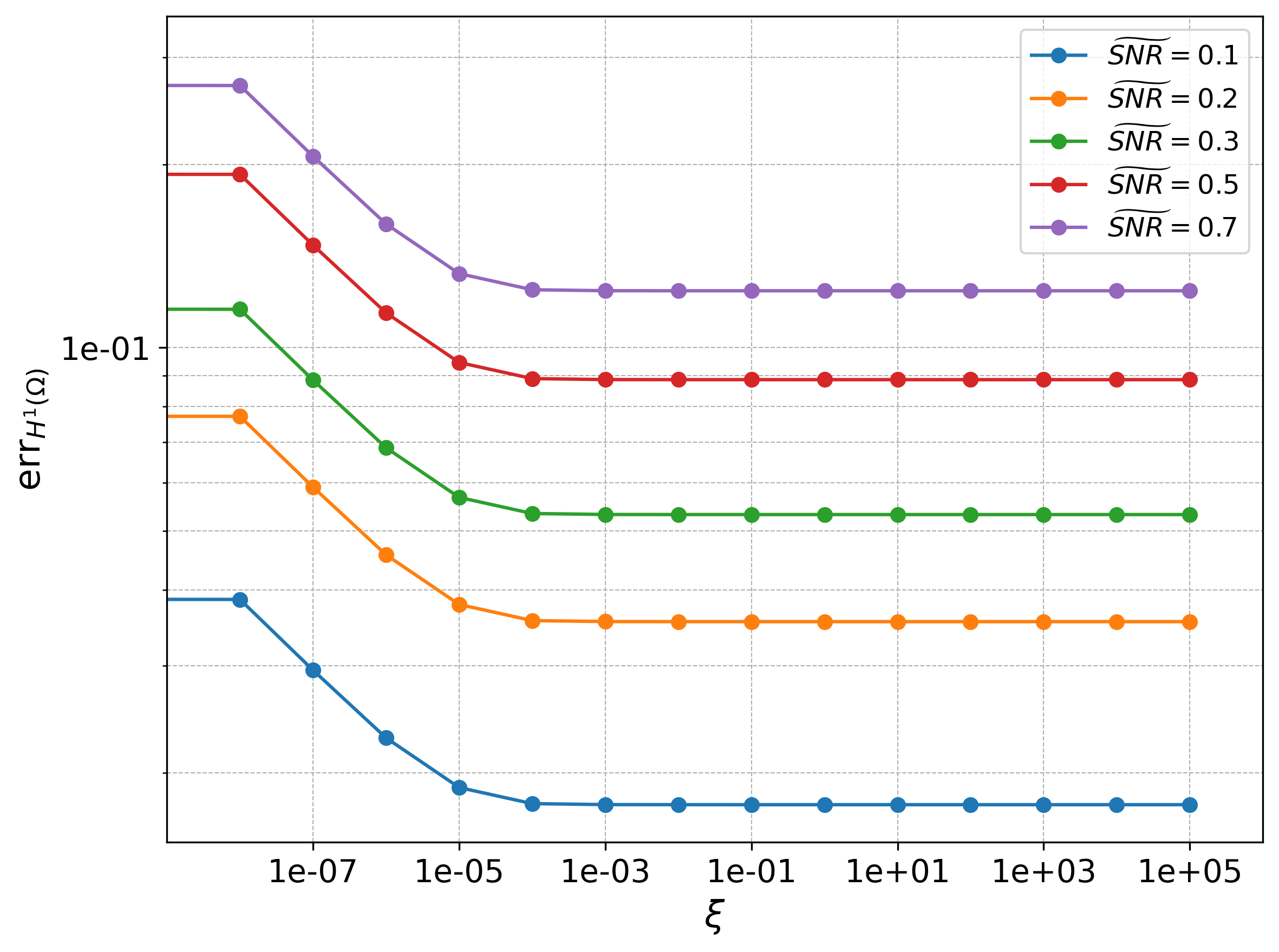}
        \caption{Trend of the relative reconstruction error in $H^1(\Omega)$ for varying $\xi$. Each line refers to data enriched with noise of different $\widetilde{\mathrm{SNR}}$.}
        \label{fig:xi_trend}
    \end{subfigure}
    \caption{Result of $K$-fold cross-validation.}
    \label{fig:5fold_total}
\end{figure*}

For estimating an optimal value of $\xi$, we perform a sensitivity analysis on the hyperparameter $\xi$.
First, to ensure that we are not biased against the constructed \gls{rb}, we carry out a 5-fold cross-validation for a fixed $\xi=30$.
We exploit a data set consisting of 150 snapshots, randomly selecting 100 to construct the \gls{rbs} and 50 as a testing set. 
In particular, each solution in the test set is polluted with noise, according to $\widetilde{\mathrm{SNR}}=0.1$ and sparsified, as explained in Sect.~\ref{subsec:sparse}. 
The distribution of the relative $H^1(\Omega)$ reconstruction errors, reported in Fig.~\ref{fig:5fold}, is very similar for the different 5 folds. 
Thus, we conclude the independence of the reconstructions of the \gls{pod}. 
Next, we sample 15 equally spaced values for $\xi\in [0, 10^5]$
, spanning a sufficiently wide range to assess the effect on the quality of the reconstructions. 
In Fig.~\ref{fig:xi_trend} we report the trend of the relative reconstruction errors in $H^1(\Omega)$ for different values of $\xi$. 
The presence of a drop in errors when $\xi$ takes nonzero values proves that $\xi$ has a sensible effect when the reconstruction deals with noisy data. 
However, in this case, as an improvement, one can only halve the reconstruction error by adopting an optimal value for $\xi$. 
This is related to the variability displayed by the snapshots, which, in turn, comes from the physics described by the \gls{ppde}~\eqref{eq:cardiac_pde} and the variability of its coefficients.

\subsection{\textcolor{myblue}{Spectral content}}
\label{sec:spec_content}

\begin{figure}[!htbp]
    \centering
    \includegraphics[width=\columnwidth]{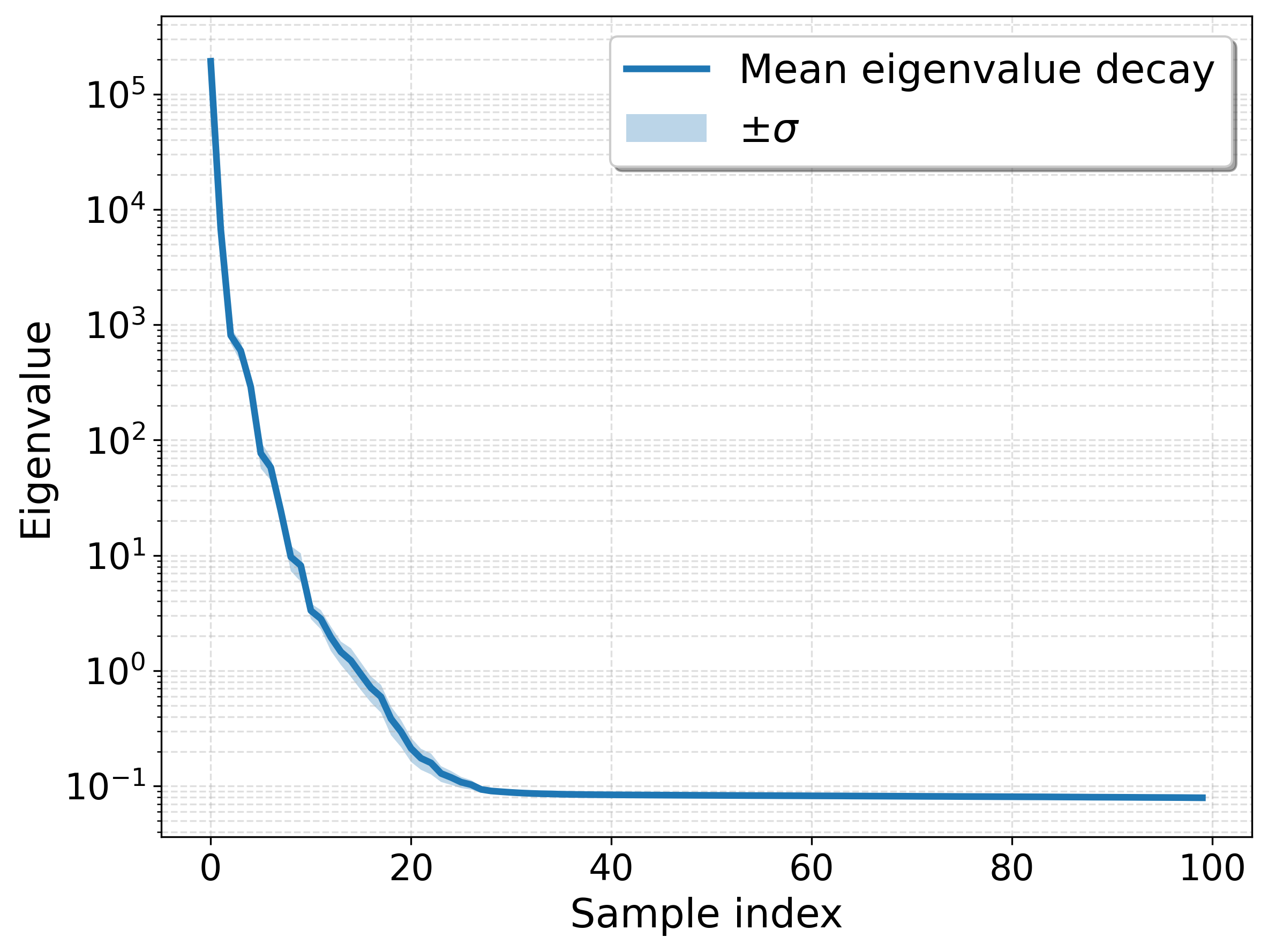}
    \caption{\textcolor{myblue}{Distribution of the eigenvalues decay of the snapshot matrices for $N_\mathrm{samples}=100,\ldots,300$. The minimal deviation from the mean decay indicates that the dominant spectral content is already captured with the smallest dataset, $i.e.$ 100 snapshots.}}
    \label{fig:eigs}
\end{figure}

\textcolor{myblue}{Figure~\ref{fig:eigs} shows the eigenvalue decay of the snapshot matrices for datasets ranging from $N_\mathrm{samples}=100$ to $300$ in steps of $25$. All curves exhibit nearly identical decay profiles: rapid attenuation of the leading modes followed by a long tail of negligible values, with minimal deviation from the mean. This convergence demonstrates that the dominant subspace is already fully captured with $100$ snapshots---increasing the dataset size does not introduce additional energetic modes. The effective rank of the snapshot set is therefore essentially independent of $N_\mathrm{samples}$ in the tested range, confirming that $100$ snapshots suffice to represent the solution manifold.}

\textcolor{myblue}{Having established that the snapshot basis adequately captures the solution manifold for the nominal parameter configuration, we now examine reconstruction performance when the true material parameters deviate from those used in training.}

\subsection{\textcolor{myblue}{Reconstruction in presence of model error}}
\label{sec:bkerr}

\textcolor{myblue}{As noted in Section~\ref{subsec:pbdw_framework}, we fixed the Guccione anisotropy parameters during training. A known limitation of \gls{pbdw} is its sensitivity to model mismatch: when the true solution lies outside the span of the background space $\mathcal{Z}_N$, reconstruction accuracy degrades proportionally to this approximation deficit~\cite{Maday2014,Gong2019}. We now quantify this sensitivity by generating test cases where the anisotropy parameters $b_f$, $b_t$, $b_{fs}$ differ from their training values, while $(p, \alpha)$ remain within the training distribution.}

\textcolor{myblue}{We keep $\kappa$ fixed throughout, as the bulk modulus primarily enforces near-incompressibility and has minimal influence on displacement compared to stiffness~\cite{Liu2021}. In contrast, the anisotropy parameters govern the directional mechanical response and can vary substantially across tissue types and disease states~\cite{caforio2021coupling,Augustin2016,marx2021efficient}.}
\textcolor{myblue}{Rather than perturbing $b_f$, $b_t$, and $b_{fs}$ independently, which may yield physically inconsistent combinations, we adopt an anisotropy ratio formulation~\cite{Xi2013, Nasopoulou2017}. Defining the total anisotropy $\beta = b_f + b_t + b_{fs}$, we introduce the ratios
\begin{equation}
      r_f = \frac{b_f}{\beta}, \qquad r_t = \frac{b_t}{\beta}, \qquad r_{fs} = \frac{b_{fs}}{\beta},
\end{equation}
which satisfy the constraint $r_f + r_t + r_{fs} = 1$. Given $\beta$ and any two ratios, the Guccione parameters are recovered as $b_f = r_f \beta$, $b_t = r_t \beta$, and $b_{fs} = r_{fs} \beta$. This parameterisation ensures that perturbations remain physically consistent with the anisotropic tissue structure.}

\textcolor{myblue}{The training snapshots were generated with default values $b_f = \num{18.48}$, $b_t = \num{3.58}$, and $b_{fs} = \num{1.627}$, yielding a baseline total anisotropy $\beta \approx 24$ and ratios $r_f \approx \num{0.780}$, $r_t \approx \num{0.151}$, and $r_{fs} \approx \num{0.069}$. For the model mismatch study, we sample perturbations using \gls{lhs} over three parameters: the total anisotropy $\beta \in [21, 26]$, and offsets to $r_f$ and $r_t$ of $\pm 10\%$ relative to their baseline values. Samples are accepted only if the constraints $r_f \geq 0$, $r_t \geq 0$, and $r_{fs} = 1 - r_f - r_t \geq 0$ are satisfied. The variable parameters $(p, \alpha)$ are fixed to values within the training distribution. This experimental design isolates the effect of anisotropy mismatch on reconstruction accuracy.}

\textcolor{myblue}{To characterise reconstruction performance, we distinguish two sources of variability: (i) differences across parameter samples, and (ii) sensitivity to noise realisations within each sample. For each of $N_\mathrm{recon}$ parameter configurations, we perform $N_\mathrm{random}$ reconstructions with independent noise and compute the field-wise mean error $\bar{e}_i$ and standard deviation $\sigma_i$. We then report:
\begin{equation}
    \label{eq:stats}
    \begin{aligned}
        \mu&=\frac{1}{N_\mathrm{recon}}\sum_{i=1}^{N_\mathrm{recon}}\bar{e}_i,\\
        \sigma_\mathrm{across}&=\mathrm{std}\big(\{\bar{e}_i\}_{i=1}^{N_\mathrm{recon}}\big),\\
        \sigma_\mathrm{within}&=\frac{1}{N_\mathrm{recon}}\sum_{i=1}^{N_\mathrm{recon}}\sigma_i.
    \end{aligned}
\end{equation}
Here $\sigma_\mathrm{across}$ quantifies how reconstruction error varies with the physical parameters---indicating whether certain configurations are intrinsically harder to reconstruct---while $\sigma_\mathrm{within}$ captures sensitivity to measurement noise for a fixed configuration. A method with small $\mu$ but large $\sigma_\mathrm{across}$ depends strongly on the parameter regime; one with small $\sigma_\mathrm{across}$ but large $\sigma_\mathrm{within}$ is susceptible to noise. Together, these metrics provide a complete picture of reconstruction reliability.}

\textcolor{myblue}{Table~\ref{tab:bkerr} reports reconstruction errors across $84$ valid parameter samples, revealing three key findings. First, in the noise-free, fully-observed case (row~1), the mean relative $L^2(\Omega)$ error reaches $\mu \approx 32\%$---a three-order-of-magnitude increase compared to $<0.01\%$ when parameters match the training configuration (Tab.~\ref{tab:noiselesstable}). This confirms that \gls{pbdw} accuracy is fundamentally limited by how well $\mathcal{Z}_N$ represents the true solution.}

\textcolor{myblue}{Second, comparing rows~1 and~2 shows that adding $10\%$ measurement noise leaves the mean error virtually unchanged ($32.0\%$ vs $31.9\%$), while $\sigma_\mathrm{within} \approx 2\%$ remains an order of magnitude smaller than $\mu$. Model mismatch, not measurement noise, is the dominant error source.}

\textcolor{myblue}{Third, introducing spatial sparsity (row~3: 1:8 pattern, $x,y$-components only) doubles the error to $\mu \approx 64\%$. With fewer measurements, the update term $\bm{\eta}^*$ has limited capacity to compensate for the inadequacy of $\mathcal{Z}_N$. The increased $\sigma_\mathrm{across} \approx 3.5\%$ indicates greater sensitivity to the specific parameter configuration under sparse observation.}

\textcolor{myblue}{These results quantify a fundamental trade-off: \gls{pbdw} achieves excellent accuracy when the background space adequately represents the solution manifold, but degrades substantially under model mismatch. For clinical applications where anisotropy parameters are uncertain, future work should incorporate this variability directly into snapshot generation by treating $(\beta, r_f, r_t)$ as additional parameters in $\mathcal{P}^\mathrm{bk}$, at the cost of increased offline effort.}


\begin{table*}[ht!]
    \caption{\textcolor{myblue}{Reconstruction errors under model mismatch.
    Test cases were generated by perturbing the material anisotropy parameters:
    total anisotropy $\beta = b_f + b_t + b_{fs}$ was sampled in $[21,26]$,
    while ratios $r_f$ and $r_t$ were perturbed by $\pm10\%$ around baseline values,
    subject to $r_{fs}=1-r_f-r_t \ge 0$.
    Parameters $(p,\alpha)$ were kept inside the training distribution and $\kappa$ fixed.
    Columns left to right: sparsity pattern, noise level, number of sensors,
    relative reconstruction errors (mean $\mu$, across-field std $\sigma_\mathrm{across}$, 
    and within-field std $\sigma_\mathrm{within}$, as explained in Eq.~\eqref{eq:stats}), measured in 
    $L^2(\Omega)$, $H^1(\Omega)$, and $L^\infty(\Omega)$ norms, respectively, and online computational time.
    For the noise-free case, only $(\mu, \sigma_\mathrm{across})$ are shown.}}
    \label{tab:bkerr}
    \centering

    \begin{tabular}{ccccccc}
    \toprule
    $z$-sparsity 
    & $\widetilde{\mathrm{SNR}}$ 
    & $N_\mathrm{sens}$ 
    & $\mathrm{err}_{L^2(\Omega)}$ 
    & $\mathrm{err}_{H^1(\Omega)}$ 
    & $\mathrm{err}_{L^\infty(\Omega)}$ 
    & $t_\mathrm{online}$ \\
    \midrule

    None & 0 & $\num{3315}\cdot 3$ &
    \begin{tabular}{lr}
        $\mu$ & \num{3.19e-01} \\
        $\sigma_\mathrm{across}$ & \num{9.06e-03}
    \end{tabular} &
    \begin{tabular}{lr}
        $\mu$ & \num{3.25e-01} \\
        $\sigma_\mathrm{across}$ & \num{8.93e-03}
    \end{tabular} &
    \begin{tabular}{lr}
        $\mu$ & \num{3.62e-01} \\
        $\sigma_\mathrm{across}$ & \num{2.25e-03}
    \end{tabular} &
    \qty{2.91e-02}{\second}
    \\[0.4em]
    \midrule
    \rule{0pt}{2.1em}
    None & 0.1 & $\num{3315}\cdot 3$ &
    \begin{tabular}{lr}
        $\mu$ & \num{3.20e-01} \\
        $\sigma_\mathrm{across}$ & \num{8.93e-03} \\
        $\sigma_\mathrm{within}$ & \num{2.14e-02}
    \end{tabular} &
    \begin{tabular}{lr}
        $\mu$ & \num{3.27e-01} \\
        $\sigma_\mathrm{across}$ & \num{8.80e-03} \\
        $\sigma_\mathrm{within}$ & \num{2.15e-02}
    \end{tabular} &
    \begin{tabular}{lr}
        $\mu$ & \num{3.62e-01} \\
        $\sigma_\mathrm{across}$ & \num{2.24e-03} \\
        $\sigma_\mathrm{within}$ & \num{2.69e-02}
    \end{tabular} &
    \qty{2.52e-02}{\second}
    \\[0.4em]
    \midrule 
    \rule{0pt}{2.1em}
    1:8 & 0.1 & $\num{2202}\cdot 2$ &
    \begin{tabular}{lr}
        $\mu$ & \num{6.41e-01} \\
        $\sigma_\mathrm{across}$ & \num{3.51e-02} \\
        $\sigma_\mathrm{within}$ & \num{1.99e-02}
    \end{tabular} &
    \begin{tabular}{lr}
        $\mu$ & \num{6.49e-01} \\
        $\sigma_\mathrm{across}$ & \num{3.46e-02} \\
        $\sigma_\mathrm{within}$ & \num{1.97e-02}
    \end{tabular} &
    \begin{tabular}{lr}
        $\mu$ & \num{6.41e-01} \\
        $\sigma_\mathrm{across}$ & \num{3.68e-02} \\
        $\sigma_\mathrm{within}$ & \num{2.10e-02}
    \end{tabular} &
    \qty{4.20e-02}{\second}
    \\
    \botrule
    \end{tabular}
\end{table*}

\section{Discussion}
\label{sec:disc}

The results presented in this work demonstrate the effectiveness and efficiency of our non-intrusive \gls{pbdw} approach for reconstructing cardiac displacement fields across various scenarios of increasing complexity. The cardiac problem provides a rigorous test case due to the Guccione model's \textcolor{myblue}{non-linearity} and the presence of stiff scar regions.

In this section, we analyse our findings from methodological, computational, and application perspectives, while addressing limitations and identifying potential avenues for further development.

\subsection{Error Analysis and Reconstruction Quality}

The spatial distribution of reconstruction errors provides valuable insights into both the method's capabilities and the underlying cardiac mechanics. As demonstrated in Figures \ref{fig:noisy_mag}-\ref{fig:noise_z}, errors concentrate predominantly in the scar region where tissue stiffness is ten times greater than the surrounding myocardium. This behaviour is expected since the discontinuity in material properties creates more complex deformation patterns that are inherently more challenging to capture in a reduced basis. The relative errors remain remarkably low despite these challenges, with $L^2(\Omega)$ error of $2.27 \times 10^{-2}$ even in the presence of $10\%$ noise (Tab.~\ref{tab:noisytable}).

Analysis of the component-wise reconstruction (Figures \ref{fig:noisy_x}-\ref{fig:noise_z}) reveals anisotropic error distribution, with significantly higher accuracy in the $z$-component compared to $x$ and $y$. This aligns with the inflation mechanics of our problem, where deformation primarily occurs in the radial and circumferential directions. Interestingly, this anisotropic error distribution proves advantageous in our sparse reconstruction scenario, as the components with more challenging reconstruction characteristics ($x$ and $y$) are precisely those that remain observable in typical \gls{mri} acquisitions.

The remarkable resilience of our method to measurement sparsity is particularly notable. When observing only 8 slices with $\qty{1}{\milli\metre}$ thickness and without $z$-component information, the reconstruction maintains relative $L^2(\Omega)$ errors below $1.74 \times 10^{-2}$ (Tab.~\ref{tab:sparsenoisytable}). This suggests that the physical information encoded in the \gls{rb} effectively constrains the reconstruction in regions between measurements, producing physiologically plausible displacement fields even with highly sparse observations.

\subsection{Methodological Insights}

\textcolor{myblue}{Our implementation builds upon the \gls{pbdw} framework~\cite{Maday2014,Taddei_phdthesis} while emphasising practical applicability to cardiac mechanics. By constructing both the \gls{rom} and performing \gls{ss} using only solution snapshots---without requiring residual evaluations or problem-specific error estimators---the method becomes immediately applicable across diverse computational frameworks, including commercial black-box solvers where source code access is unavailable. This is particularly valuable in cardiac mechanics, where different research groups employ various constitutive models and numerical solvers.}

The proposed $H$-minibatch enhancement to \gls{womp} provides practical improvements to \gls{ss} efficiency, particularly valuable for large-scale problems where sequential processing becomes prohibitive.
The optimal minibatch size $H \in [50, 100]$ observed in our cardiac mechanics tests is problem-dependent and may vary with the physics, mesh size, and sparsity pattern of the Riesz representers. Problems with different spatial correlation structures or measurement distributions may benefit from different batch sizes, suggesting that $H$ should be tuned for each application domain. \textcolor{myblue}{Crucially, the stopping criterion in Alg.~\ref{alg:minibatchwomp} guarantees that \gls{ss} continues until $\beta_{N,M} \geq \beta_\odot$, regardless of minibatch size $H$. Consequently, larger $H$ cannot degrade stability below the user-specified threshold---it may only require selecting more sensors $M$ to achieve the same observability guarantee, representing a trade-off between computational efficiency and sensor parsimony rather than between efficiency and reconstruction stability.}

The \gls{ss} pattern observed (Fig.~\ref{fig:sensor_selection}) merits further investigation. The preference for sensors opposite the scar region suggests that \gls{womp} identifies locations where measurements provide maximum complementary information to the \gls{pod} basis. Since the \gls{pod} modes are constructed from snapshots containing the scar, they inherently encode its mechanical influence. Measurements in regions with more variable response across parameter space may therefore provide greater discriminatory power for state reconstruction.



The memory optimisation techniques described in Sec.~\ref{subsubsec:memory} are crucial for managing the computational demands of \gls{3d} problems. By exploiting the block structure of matrices through permutation and incremental computation, we achieve substantial reductions in both memory requirements and computational complexity. These implementation details, often overlooked in theoretical presentations of \gls{da} methods, are essential for practical applications involving high-dimensional state spaces.

\subsection{Comparison with Existing Approaches}

Our approach addresses several limitations of previous \gls{da} methods for cardiac mechanics. Traditional methods often rely on direct interpolation of displacement measurements \cite{Gudbjartsson1995,CardenasBlanco2008}, which can produce physically implausible results in regions far from the measurement locations. By contrast, our physics-informed reconstruction uses \gls{rb} to ensure that reconstructed displacement fields respect the underlying mechanical principles throughout the domain.
\textcolor{myblue}{\cite{Galarce2020} applies \gls{pbdw} to fluid dynamics, a non-linear phenomenon similar to cardiac mechanics, while \cite{Galarce2023} reconstructs brain tissue displacement from \gls{mri} elastography data using a linear model.
Both works employ a fixed background mesh of voxels covering the entire imaging domain, which accommodates variable patient geometries---an advantage over our current geometry-specific implementation.
These works focus on constructing informative \gls{bk} bases, including time-dependent formulations, whereas our contribution emphasises efficient \gls{ss}. In~\cite{Galarce2020,Galarce2023}, all available sensor locations are used directly without subset selection, which is appropriate when the number of measurements is moderate.
Regarding \gls{ss} strategies (which we distinguish from \gls{sp} in Remark~\ref{rem:ss_vs_sp}), \cite{Mula2025} proposes an optimisation approach that solves an eigenvalue problem similar to that in Alg.~\ref{alg:womp}, while \cite{Galarce2025} employs a greedy algorithm resembling SGreedy~\cite{Maday2014}, combining stability and maximisation steps via eigenvalue decomposition. None of these works incorporates minibatch acceleration.
}
Compared to alternative \gls{ss} strategies such as SGreedy \cite{Maday2014}, our minibatch \gls{womp} implementation provides superior computational efficiency while maintaining comparable reconstruction accuracy. 

The computational performance of our method is particularly notable compared to \gls{fom} simulations. The online reconstruction phase requires less than $\qty{0.1}{\second}$ (depending on the scenario), while a single full \gls{fe} simulation takes approximately 15 minutes (Tabs.~\ref{tab:noiselesstable}, \ref{tab:noisytable}, \ref{tab:sparsenoisytable}). This four-order-of-magnitude acceleration enables the rapid processing of multiple reconstructions, making the approach suitable for parameter estimation and uncertainty quantification studies that require numerous evaluations.

\subsection{Clinical Relevance and Practical Considerations}

The clinical relevance of our method stems from its ability to reconstruct full displacement fields from realistic \gls{mri}-like measurements. 
It is important to note that, while our sparsity assumptions might appear to be limitations of \gls{mri} technology, they actually reflect the reality of clinical data acquisition. Modern strain-encoded cardiac \gls{mri} techniques such as DENSE and SENC can provide continuous volumetric measurements when implemented with appropriate protocols~\cite{Aletras1999}. However, the constraints of clinical practice -- including limited scan time, patient tolerability, and the need for multiple breath-holds -- mean that sparse 2D multi-slice protocols remain the clinical standard~\cite{Kramer2020,Kocaoglu2020}. Our method's ability to accurately reconstruct full displacement fields from such sparse clinical acquisitions therefore addresses a practical gap between what is theoretically possible with \gls{mri} hardware and what is routinely acquired in clinical settings.



Our implementation accommodates the Gaussian noise characteristics typical of high-\gls{snr} \gls{mri} acquisitions, as justified in Sec.~\ref{subsec:noisy}. The regularization parameter $\xi$ in the \gls{pbdw} formulation effectively balances data fidelity against model fidelity, preventing overfitting to noise while preserving the essential mechanical features of the displacement field. Future work could explore more sophisticated noise models for low-\gls{snr} scenarios, where the Rician distribution diverges significantly from Gaussian \cite{Gudbjartsson1995}.

The offline-online decomposition of our approach offers practical advantages for clinical deployment. Through elaborated linear algebra, we managed to obtain a non computationally intensive offline phase, which needs only be performed once for a given cardiac geometry and constitutes a type of ``atlas'' that can be reused across multiple patients with similar anatomical characteristics. The rapid online phase enables near-real-time reconstruction during clinical examinations, potentially providing immediate feedback to clinicians.

\subsection{Limitations and Future Work}

Despite its promising performance, our current implementation has several limitations that warrant future investigation. First, our test cases consider only quasi-static inflation of an idealised \gls{lv}. Extending the approach to dynamic cardiac cycles with time-dependent displacement fields would require additional considerations for temporal basis functions and possibly space-time separated representations to maintain computational efficiency~\cite{Benaceur2021}.

\textcolor{myblue}{Second, \gls{pbdw} takes as input the \gls{mri} data and the geometry it refers to. When dealing with variable patient geometries, the offline phase described in Sec.~\ref{subsec:workflow} must in principle be performed for each geometry, since the \gls{rbs} are constructed for a specific mesh. Two strategies could address this limitation. The first introduces a \gls{ssm}: given a sample of geometries, a \gls{ssm} extracts a reference geometry together with diffeomorphic mappings to each individual geometry, and \gls{pbdw} is then performed entirely on the reference geometry. An alternative strategy constructs Riesz representers on a fixed voxel grid defined by the \gls{mri} sensor space. When a new patient mesh becomes available, these precomputed representers could be mapped from the voxel domain to the patient-specific mesh. While conceptually elegant, this approach presents theoretical challenges: the mapping breaks the Riesz representer property (since the mapped functions no longer satisfy Eq.~\eqref{eq:var_riesz} on the new mesh), and robust inter-mesh transfers are non-trivial~\cite{Dickopf2014}. A thorough investigation of both strategies is beyond the scope of the present work.}

The observed differences in reconstruction accuracy between healthy and scarred regions highlight the challenge of capturing pathological mechanics. This suggests that patient-specific \gls{rbs} may be necessary for subjects with significant myocardial abnormalities, rather than relying on bases constructed from primarily healthy tissue mechanics \cite{Caforio2024}. Adaptive enrichment strategies that refine the basis in regions of high reconstruction error could potentially address this challenge.

While machine learning methods using autoencoders~\cite{Romor2023} show promise for problems with slow Kolmogorov $n$-width decay, our physics-based approach prioritises interpretability and generalisation beyond the training data, which is crucial for clinical applications where data scarcity and trust in predictions are paramount.

Our current model uses a rule-based approach for fiber orientation~\cite{Bayer2012}, which provides a reasonable approximation of the helical fiber architecture. Future work could incorporate patient-specific fiber distributions derived from diffusion tensor \gls{mri} to further improve the physiological fidelity of both the mechanical model and the subsequent reconstruction.


Additionally, our implementation assumes measurements in the reference configuration, aligned with standard end-diastolic imaging protocols. Future work could explore iterative approaches to simultaneously estimate both the reference configuration and displacement fields from loaded states, though this significantly increases computational complexity and is beyond the scope of the current study.

\textcolor{myblue}{Regarding stability guarantees, as noted in Section~\ref{subsubsec:pbdw_stability}, we have followed the empirical validation approach of~\cite{Gong2019,Bui2023} rather than computing the analytical stability constants. Deriving problem-specific bounds for these constants in the context of nonlinear cardiac mechanics remains an open problem.}

Finally, a more rigorous uncertainty quantification framework could provide confidence intervals for the reconstructed displacement fields, enabling clinicians to make more informed decisions based on the reliability of the results. This might involve Bayesian approaches that explicitly model measurement and model uncertainties, providing a probability distribution over possible displacement fields rather than a single deterministic reconstruction~\cite{Benaceur2024}.

\section{Conclusions}
\label{sec:concl}

In this work, we have presented a non-intrusive implementation of the \gls{pbdw} framework for cardiac displacement field reconstruction from sparse, noisy observations. The key contributions of our approach are summarized as follows:

\begin{itemize}
    \item We applied the non-intrusive \gls{pbdw} framework to \gls{3d} cardiac mechanics, demonstrating its effectiveness for displacement field reconstruction from sparse, noisy observations.

    \item We introduced an efficient $H$-size minibatch \gls{womp} algorithm for \gls{ss} that reduces computational time compared to standard sequential implementations, addressing one of the primary bottlenecks in \gls{pbdw} application to large-scale problems.
    
    \item We developed memory-efficient implementation strategies that exploit the block structure of matrices through strategic permutation, making the approach feasible for high-dimensional cardiac mechanics problems with millions of \gls{dofs}.
    
    \item We validated our approach through three test cases of increasing complexity, demonstrating exceptional accuracy in noise-free conditions (relative $L^2(\Omega)$ error of $7.52 \times 10^{-5}$), robust performance with 10\% Gaussian noise (relative $L^2(\Omega)$ error of $2.27 \times 10^{-2}$), and effective reconstruction from sparse measurements mimicking realistic \gls{mri} acquisition protocols (relative $L^2(\Omega)$ error of $1.74 \times 10^{-2}$).
\end{itemize}

The proposed framework achieves a four-order-of-magnitude acceleration in displacement field reconstruction compared to full \gls{fe} simulations, with online reconstruction times of less than one tenth of second for the sparse measurement case.

The method exhibits particular robustness to measurement sparsity, maintaining accurate reconstructions even when observations are limited to a few imaging slices without $z$-component information. This resilience to realistic clinical data limitations, combined with the method's computational efficiency, suggests strong potential for integration into cardiac diagnostic and modelling workflows.



Our non-intrusive \gls{pbdw} implementation achieves accurate cardiac displacement reconstruction from sparse \gls{mri} data while maintaining compatibility across computational platforms. The method's four-order-of-magnitude speedup and robustness to clinical data limitations make it practical for integration into existing cardiac modelling workflows, where diverse software ecosystems and time constraints currently limit advanced \gls{da} methods.

\backmatter

\bmhead{Supplementary information}
The code and data for reproducing our results will be made available upon acceptance of this manuscript.
\bmhead{Acknowledgements}
\gls{fe} simulations for this study were performed on the \gls{vsc5} under
PRACE project \#71962, which are maintained by the VSC Research Center in collaboration with the Information Technology Solutions of
TU Wien. 
FC acknowledges support from L’Oréal Austria and Unesco.
FC is member of the INdAM research group GNCS. 
FM, and EK acknowledge support from the BioTechMed-Graz Young Researcher
Grant "CICLOPS—Computational Inference of Clinical Biomarkers
from Non-Invasive Partial Data Sources". 
This project has received funding from the ERA-NET co-fund action No. 680969 (ERA-CVD SICVALVES, JTC2019) funded by the Austrian Science Fund (FWF), Grant I 4652-B to CA.






\bigskip
\begin{flushleft}%
Editorial Policies for:

\bigskip\noindent
Springer journals and proceedings: \url{https://www.springer.com/gp/editorial-policies}

\bigskip\noindent
Nature Portfolio journals: \url{https://www.nature.com/nature-research/editorial-policies}

\bigskip\noindent
\textit{Scientific Reports}: \url{https://www.nature.com/srep/journal-policies/editorial-policies}

\bigskip\noindent
BMC journals: \url{https://www.biomedcentral.com/getpublished/editorial-policies}
\end{flushleft}

\begin{appendices}

\section{Important Notations}
\label{app:notations}
In this section we summarise the most important notations used throughout the paper.
\subsection{Meshes and \gls{fe} spaces}
Let $\Omega \subset \mathbb R^3$ be a bounded domain with Lipschitz-continuous boundary.
Let us also define the Sobolev space $\mathcal{X}_{1d} = H^1(\Omega)$ consisting of all square-integrable functions over $\Omega$ with square-integrable weak gradient over $\Omega$. 
We define a decomposition of $\Omega \approx \mathcal T_M = \bigcup_{l=1}^M \tau_l$ into $M$ simplicial finite elements as \emph{mesh}. The number of unique nodes in the mesh is defined as $\mathcal N$. Provided $\Omega$ has a polyhedral boundary this decomposition is exact, see~\cite{Jung2013}.
Over the mesh $\Omega$ we define the space of piecewise linear globally continuous functions as $S_h^1(\Omega) \subset H^1(\Omega)$, with dimension $\lvert S_h^1(\Omega) \rvert = \mathcal N$.
Throughout this paper we make heavy use of the \emph{Galerkin identity} for identifying an element $v \in S_h^1(\Omega)$ with a vector $\bm v \in \mathbb R^\mathcal N$ through
\begin{equation}
    \label{eq:galerkin_id}
    \forall v \in S_h^1(\Omega)~\exists \underline{v} \in \mathbb{R}^\mathcal{N}~\text{s.t.}~v(\bm x)=\sum_{i=1}^\mathcal{N} \vec v_i \phi_i(\bm x),
\end{equation}
with $\{\phi_i(\bm x)\}_{i=1}^\mathcal{N}$ being a basis of $S_h^1(\Omega)$.
If we define the weighted $H^1$-scalar product
\begin{equation*}
    (v,w)_{\mathcal{X}_{1d}} = \int_\Omega vw~dx + L_g^2 \int_\Omega (\nabla v, \nabla w)~dx,
\end{equation*}
we can deduce the associated weight matrix $\mathbf{I}_{\mathcal{X}_{1d}}$ as
\begin{align}
    \mathbf{I}_{\mathcal{X}_{1d}} &= \mathbf{M} + L_g^2 \mathbf{K},\\
    \mathbf{M}_{ij} &= \int_{\Omega} \phi_i \phi_j~dx,\\
    \mathbf{K}_{ij} &= \int_{\Omega} (\nabla \phi_i, \nabla \phi_j)~dx,
\end{align}
with the mass matrix $\mathbf{M} \in \mathbb R^{\mathcal N \times \mathcal N}$ and stiffness matrix $\mathbf{K}  \in \mathbb R^{\mathcal N \times \mathcal N}$.
Using Eq.~\eqref{eq:galerkin_id}, it holds for any $v \in S_h^1(\Omega)$: $$ \lVert v\rVert_{\mathcal{X}_{1d}}^2 = (\mathbf{I}_{\mathcal{X}_{1d}} \vec v, \vec v).$$
The extension to vector-valued problems is straightforward.
In Sec.~\ref{sec:methods} we introduced $\mathcal{X} = [H^1(\Omega)]^3$.
Similarly, we can define $[S_h^1(\Omega)]^3 \subset \mathcal X$ and the weighted scalar product
\begin{equation*}
    (\bm v,\bm w)_{\mathcal{X}} = \int_\Omega (\bm v, \bm w)~dx + L_g^2 \int_\Omega (\bm\nabla \bm v, \bm\nabla \bm w)~dx.
\end{equation*}
The equivalent Galerkin identity for vector-valued problems reads as
\begin{multline}
    \label{eq:galerkin_id_vec}
    \forall \bm v \in [S_h^1(\Omega)]^3~\exists
    \vec{v} \in \mathbb{R}^{3\mathcal{N}}\\\text{s.t.}~\bm v(\bm x)=\sum_{i=0}^{\mathcal N-1}\sum_{j=0}^2 \underline{v}_{j\mathcal N + i} \phi_i(\bm x) \vec e_j.
\end{multline}
Using Eq.~\eqref{eq:galerkin_id_vec} we can deduce the associated weight matrix $\mathbf{I}_\mathcal{X}$ as
\begin{align}
    \label{eq:weight_matrix_vec}
    \mathbf{I}_\mathcal{X} &= \mathbf{I}_{\mathcal{X}_{1d}} \otimes \mathbf{I}_3 \in \mathbb R^{3 \mathcal N \times 3 \mathcal N},\\
    \lVert \bm v\rVert_{\mathcal X}^2 &= (\mathbf{I}_\mathcal{X} \vec v, \vec v)
\end{align}
with $\mathbf{I}_3$ being the three dimensional identity matrix, and $\otimes$ denotes the Kronecker product.

\subsection{Necessary Elements from Functional Calculus}
The Riesz representer $r = \mathcal R_{\ell} \in \mathcal X_{1d}$ of any linear functional $\ell \in \mathcal X'_{1d}$ is guaranteed by the Riesz representation theorem~\cite{Rudin1990} and fulfills
\begin{equation*}
\begin{aligned}
    (v, \mathcal R_{\ell})_{\mathcal X} &= \ell(v),\\
    \lVert R_\ell \rVert_{\mathcal X} &= \lVert \ell \rVert_{\mathcal X'},
\end{aligned}
\end{equation*}
for all $v \in \mathcal X$.
Furthermore, for a Hilbert Space $\mathcal H$ we define the orthogonal projection of $v \in \mathcal H$ onto a subspace $\mathcal P \subset \mathcal H$ as
\begin{equation*}
    \Pi_{\mathcal P} v := \sum_i (v, \xi_i) \xi_i,
\end{equation*}
with an orthonormal basis $\{\xi_i\}_i \subset \mathcal P$.
Additionally, we can define the projection onto the orthogonal complement of $\mathcal P$ as
\begin{equation*}
    \Pi_{\mathcal P^\perp} v := v - \sum_i (v, \xi_i)\xi_i.
\end{equation*}

\section{Derivation of \gls{pbdw} Saddle Point System}
\label{app:pbdw_derivation}

Here we provide details on the derivation of the saddle point system~\eqref{eq:pbdw_system_cardiac}.
For ease of presentation, we provide the derivation for scalar-valued reconstructions; however, the same arguments apply to vector-valued problems.

We denote the collection of sensor functionals as $\mathcal U_M =\{q_m=R_{\ell_m}\}_{m=1}^M \subset \mathcal X_{1d}$, and we define the $\mathcal X_{1d}$-orthonormal bases $\mathcal Z_N=\{\zeta_n\}_{n=1}^N$ and $\mathcal U_M = \{\tau_m\}_{m=1}^M$.

%
%
%
Let $\mathbf{Z}\in\mathbb{R}^{N\times\mathcal{N}}$ and $\mathbf{Q}\in\mathbb{R}^{M\times\mathcal{N}}$ be the \gls{fe} coefficient matrices of $\mathcal{Z}_N$ and $\mathcal{U}_M$, respectively. 
Moreover, we define the matrices $\mathbf{T} \in \mathbb{R}^{M \times M}$ through $T_{kl}=\ell_k(\tau_l) = (q_k, \tau_l)_{\mathcal{X}_{1d}}$ and $\mathbf{P}=\mathbf{Z} \mathbf{I}_{\mathcal{X}_{1d}} \mathbf{Q}^\top \in \mathbb{R}^{N \times M}$.
From\textcolor{myblue}{~\cite[Prop.~2.2.2]{Taddei_phdthesis}} we know that the minimisation problem~\eqref{eq:pbdw_minimization_cardiac} solves
\begin{align}
\label{eq:saddlepointTaddei:1} \xi (\eta_\xi^*,q)_\mathcal X + \frac{1}{M}\sum_{m=1}^M (\ell_m(z_\xi^*+\eta_\xi^*)-\vec y^{\mathrm{obs}}_m)\ell_m(q) &= 0,\\ 
\label{eq:saddlepointTaddei:2} (\eta_\xi^*,p)_\mathcal X &= 0,
\end{align}
for all $(q, p) \in \mathcal{U}_M \times \mathcal{Z}_N$.
The state reconstruction $u^*_\xi$ can be obtained by applying two linear combinations: 
\begin{equation}
    \label{eq:recscal}
    u_\xi^*=z_\xi^*+\eta_\xi^*=\sum_{n=1}^N \vec z_{\xi,n}^*\zeta_n + \sum_{m=1}^M \vec \eta_{\xi,m}^* \tau_m.
\end{equation}

Next, we show how to retrieve~\eqref{eq:pbdw_system_cardiac} starting from~\eqref{eq:saddlepointTaddei:1}--\eqref{eq:saddlepointTaddei:2}. 
First, we substitute the expressions for $z_\xi^*,\eta_\xi^*$, given in \eqref{eq:recscal}, into~\eqref{eq:saddlepointTaddei:1}--\eqref{eq:saddlepointTaddei:2}, by also considering  $q=\tau_s\in\mathcal{U}_M$ for some $s\in[1,\ldots,M]$ and $p=\zeta_r\in\mathcal{Z}_N$ for some $r\in[1,\ldots,N]$. For the first term in~\eqref{eq:saddlepointTaddei:1} we have
\begin{equation*}
    \begin{aligned}
    \xi \left(\eta_\xi^*,\tau_s\right)_\mathcal X &= \xi \left(\sum_{m=1}^M \vec\eta_{\xi,m}^* \tau_m, \tau_s\right)_\mathcal X\\
    &= \xi \sum_{m=1}^M \vec\eta_{\xi,m}(\tau_m, \tau_s)_\mathcal X \\
    &= \xi \sum_{m=1}^M \delta_{ms} \vec\eta_{\xi,m}^* = \xi \vec\eta_{\xi,s}^*.
    \end{aligned}
\end{equation*}
For the first part of the second term in~\eqref{eq:saddlepointTaddei:1} we have (summation over double indices applies):
\begin{equation*}
    \begin{aligned}
     \frac{1}{M} \ell_m(\vec z_{\xi,n}^*\zeta_n)\ell_m(\tau_s) &= \vec z_{\xi,n}^* \ell_m(\zeta_n) \ell_m(\tau_s)\\
     &=\frac{1}{M}\vec z_{\xi,n}^* T_{ms} \tilde{P}_{mn} \\
     &=\frac{1}{M}T_{sm}^\top \tilde{P}_{mn} \vec z_{\xi,n}^* \\
     &= \frac{1}{M}(\mathbf{T}^\top \tilde{\mathbf{P}} \vec z_\xi^*)_s,
    \end{aligned}
\end{equation*}
with 
\begin{equation*}
   \tilde{P}_{mn}=\ell_m(\zeta_n)=(q_m, \zeta_n)_\mathcal X. 
\end{equation*}
With a similar argument, it follows for the second part of the second term in~\eqref{eq:saddlepointTaddei:1} (summation over double indices applies):
\begin{equation*}
    \begin{aligned}
        \frac{1}{M} \ell_m (\vec \eta_{\xi,m'}^*\tau_{m'}) \ell_m(\tau_s)&=\frac{1}{M} T_{sm}^\top T_{mm'} \vec\eta_{\xi,m'}^* \\
        &= \frac{1}{M} (\mathbf{T}^\top \mathbf{T} \vec\eta_\xi^*)_s.
    \end{aligned}
\end{equation*}
For the third term in~\eqref{eq:saddlepointTaddei:1} it follows:
\begin{equation*}
    \begin{aligned}
       -\frac{1}{M} \sum_{m=1}^M \vec y^{\mathrm{obs}}_m \ell_m(\tau_s) &= -\frac{2}{M} \sum_{m=1}^M \vec y^{\mathrm{obs}}_m T_{ms}\\
       &= - \frac{1}{M} \sum_{m=1}^M T_{sm}^\top \vec y^{\mathrm{obs}}_m\\
       &= -\frac{1}{M} (\mathbf{T}^\top \vec y^{\mathrm{obs}})_s. 
    \end{aligned}
\end{equation*}
Finally, for~\eqref{eq:saddlepointTaddei:2} we have:
\begin{equation*}
    \begin{aligned}
         (\eta_\xi^*,\zeta_r)_\mathcal X &= \left(\sum_{m=1}^M \vec\eta_{\xi,m}^* \tau_m,\zeta_r\right)_\mathcal X\\
         &= \sum_{m=1}^M (\tau_m,\zeta_r)_\mathcal X \vec\eta_{\xi,m}^*\\
         &= \sum_{m=1}^M P_{rm} \vec\eta_{\xi,m}^* = (\mathbf{P} \vec\eta_\xi^*)_r.
    \end{aligned}
\end{equation*}
%
%
Collecting these results we obtain the following linear system:
\begin{equation}
    \label{eq:linsys3}
    \begin{bmatrix}
        M\xi\mathbf{I} + \mathbf{T}^\top \mathbf{T} & \quad\mathbf{T}^\top \tilde{\mathbf{P}} \\
        \mathbf{P} & \mathbf{0}
    \end{bmatrix}
    \begin{bmatrix}
        \vec{\eta}_\xi^* \\ \vec{z}_\xi^*
    \end{bmatrix}
    =
    \begin{bmatrix}
        \mathbf{T}^\top  \vec{y}^\mathrm{obs} \\ \mathbf{0}
    \end{bmatrix}.
\end{equation}
Using the fact that
\begin{equation}
    \color{myblue}
    \label{eq:P_TP}
    \begin{aligned}
    \tilde{P}_{mn}&=(q_m,\zeta_n)_\mathcal X=\left(\sum_{j=1}^M T_{mj}\tau_j,\zeta_n\right)_\mathcal X\\
    &=\sum_{j=1}^M T_{mj}(\tau_j,\zeta_n)_\mathcal X=\sum_{j=1}^M T_{mj}P_{nj}\\
    &=\sum_{j=1}^M T_{mj}P_{jn}^\top,
    \end{aligned}
\end{equation}
we conclude.

\textcolor{myblue}{
\begin{remark}[Computational complexity]
    Solving the saddle point system~\eqref{eq:pbdw_system_cardiac} directly requires $\mathcal{O}((N+M)^3)$ operations. Bui et al.~\cite{Bui2023} propose an algebraic reformulation that reduces this to $\mathcal{O}(N^3+M^3)$, which can be beneficial when both $N$ and $M$ are moderately large.
\end{remark}
}



\section{\gls{pod} Implementation Details}
\label{app:pod_implementation}

For completeness, we provide the \gls{pod} algorithm used in our implementation and refer to~\cite{Carlberg2008, Carlberg2010} for details. 
We define $\mathbf{W}=[\vec{u}_1^\mathrm{bk},\ldots,\vec{u}_{N_s}^\mathrm{bk}]^\top\in\mathbb{R}^{N_s\times3\mathcal{N}}$ as the snapshot matrix, where $N_s$ is the number of snapshots. We assume that $N_s<<\mathcal{N}$, which is typical for $d=3$. 
The \gls{svd} is applied to $\mathbf{W}^\top\mathbf{I}_{\mathcal X}\mathbf{W}$ and the chosen \gls{rb} corresponds to a truncation of the rescaled eigenvectors in $\mathbf{V}$.

\begin{algorithm}
    \caption{Eigenvalue decomposition-based computation of \gls{pod} basis}
    \label{alg:pod}
    \begin{algorithmic}[1]
        \Require Snapshot matrix $\mathbf{W}$, metric matrix $\mathbf{I}_{\mathcal X}$
        \Ensure $\mathbf{Z}_{n_\phi}$
        \State $\bar{\mathbf{W}}=\mathbf{W}^\top \mathbf{I}_{\mathcal X} \mathbf{W}$
        \State Compute symmetric eigenvalue decomposition $\bar{\mathbf{W}}=\mathbf{V} \boldsymbol{\Sigma}^2 \mathbf{V}^\top$
        \State Choose dimension of truncated basis $n_\phi \in \{1,...,\mathrm{min}(N_s, \mathrm{rank}(\mathbf{I}_\mathcal{X}))\}$ based on singular value decay in Eq.~\eqref{eq:pod_energy}
        \State $\mathbf{Z}_{n_\phi}=\mathbf{W} \begin{bmatrix}
            \frac{1}{\sigma_1}\vec{v}_1 & ... & \frac{1}{\sigma_{n_\phi}}\vec{v}_{n_\phi}
        \end{bmatrix}$ where $\mathbf{V} = [\vec v_1, \ldots, \vec v_{N_s}]$, and $\mathbf{\Sigma} = \mathrm{diag}(\sigma_1, \ldots, \sigma_{N_s})$.
    \end{algorithmic}
\end{algorithm}

The truncation parameter $n_\phi$ is selected based on the decay of singular values $\{\sigma_i\}_{i=1}^{N_s}$, typically retaining modes that capture a prescribed fraction $v$ of the total energy, i.e.
\begin{equation}
\label{eq:pod_energy}
    n_\phi = \underset{n \in (1,\ldots,N_s)}{\mathrm{min}} \left\{ n : \frac{\sum_{i=n+1}^{N_s} \sigma_i^2}{\sum_{i=1}^{N_s} \sigma_i^2} < 1-v^2\right\}
\end{equation}
for $v \in (0, 1)$.
\section{Sequential \gls{womp}}
\label{subsec:ss}
Sensor selection aims at identifying and selecting from a library of available linear functionals, which contains the most informative ones. 
The literature presents different alternatives: SGreedy~\cite{Maday2014}, \gls{womp}~\cite{Binev2018} and \gls{comp}~\cite{Binev2018}. 
Within the \gls{pbdw} formulation presented in this work, the major difference between these algorithms is given by the fact that the former yields only the selection, whereas the latter two, \gls{womp} or \gls{comp}, return the orthonormal basis $\{\bm{\tau}_m\}_{m=1}^M$ of $\mathcal{U}_M$ along with the selected functionals.
For a detailed analysis of \gls{womp} we refer to~\cite{Binev2018}.
This work also concluded that \gls{comp} yields worse reconstructions on average than \gls{womp}, thus we focus on \gls{womp}, whose algorithm is depicted in Alg.~\ref{alg:womp} for scalar valued problems.
The extension to vector-valued problems is straightforward.
\begin{algorithm}
    \caption{\gls{womp} scalar-valued}
    \label{alg:womp}
    \begin{algorithmic}
        \Require $\mathcal{Z}_N,\mathcal{L}_K, \beta_\odot \in (0,1)$
        \Ensure $\mathcal{U}_M,\mathbf{Q}$ s.t. $\beta_{N,M} \geq \beta_\odot$
        \State $M=0$
        \State $q_\beta^\perp  \gets \zeta_1$
        \While{$\beta_{N,M}<\beta_\odot$}
        \State $R_{\ell_{M+1}} = \underset{R_{\ell} \in \mathcal{L}_K}{\arg\max} \frac{\lvert(R_{\ell}, q_{\beta}^\perp)_{\mathcal X}\rvert}{\lVert R_{\ell\rVert_{\mathcal X}}}$
        \State $\tau_{M+1}= \frac{\Pi_{\mathcal{U}_M^\perp} \mathcal{R}_{\ell_{M+1}}}{\lVert\Pi_{\mathcal{U}_M^\perp} \mathcal{R}_{\ell_{M+1}}\rVert_\mathcal{X}}$
        \State $\mathcal{U}_M=\mathrm{span}\{\tau_1,...,\tau_{M+1}\}$
        \State $\mathbf{Q}=[\vec{\tau}_1,...,\vec{\tau}_{M+1}]^\top$
        \State $\mathbf{P}\mathbf{P}^\top \vec{v}_\beta = \lambda_\beta \vec{v}_\beta$
        \State $\beta_{N,M}=\sqrt{\lambda_{\beta,\mathrm{min}}}$
        \State $\phi_\beta=\sum_{n=1}^N(\vec{v}_\beta)_n \zeta_n$
        \State $q_\beta^\perp=\phi_\beta-\Pi_{\mathcal{U}_M}\phi_\beta$
        \State $q_\beta^\perp=\frac{q_\beta^\perp}{\lVert q_\beta^\perp\rVert_{\mathcal{X}}}$
        \State $M=M+1$
        \EndWhile
    \end{algorithmic}
\end{algorithm}
\noindent
Note that in Alg.~\ref{alg:womp} $\phi_\beta$ can also be computed as $\vec{\phi}_\beta=(\mathbf{Z}^\top-\mathbf{Q}^\top\mathbf{P}^\top)\vec{v}_\beta$. The need for efficiency comes from the fact that \gls{womp} algorithm is intrinsically sequential, which makes it unsuitable for acceleration based on parallelism.
%

\section{Implementation details}
\label{sec:impl}

The implementation of this work is based on a mix of Python and C++ languages. We rely on C++ to efficiently compute the Riesz elements from the linear functionals. This is done by first constructing a bounding box containing the target geometry $\Omega$. We divide the box into $K$ parallelepipedons $\{\Omega_k\}_{k=1}^K$ with fixed dimensions $\Omega_k=[a,b,c]\quad\forall k=1,\ldots,L$, discarding all those that have a null intersection with the domain $\Omega$. Therefore, the number of sensors is $N_\mathrm{sens} = K$ and each one sees in $\Omega_k=[a,b,c]\quad\forall k=1,\ldots,K$. Then, we identify, among the mesh elements (tethrahedra) $T=\{\mathcal{T}_n\}_{n=1}^{N_\mathrm{tets}}$, those which are contained (even partially) in each $\Omega_k$, $e.g.\,G_k\subset T$, and consider $G_k$ as the observational voxel of the $k^\text{th}$ \gls{mr} sensor. 
Finally, we solve $K$ problems in order to compute the Riesz elements. The computation is efficient since the \gls{fe} system is assembled only once, whereas the right-hand side is constructed for each problem separately. 
The Python code takes care of the whole reconstruction, given the \gls{fe} approximation of the Riesz elements $\{\mathcal{R}_{\ell_k}\}_{k=1}^K$ and the snapshots $\{\underline{u}^\mathrm{bk}_n\}_{n=1}^{N_s}$. 
We introduce a JSON file to set all relevant parameters in \gls{pbdw}:
\begin{itemize}
    \item Choice of the \gls{snr} definition (\eqref{eq:sigma_ld} or \eqref{eq:sigma_snr}) and its value (and other parameters related to Guassian noise);
    \item Choice of the \gls{rom} and its parameters;
    \item Choice of the \gls{ss} algorithm and its parameters.
\end{itemize}
\section{Memory Saving Optimisations for Vector-Valued Problems}
\label{sec:vec_efficient}

Here we give a detailed overview of how we exploit the block structure of the Riesz representers to minimise computational overhead.
To this end:
\begin{itemize}
    \item $\mathbf{R}_{1d} \in \mathbb R^{K\times \mathcal N}$ \textcolor{myblue}{is} the \gls{fe} coefficients of the Riesz elements $\{\mathcal{R}_{\ell_k}\}_{k=0}^{K-1}$ corresponding to the functional
    \begin{equation*}
        \ell_k(u):= \frac{1}{\lvert \Omega_k\rvert}\int\limits_{\Omega_k} u~dx,\quad u\in \mathcal{X}_{1d}.
    \end{equation*}
    \item The corresponding matrix of \gls{fe} coefficients for the vector-valued Riesz representers is then given by $$\mathbf{R}_{3d} = \mathbf{R}_{1d} \otimes \mathbf{I}_3 \in \mathbb{R}^{3K \times 3\mathcal{N}},$$ using the Kronecker product.
\end{itemize}

We can avoid assembling $\mathbf{R}_{3d}$ by referencing its values through $\mathbf{R}_{1d}$.
In particular this means that given
\begin{equation*}
    \mathbf{R}_{1d} = 
    \begin{bmatrix}
    \vec{r}_0^\top\\
    \vec{r}_1^\top\\
    \vdots \\
    \vec{r}_{K-1}^\top
    \end{bmatrix}
\end{equation*}
with $\vec{r}_k \in \mathbb{R}^\mathcal{N}$ denoting the \gls{fe} coefficients of the $k$-th Riesz representer we can write
\begin{align*}
    \mathbf{R}_{3d} &= \mathbf{R}_{1d} \otimes \mathbf{I}_3 \\
    &= \begin{bmatrix}
        \vec{r}_0^\top & \vec{0}^\top & \vec{0}^\top \\
        \vec{0}^\top & \vec{r}_0^\top & \vec{0}^\top \\
        \vec{0}^\top & \vec{0}^\top & \vec{r}_0^\top \\
        & \vdots & \\
        \vec{r}_{K-1}^\top & \vec{0}^\top & \vec{0}^\top \\
        \vec{0}^\top & \vec{r}_{K-1}^\top & \vec{0}^\top \\
        \vec{0}^\top & \vec{0}^\top & \vec{r}_{K-1}^\top 
    \end{bmatrix}
\end{align*}
    %
    %


%
Now, we can construct an index map from $\mathbf{R}_{3d}$ to $\mathbf{R}_{1d}$ by considering the row index $i$ of the vector in $\mathbf{R}_{3d}$ and exploiting the results of an integer division $(q, r) = (\lfloor\frac{i}{3}\rfloor, i \bmod 3)$ as stated in~Eq.\eqref{eq:index}.
In fact, the quotient $q$ indicates the correct element of $\mathbf{R}_{1d}$, whereas
the rest $r$ gives the component, where $x=0,y=1,z=2$. 
This trick is exploited throughout the code to save memory allocation by a factor of nine.
In the same manner, we account for the selection of the sensors, too. 
Indeed, \gls{womp} returns a list of indices corresponding to the sensors chosen in $\mathcal{L}_K$, and therefore the Riesz elements in $\mathbf{R}_{3d}$, which can be referenced as in \textcolor{myblue}{Expr.}~\eqref{eq:index} to obtain the corresponding Riesz elements in $\mathbf{R}_{1d}$. 
In Eq.~\eqref{eq:pbdw_system_cardiac} one has to assemble $\mathbf{T}^\top \mathbf{T} \mathbf{P}^\top$ and $\mathbf{T}^\top\vec{\ell}^\mathrm{obs}$, where both $\mathbf{T},\mathbf{Q}$ undergo the selection. In particular, we can write
\begin{equation*}
\begin{aligned}
    \mathbf{T}^\top \mathbf{T} \mathbf{P}^\top &= \mathbf{T}^\top \mathbf{T} (\mathbf{Z} \mathbf{I}_\mathcal{X} \mathbf{Q}^\top)^\top\\ &= \mathbf{T}^\top \mathbf{T} \mathbf{Q} \mathbf{I}_\mathcal{X}^\top \mathbf{Z}^\top = \mathbf{T}^\top \mathbf{T} \mathbf{Q} \mathbf{I}_\mathcal{X} \mathbf{Z}^\top .
\end{aligned}
\end{equation*}
We can construct $T_{kl}=\vec{\ell}_k(\bm{\tau}_l)$ as $\mathbf{T}=\mathbf{R}_{3d}^f \mathbf{I}_{\mathcal{X}} \mathbf{Q}^\top$, where $\mathbf{R}_{3d}^f = \mathbf C \mathbf{R}_{3d} \in\mathbb{R}^{M\times3\mathcal{N}}$ is obtained by selecting a subset of rows of $\mathbf{R}_{3d}$ according to \gls{womp}. The choice matrix $\mathbf C$ above is defined as $C_{i j} = 1$ when $j=s_i$ and zero elsewhere, with $\{s_i\}_{i=1}^M$ being the list of indices returned by \gls{womp}.
By reordering the rows of $\mathbf{C}$, using a permutation matrix $\mathbf{P}_\mathrm{ord}\in\mathbb{R}^{M\times M}$, we can define $\tilde{\mathbf{R}}_{3d}^f=\mathbf{P}_\mathrm{ord} \mathbf{R}_{3d}^f$ which is a block sparse matrix:
\begin{equation*}
    \mathbf{P}_\mathrm{ord}\mathbf{R}_{3d}^f =
    \operatorname{diag}\!
    \begin{bmatrix}
        \tilde{\mathbf{R}}_{3d,x}^f \\
        \tilde{\mathbf{R}}_{3d,y}^f\\
        \tilde{\mathbf{R}}_{3d,z}^f
    \end{bmatrix}
    =
    \begin{bmatrix}
        \tilde{\mathbf{R}}_{3d,x}^f & 0 & 0 \\
        0 & \tilde{\mathbf{R}}_{3d,y}^f & 0 \\
        0 & 0 & \tilde{\mathbf{R}}_{3d,z}^f
    \end{bmatrix}, 
\end{equation*}

with three blocks \textcolor{myblue}{$\tilde{\mathbf{R}}_{3d,j}^f\in\mathbb{R}^{M_j\times\mathcal{N}}\quad\forall j=x,y,z$ such that $\sum_{j=x,y,z}M_j=M$.} The same reasoning can be applied to $\mathbf{Q}$, writing $\tilde{\mathbf{Q}}=\mathbf{P}_\mathrm{ord}\mathbf{Q}$. We obtain three blocks in $\tilde{\mathbf{Q}}$, namely \textcolor{myblue}{$\tilde{\mathbf{Q}}_j\in\mathbb{R}^{M_j\times\mathcal{N}}\quad\forall j=x,y,z$.}
Next, using the fact that the inverse of any permutation matrix is its transpose we have that
\begin{align*}
    \mathbf{Q}&=\mathbf{P}_\mathrm{ord}^\top\tilde{\mathbf{Q}} \\
    \mathbf{R}_{3d}^f&=\mathbf{P}_\mathrm{ord}^\top\tilde{\mathbf{R}}_{3d}^f
\end{align*}
Using this we can compute $\mathbf T$ as
\begin{equation}
    \label{eq:T}
    \begin{aligned}
    &\mathbf{T} =\mathbf{P}_\mathrm{ord}^\top \tilde{\mathbf{R}}_{3d}^f \mathbf{I}_{\mathcal{X}} \tilde{\mathbf{Q}}^\top\mathbf{P}_\mathrm{ord}\\ 
    & =\mathbf{P}_\mathrm{ord}^\top
       \operatorname{diag}\!
         \begin{bmatrix}
           \tilde{\mathbf{R}}_{3d,x}^f \mathbf{I}_{\color{myblue}{\mathcal{X}_{1d}}} \tilde{\mathbf{Q}}_x^\top \\
           \tilde{\mathbf{R}}_{3d,y}^f \mathbf{I}_{\color{myblue}{\mathcal{X}_{1d}}} \tilde{\mathbf{Q}}_y^\top \\
           \tilde{\mathbf{R}}_{3d,y}^f \mathbf{I}_{\color{myblue}{\mathcal{X}_{1d}}} \tilde{\mathbf{Q}}_y^\top
         \end{bmatrix}
    \mathbf{P}_\mathrm{ord}.
    \end{aligned}
\end{equation}
In this way, we avoid computing dense matrix-matrix products.
Now, we consider that on the left-hand side, in the top row, the factor $\mathbf{T}^\top \mathbf{T}$ appears in both entries, but we can exploit the expression of $\mathbf{T}$ that we have obtained above:
\begin{align*}
    & \mathbf{T}^\top \mathbf{T} \\
    &=( \mathbf{P}_\mathrm{ord}^\top \tilde{\mathbf{R}}_{3d}^f \mathbf{I}_{\mathcal{X}} \tilde{\mathbf{Q}}^\top\mathbf{P}_\mathrm{ord} )^\top \mathbf{P}_\mathrm{ord}^\top \tilde{\mathbf{R}}_{3d}^f \mathbf{I}_{\mathcal{X}} \tilde{\mathbf{Q}}^\top\mathbf{P}_\mathrm{ord} \\
    &= 
    \mathbf{P}_\mathrm{ord}^\top \tilde{\mathbf{Q}} \mathbf{I}_{\mathcal{X}}^\top (\tilde{\mathbf{R}}_{3d}^f)^\top \mathbf{P}_\mathrm{ord} \mathbf{P}_\mathrm{ord}^\top \tilde{\mathbf{R}}_{3d}^f \mathbf{I}_{\mathcal{X}} \tilde{\mathbf{Q}}^\top\mathbf{P}_\mathrm{ord}\\
    &= \mathbf{P}_\mathrm{ord}^\top \tilde{\mathbf{Q}} \mathbf{I}_{\mathcal{X}} (\tilde{\mathbf{R}}_{3d}^f)^\top \tilde{\mathbf{R}}_{3d}^f \mathbf{I}_{\mathcal{X}} \tilde{\mathbf{Q}}^\top\mathbf{P}_\mathrm{ord} \\
    &= 
    \mathbf{P}_\mathrm{ord}^\top
    \operatorname{diag}\!
    \begin{bmatrix}
       \tilde{\mathbf{R}}_{3d,x}^f \mathbf{I}_{\color{myblue}{\mathcal{X}_{1d}}} \tilde{\mathbf{Q}}_x^\top \\
       \tilde{\mathbf{R}}_{3d,y}^f \mathbf{I}_{\color{myblue}{\mathcal{X}_{1d}}} \tilde{\mathbf{Q}}_y^\top \\
       \tilde{\mathbf{R}}_{3d,z}^f \mathbf{I}_{\color{myblue}{\mathcal{X}_{1d}}} \tilde{\mathbf{Q}}_z^\top \\
   \end{bmatrix}^\top
   \operatorname{diag}\!
   \begin{bmatrix}
       \tilde{\mathbf{R}}_{3d,x}^f \mathbf{I}_{\color{myblue}{\mathcal{X}_{1d}}} \tilde{\mathbf{Q}}_x^\top \\
       \tilde{\mathbf{R}}_{3d,y}^f \mathbf{I}_{\color{myblue}{\mathcal{X}_{1d}}} \tilde{\mathbf{Q}}_y^\top \\
       \tilde{\mathbf{R}}_{3d,z}^f \mathbf{I}_{\color{myblue}{\mathcal{X}_{1d}}} \tilde{\mathbf{Q}}_z^\top 
   \end{bmatrix}
    \mathbf{P}_\mathrm{ord}\\
    &= \mathbf{P}_\mathrm{ord}^\top 
       \operatorname{diag}\!
         \begin{bmatrix}
           \tilde{\mathbf{Q}}_x \mathbf{I}_{\color{myblue}{\mathcal{X}_{1d}}} 
             (\tilde{\mathbf{R}}_{3d,x}^f)^\top 
             \tilde{\mathbf{R}}_{3d,x}^f 
             \mathbf{I}_{\color{myblue}{\mathcal{X}_{1d}}} \tilde{\mathbf{Q}}_x^\top \\
           \tilde{\mathbf{Q}}_y \mathbf{I}_{\color{myblue}{\mathcal{X}_{1d}}} 
             (\tilde{\mathbf{R}}_{3d,y}^f)^\top 
             \tilde{\mathbf{R}}_{3d,y}^f 
             \mathbf{I}_{\color{myblue}{\mathcal{X}_{1d}}} \tilde{\mathbf{Q}}_y^\top \\
           \tilde{\mathbf{Q}}_z \mathbf{I}_{\color{myblue}{\mathcal{X}_{1d}}} 
             (\tilde{\mathbf{R}}_{3d,z}^f)^\top 
             \tilde{\mathbf{R}}_{3d,z}^f 
             \mathbf{I}_{\color{myblue}{\mathcal{X}_{1d}}} \tilde{\mathbf{Q}}_z^\top
         \end{bmatrix}
    \mathbf{P}_\mathrm{ord}.
\end{align*}
We note that $\mathbf{I}_{\mathcal{X}}^\top = \mathbf{I}_{\mathcal{X}}$ since $\mathbf{I}_{\mathcal{X}}$ is block diagonal and $\mathbf{I}_{\color{myblue}{\mathcal{X}_{1d}}}$ is symmetric. Furthermore, since $\tilde{\mathbf{R}}_{3d,j}^f\in\mathbb{R}^{M_i\times\mathcal{N}} \quad \forall j=x,y,z$,  $(\tilde{\mathbf{R}}_{3d,j}^f)^\top \tilde{\mathbf{R}}_{3d,j}^f\in\mathbb{R}^{\mathcal{N}\times\mathcal{N}} \quad \forall j=x,y,z$ and also $\mathbf{I}_{\color{myblue}{\mathcal{X}_{1d}}}(\tilde{\mathbf{R}}_{3d,j}^f)^\top \tilde{\mathbf{R}}_{3d,j}^f \mathbf{I}_{\color{myblue}{\mathcal{X}_{1d}}} \in\mathbb{R}^{\mathcal{N}\times\mathcal{N}} \quad \forall j=x,y,z$. \\ 
Let us focus now on the right-hand side. 
As in 
\begin{alignat*}{2}
    & \vec{y}^\mathrm{obs} = \{\vec\ell_m(\bm{u}_\mathrm{true})\}_{m=1}^M+\{\vec\varepsilon_m\}_{m=1}^M\\ 
    & =\mathbf{R}_{3d}^f \mathbf{I}_{\mathcal{X}} \vec{u}_\mathrm{true}+\vec{\varepsilon} = \mathbf{P}_\mathrm{ord}^\top \tilde{\mathbf{R}}_{3d}^f \mathbf{I}_{\mathcal{X}} \vec{u}_\mathrm{true}+\vec{\varepsilon},
\end{alignat*}
\textcolor{myblue}{a similar simplification occurs also in this case:}
\begin{alignat*}{2}
    & \mathbf{T}^\top \vec{y}^\mathrm{obs} \\
    & = ( \mathbf{P}_\mathrm{ord}^\top \tilde{\mathbf{R}}_{3d}^f \mathbf{I}_{\mathcal{X}} \tilde{\mathbf{Q}}^\top\mathbf{P}_\mathrm{ord} )^\top (\mathbf{P}_\mathrm{ord}^\top \tilde{\mathbf{R}}_{3d}^f \mathbf{I}_{\mathcal{X}} \vec{u}_\mathrm{true}+\vec{\varepsilon}) \\
    & = (\mathbf{P}_\mathrm{ord}^\top \tilde{\mathbf{Q}} \mathbf{I}_{\mathcal{X}}^\top (\tilde{\mathbf{R}}_{3d}^f)^\top \mathbf{P}_\mathrm{ord})(\mathbf{P}_\mathrm{ord}^\top \tilde{\mathbf{R}}_{3d}^f \mathbf{I}_{\mathcal{X}} \vec{u}_\mathrm{true}+\vec{\varepsilon}) \\
    & = \mathbf{P}_\mathrm{ord}^\top \tilde{\mathbf{Q}} \mathbf{I}_{\mathcal{X}} (\tilde{\mathbf{R}}_{3d}^f)^\top \tilde{\mathbf{R}}_{3d}^f \mathbf{I}_{\mathcal{X}} \vec{u}_\mathrm{true} \\
    & + \mathbf{P}_\mathrm{ord}^\top \tilde{\mathbf{Q}} \mathbf{I}_{\mathcal{X}}^\top (\tilde{\mathbf{R}}_{3d}^f)^\top \mathbf{P}_\mathrm{ord}\vec{\varepsilon}\\
    & = \mathbf{P}_\mathrm{ord}^\top
    \operatorname{diag}\!
         \begin{bmatrix}
           \tilde{\mathbf{Q}}_x\mathbf{I}_{\color{myblue}{\mathcal{X}_{1d}}} (\tilde{\mathbf{R}}_{3d,x}^f)^\top \tilde{\mathbf{R}}_{3d,x}^f \mathbf{I}_{\color{myblue}{\mathcal{X}_{1d}}}\\
           \tilde{\mathbf{Q}}_y \mathbf{I}_{\color{myblue}{\mathcal{X}_{1d}}} (\tilde{\mathbf{R}}_{3d,y}^f)^\top \tilde{\mathbf{R}}_{3d,y}^f \mathbf{I}_{\color{myblue}{\mathcal{X}_{1d}}} \\
           \tilde{\mathbf{Q}}_z \mathbf{I}_{\color{myblue}{\mathcal{X}_{1d}}} (\tilde{\mathbf{R}}_{3d,z}^f)^\top \tilde{\mathbf{R}}_{3d,z}^f \mathbf{I}_{\color{myblue}{\mathcal{X}_{1d}}}
         \end{bmatrix}
    \begin{bmatrix}
        \vec{u}_{\mathrm{true},x} \\ \vec{u}_{\mathrm{true},y} \\ \vec{u}_{\mathrm{true},z} \\
    \end{bmatrix}\\
    & + \mathbf{P}_\mathrm{ord}^\top
    \operatorname{diag}\!
         \begin{bmatrix}
           \tilde{\mathbf{Q}}_x\mathbf{I}_{\color{myblue}{\mathcal{X}_{1d}}} (\tilde{\mathbf{R}}_{3d,x}^f)^\top\\
           \tilde{\mathbf{Q}}_y \mathbf{I}_{\color{myblue}{\mathcal{X}_{1d}}} (\tilde{\mathbf{R}}_{3d,y}^f)^\top \\
           \tilde{\mathbf{Q}}_z \mathbf{I}_{\color{myblue}{\mathcal{X}_{1d}}} (\tilde{\mathbf{R}}_{3d,z}^f)^\top
         \end{bmatrix}
    \mathbf{P}_\mathrm{ord}
    \begin{bmatrix}
        \vec{\varepsilon}_x \\ \vec{\varepsilon}_y \\ \vec{\varepsilon}_z \\
    \end{bmatrix}\\    
    & = \mathbf{P}_\mathrm{ord}^\top (
    \tilde{\mathbf{Q}}_x\mathbf{I}_{\color{myblue}{\mathcal{X}_{1d}}} (\tilde{\mathbf{R}}_{3d,x}^f)^\top \tilde{\mathbf{R}}_{3d,x}^f \mathbf{I}_{\color{myblue}{\mathcal{X}_{1d}}} \vec{u}_{\mathrm{true},x}\\
    & + \tilde{\mathbf{Q}}_y \mathbf{I}_{\color{myblue}{\mathcal{X}_{1d}}} (\tilde{\mathbf{R}}_{3d,y}^f)^\top \tilde{\mathbf{R}}_{3d,y}^f \mathbf{I}_{\color{myblue}{\mathcal{X}_{1d}}} \vec{u}_{\mathrm{true},y}\\
    & + \tilde{\mathbf{Q}}_z \mathbf{I}_{\color{myblue}{\mathcal{X}_{1d}}} (\tilde{\mathbf{R}}_{3d,z}^f)^\top \tilde{\mathbf{R}}_{3d,z}^f \mathbf{I}_{\color{myblue}{\mathcal{X}_{1d}}} \vec{u}_{\mathrm{true},z}\\
    & +
    \tilde{\mathbf{Q}}_{x}\mathbf{I}_{\color{myblue}{\mathcal{X}_{1d}}} (\tilde{\mathbf{R}}_{3d,x}^f)^\top \tilde{\vec{\varepsilon}}_x + \tilde{\mathbf{Q}}_{y}\mathbf{I}_{\color{myblue}{\mathcal{X}_{1d}}} (\tilde{\mathbf{R}}_{3d,y}^f)^\top \tilde{\vec{\varepsilon}}_y \\
    & + \tilde{\mathbf{Q}}_{z}\mathbf{I}_{\color{myblue}{\mathcal{X}_{1d}}} (\tilde{\mathbf{R}}_{3d,z}^f)^\top \tilde{\vec{\varepsilon}}_z ).
\end{alignat*}
Note that $\tilde{\vec{\varepsilon}}=\mathbf{P}_\mathrm{ord}\vec{\varepsilon}$. However, since $\vec{\varepsilon}$ is a random vector, permuting its rows returns another random vector. Therefore, we can say $\tilde{\vec{\varepsilon}}\approx\vec{\varepsilon}$ and we can avoid permuting $\vec{\varepsilon}$. Now we analyse $\mathbf{P}$ and, consequently, $\mathbf{T}^\top\mathbf{T}\mathbf{P}^\top$:
\begin{alignat*}{2}
    & \mathbf{P} && = \mathbf{Z}\mathbf{I}_{\mathcal{X}}\mathbf{Q}^\top = \mathbf{Z}\mathbf{I}_{\mathcal{X}} \tilde{\mathbf{Q}}^\top \mathbf{P}_\mathrm{ord} \\
    & && =
    \begin{bmatrix}
        \mathbf{Z}_x & \mathbf{Z}_y & \mathbf{Z}_z
    \end{bmatrix}
    \operatorname{diag}\!
    \begin{bmatrix}
        \mathbf{I}_{\color{myblue}{\mathcal{X}_{1d}}}\\
        \mathbf{I}_{\color{myblue}{\mathcal{X}_{1d}}}\\
        \mathbf{I}_{\color{myblue}{\mathcal{X}_{1d}}}
    \end{bmatrix}
    \operatorname{diag}\!
    \begin{bmatrix}
        \tilde{\mathbf{Q}}_x^\top\\
        \tilde{\mathbf{Q}}_y^\top\\
        \tilde{\mathbf{Q}}_z^\top\\
    \end{bmatrix}
    \mathbf{P}_\mathrm{ord}, \\
    & && =\begin{bmatrix}
        \mathbf{Z}_x\mathbf{I}_{\color{myblue}{\mathcal{X}_{1d}}}\tilde{\mathbf{Q}}_x^\top & \mathbf{Z}_y\mathbf{I}_{\color{myblue}{\mathcal{X}_{1d}}}\tilde{\mathbf{Q}}_y^\top & \mathbf{Z}_z\mathbf{I}_{\color{myblue}{\mathcal{X}_{1d}}}\tilde{\mathbf{Q}}_z^\top
    \end{bmatrix}
    \mathbf{P}_\mathrm{ord}
\end{alignat*}
\begin{alignat*}{2}
    & \mathbf{T}^\top\mathbf{T}\mathbf{P}^\top \\
    & = \mathbf{P}_\mathrm{ord}^\top \tilde{\mathbf{Q}} \mathbf{I}_{\mathcal{X}} (\tilde{\mathbf{R}}_{3d}^f)^\top \tilde{\mathbf{R}}_{3d}^f \mathbf{I}_{\mathcal{X}} \tilde{\mathbf{Q}}^\top\mathbf{P}_\mathrm{ord} \mathbf{P}_\mathrm{ord}^\top \tilde{\mathbf{Q}} \mathbf{I}_{\mathcal{X}}\mathbf{Z}^\top \\
    & = \mathbf{P}_\mathrm{ord}^\top
    \operatorname{diag}\!
        \begin{bmatrix}
           \tilde{\mathbf{Q}}_x\mathbf{I}_{\color{myblue}{\mathcal{X}_{1d}}} (\tilde{\mathbf{R}}_{3d,x}^f)^\top \tilde{\mathbf{R}}_{3d,x}^f \mathbf{I}_{\color{myblue}{\mathcal{X}_{1d}}} \tilde{\mathbf{Q}}_x^\top \\
           \tilde{\mathbf{Q}}_y \mathbf{I}_{\color{myblue}{\mathcal{X}_{1d}}} (\tilde{\mathbf{R}}_{3d,y}^f)^\top \tilde{\mathbf{R}}_{3d,y}^f \mathbf{I}_{\color{myblue}{\mathcal{X}_{1d}}} \tilde{\mathbf{Q}}_y^\top \\
           \tilde{\mathbf{Q}}_z \mathbf{I}_{\color{myblue}{\mathcal{X}_{1d}}} (\tilde{\mathbf{R}}_{3d,z}^f)^\top \tilde{\mathbf{R}}_{3d,z}^f \mathbf{I}_{\color{myblue}{\mathcal{X}_{1d}}} \tilde{\mathbf{Q}}_z^\top \\
       \end{bmatrix}
    \begin{bmatrix}
        \tilde{\mathbf{Q}}_x\mathbf{I}_{\color{myblue}{\mathcal{X}_{1d}}}\mathbf{Z}_x^\top \\ \tilde{\mathbf{Q}}_y\mathbf{I}_{\color{myblue}{\mathcal{X}_{1d}}}\mathbf{Z}_y^\top \\ \tilde{\mathbf{Q}}_z\mathbf{I}_{\color{myblue}{\mathcal{X}_{1d}}}\mathbf{Z}_z^\top
    \end{bmatrix} \\
    & = \mathbf{P}_\mathrm{ord}^\top
    \begin{bmatrix}
        \tilde{\mathbf{Q}}_x\mathbf{I}_{\color{myblue}{\mathcal{X}_{1d}}} (\tilde{\mathbf{R}}_{3d,x}^f)^\top \tilde{\mathbf{R}}_{3d,x}^f \mathbf{I}_{\color{myblue}{\mathcal{X}_{1d}}} \tilde{\mathbf{Q}}_x^\top\tilde{\mathbf{Q}}_x\mathbf{I}_{\color{myblue}{\mathcal{X}_{1d}}}\mathbf{Z}_x^\top \\
        \tilde{\mathbf{Q}}_y\mathbf{I}_{\color{myblue}{\mathcal{X}_{1d}}} (\tilde{\mathbf{R}}_{3d,x}^f)^\top \tilde{\mathbf{R}}_{3d,x}^f \mathbf{I}_{\color{myblue}{\mathcal{X}_{1d}}} \tilde{\mathbf{Q}}_y^\top\tilde{\mathbf{Q}}_y\mathbf{I}_{\color{myblue}{\mathcal{X}_{1d}}}\mathbf{Z}_y^\top \\
        \tilde{\mathbf{Q}}_z\mathbf{I}_{\color{myblue}{\mathcal{X}_{1d}}} (\tilde{\mathbf{R}}_{3d,x}^f)^\top \tilde{\mathbf{R}}_{3d,x}^f \mathbf{I}_{\color{myblue}{\mathcal{X}_{1d}}} \tilde{\mathbf{Q}}_z^\top\tilde{\mathbf{Q}}_z\mathbf{I}_{\color{myblue}{\mathcal{X}_{1d}}}\mathbf{Z}_z^\top
    \end{bmatrix}.
\end{alignat*}
\textcolor{myblue}{In the first equation $$(M\xi\mathbf{I}_M + \mathbf{T}^\top \mathbf{T})\vec{\eta}_\xi^* + \mathbf{T}^\top \mathbf{T}\mathbf{P}^\top \vec{z}_\xi^* = \mathbf{T}^\top \vec{y}^\mathrm{obs}$$ of System ~\eqref{eq:pbdw_system_cardiac}, every term can be written with a left factor $\mathbf{P}_\mathrm{ord}^\top$, except for the term $M\xi\mathbf{I}_M$, which remains unaffected.}
Thus, we obtain:
\begin{align*}
    & (M\xi\mathbf{I}_M + \mathbf{P}_\mathrm{ord}^\top \tilde{\mathbf{Q}} \mathbf{I}_{\mathcal{X}} (\tilde{\mathbf{R}}_{3d}^f)^\top \tilde{\mathbf{R}}_{3d}^f \mathbf{I}_{\mathcal{X}} \tilde{\mathbf{Q}}^\top\mathbf{P}_\mathrm{ord})\vec{\eta}_\xi^* \\& + \mathbf{P}_\mathrm{ord}^\top \tilde{\mathbf{Q}} \mathbf{I}_{\mathcal{X}} (\tilde{\mathbf{R}}_{3d}^f)^\top \tilde{\mathbf{R}}_{3d}^f \mathbf{I}_{\mathcal{X}} \tilde{\mathbf{Q}}^\top\tilde{\mathbf{Q}} \mathbf{I}_{\mathcal{X}}\mathbf{Z}^\top \vec{z}_\xi^* \\ 
    & = \mathbf{P}_\mathrm{ord}^\top (
    \tilde{\mathbf{Q}}_x\mathbf{I}_{\color{myblue}{\mathcal{X}_{1d}}} (\tilde{\mathbf{R}}_{3d,x}^f)^\top \tilde{\mathbf{R}}_{3d,x}^f \mathbf{I}_{\color{myblue}{\mathcal{X}_{1d}}} \vec{u}_{\mathrm{true},x} \\
    & + \tilde{\mathbf{Q}}_y \mathbf{I}_{\color{myblue}{\mathcal{X}_{1d}}} (\tilde{\mathbf{R}}_{3d,y}^f)^\top \tilde{\mathbf{R}}_{3d,y}^f \mathbf{I}_{\color{myblue}{\mathcal{X}_{1d}}} \vec{u}_{\mathrm{true},y} \\ 
    & + \tilde{\mathbf{Q}}_z \mathbf{I}_{\color{myblue}{\mathcal{X}_{1d}}} (\tilde{\mathbf{R}}_{3d,z}^f)^\top \tilde{\mathbf{R}}_{3d,z}^f \mathbf{I}_{\color{myblue}{\mathcal{X}_{1d}}} \vec{u}_{\mathrm{true},z} \\
    & + \tilde{\mathbf{Q}}_{x}\mathbf{I}_{\color{myblue}{\mathcal{X}_{1d}}} (\tilde{\mathbf{R}}_{3d,x}^f)^\top \tilde{\vec{\varepsilon}}_x + \tilde{\mathbf{Q}}_{y}\mathbf{I}_{\color{myblue}{\mathcal{X}_{1d}}} (\tilde{\mathbf{R}}_{3d,y}^f)^\top \tilde{\vec{\varepsilon}}_y \\
    & + \tilde{\mathbf{Q}}_{z}\mathbf{I}_{\color{myblue}{\mathcal{X}_{1d}}} (\tilde{\mathbf{R}}_{3d,z}^f)^\top \tilde{\vec{\varepsilon}}_z ).
\end{align*}
\textcolor{myblue}{Therefore, we multiply both sides on the left with $\mathbf{P}_\mathrm{ord}$ to take advantage of $\mathbf{P}_\mathrm{ord}\mathbf{P}_\mathrm{ord}^\top=\mathbf{I}_M$.}
\begin{align*}
    & (M\xi\mathbf{P}_\mathrm{ord}\mathbf{I}_M + \tilde{\mathbf{Q}} \mathbf{I}_{\mathcal{X}} (\tilde{\mathbf{R}}_{3d}^f)^\top \tilde{\mathbf{R}}_{3d}^f \mathbf{I}_{\mathcal{X}} \tilde{\mathbf{Q}}^\top\mathbf{P}_\mathrm{ord})\vec{\eta}_\xi^* \\ 
    & +\tilde{\mathbf{Q}} \mathbf{I}_{\mathcal{X}} (\tilde{\mathbf{R}}_{3d}^f)^\top \tilde{\mathbf{R}}_{3d}^f \mathbf{I}_{\mathcal{X}} \tilde{\mathbf{Q}}^\top\tilde{\mathbf{Q}} \mathbf{I}_{\mathcal{X}}\mathbf{Z}^\top \vec{z}_\xi^* \\ 
    & = \tilde{\mathbf{Q}}_x\mathbf{I}_{\color{myblue}{\mathcal{X}_{1d}}} (\tilde{\mathbf{R}}_{3d,x}^f)^\top \tilde{\mathbf{R}}_{3d,x}^f \mathbf{I}_{\color{myblue}{\mathcal{X}_{1d}}} \vec{u}_{\mathrm{true},x} \\&+
   \tilde{\mathbf{Q}}_y \mathbf{I}_{\color{myblue}{\mathcal{X}_{1d}}} (\tilde{\mathbf{R}}_{3d,y}^f)^\top \tilde{\mathbf{R}}_{3d,y}^f \mathbf{I}_{\color{myblue}{\mathcal{X}_{1d}}} \vec{u}_{\mathrm{true},y} \\&+
    \tilde{\mathbf{Q}}_z \mathbf{I}_{\color{myblue}{\mathcal{X}_{1d}}} (\tilde{\mathbf{R}}_{3d,z}^f)^\top \tilde{\mathbf{R}}_{3d,z}^f \mathbf{I}_{\color{myblue}{\mathcal{X}_{1d}}} \vec{u}_{\mathrm{true},z} \\
    & + \tilde{\mathbf{Q}}_{x}\mathbf{I}_{\color{myblue}{\mathcal{X}_{1d}}} (\tilde{\mathbf{R}}_{3d,x}^f)^\top \tilde{\vec{\varepsilon}}_x + \tilde{\mathbf{Q}}_{y}\mathbf{I}_{\color{myblue}{\mathcal{X}_{1d}}} (\tilde{\mathbf{R}}_{3d,y}^f)^\top \tilde{\vec{\varepsilon}}_y \\
    & + \tilde{\mathbf{Q}}_{z}\mathbf{I}_{\color{myblue}{\mathcal{X}_{1d}}} (\tilde{\mathbf{R}}_{3d,z}^f)^\top \tilde{\vec{\varepsilon}}_z.
\end{align*}
The second equation $$\mathbf{P}\vec{\eta}_\xi^*=\vec{0}$$ from Sys.~\eqref{eq:pbdw_system_cardiac} is as follows:
\begin{equation*}
    \mathbf{Z}\mathbf{I}_{\mathcal{X}} \tilde{\mathbf{Q}}^\top \mathbf{P}_\mathrm{ord}\vec{z}_\xi^*=\vec{0}.
\end{equation*}
In this manner we have reduced the computations involving $\mathbf{P}_\mathrm{ord}$ to three only.
Finally, we elaborate on a couple of matrix-matrix products appearing in \gls{womp}:
\begin{alignat*}{2}
    & \mathbf{P}\mathbf{P}^\top = \mathbf{Z}\mathbf{I}_{\mathcal{X}} \tilde{\mathbf{Q}}^\top \mathbf{P}_\mathrm{ord} (\mathbf{Z}\mathbf{I}_{\mathcal{X}} \tilde{\mathbf{Q}}^\top \mathbf{P}_\mathrm{ord})^\top\\ 
    & = \mathbf{Z}\mathbf{I}_{\mathcal{X}} \tilde{\mathbf{Q}}^\top \mathbf{P}_\mathrm{ord} \mathbf{P}_\mathrm{ord}^\top \tilde{\mathbf{Q}} \mathbf{I}_{\mathcal{X}} \mathbf{Z}^\top\\
    & =
    \begin{bmatrix}
        \mathbf{Z}_x\mathbf{I}_{\color{myblue}{\mathcal{X}_{1d}}}\tilde{\mathbf{Q}}_x^\top & \mathbf{Z}_y\mathbf{I}_{\color{myblue}{\mathcal{X}_{1d}}}\tilde{\mathbf{Q}}_y^\top & \mathbf{Z}_z\mathbf{I}_{\color{myblue}{\mathcal{X}_{1d}}}\tilde{\mathbf{Q}}_z^\top
    \end{bmatrix}
    \begin{bmatrix}
        \tilde{\mathbf{Q}}_x\mathbf{I}_{\color{myblue}{\mathcal{X}_{1d}}}\mathbf{Z}_x^\top \\ \tilde{\mathbf{Q}}_y\mathbf{I}_{\color{myblue}{\mathcal{X}_{1d}}}\mathbf{Z}_y^\top \\ \tilde{\mathbf{Q}}_z\mathbf{I}_{\color{myblue}{\mathcal{X}_{1d}}}\mathbf{Z}_z^\top
    \end{bmatrix} \\   
    & = \mathbf{Z}_x\mathbf{I}_{\color{myblue}{\mathcal{X}_{1d}}}\tilde{\mathbf{Q}}_x^\top\tilde{\mathbf{Q}}_x\mathbf{I}_{\color{myblue}{\mathcal{X}_{1d}}}\mathbf{Z}_x^\top + \mathbf{Z}_y\mathbf{I}_{\color{myblue}{\mathcal{X}_{1d}}}\tilde{\mathbf{Q}}_y^\top\tilde{\mathbf{Q}}_y\mathbf{I}_{\color{myblue}{\mathcal{X}_{1d}}}\mathbf{Z}_y^\top\\ 
    & + \mathbf{Z}_z\mathbf{I}_{\color{myblue}{\mathcal{X}_{1d}}}\tilde{\mathbf{Q}}_z^\top\tilde{\mathbf{Q}}_z\mathbf{I}_{\color{myblue}{\mathcal{X}_{1d}}}\mathbf{Z}_z^\top,
\end{alignat*}
\begin{alignat*}{2}
    & (\mathbf{Z}^\top - \mathbf{Q}^\top\mathbf{P}^\top)\vec{v}_\beta \\
    & = (\mathbf{Z}^\top - \tilde{\mathbf{Q}}^\top\mathbf{P}_\mathrm{ord} (\mathbf{Z}\mathbf{I}_{\mathcal{X}} \tilde{\mathbf{Q}}^\top \mathbf{P}_\mathrm{ord})^\top)\vec{v}_\beta \\
    & = (\mathbf{Z}^\top - \tilde{\mathbf{Q}}^\top\mathbf{P}_\mathrm{ord} \mathbf{P}_\mathrm{ord}^\top\tilde{\mathbf{Q}}\mathbf{I}_{\mathcal{X}} \mathbf{Z}^\top )\vec{v}_\beta \\
    & = (\mathbf{I}_{3\mathcal{N}} - \tilde{\mathbf{Q}}^\top\tilde{\mathbf{Q}}\mathbf{I}_{\mathcal{X}})\mathbf{Z}^\top\vec{v}_\beta \\
    & = 
    \begin{bmatrix}
        \mathbf{I}_\mathcal{N} - \tilde{\mathbf{Q}}_x^\top\tilde{\mathbf{Q}}_x\mathbf{I}_{\color{myblue}{\mathcal{X}_{1d}}} \\
        \mathbf{I}_\mathcal{N} - \tilde{\mathbf{Q}}_y^\top\tilde{\mathbf{Q}}_y\mathbf{I}_{\color{myblue}{\mathcal{X}_{1d}}} \\
        \mathbf{I}_\mathcal{N} - \tilde{\mathbf{Q}}_z^\top\tilde{\mathbf{Q}}_z\mathbf{I}_{\color{myblue}{\mathcal{X}_{1d}}}
    \end{bmatrix}
    \mathbf{Z}^\top\vec{v}_\beta \\
    & = 
    \begin{bmatrix}
        (\mathbf{I}_\mathcal{N} - \tilde{\mathbf{Q}}_x^\top\tilde{\mathbf{Q}}_x\mathbf{I}_{\color{myblue}{\mathcal{X}_{1d}}})\mathbf{Z}_x^\top \\
        (\mathbf{I}_\mathcal{N} - \tilde{\mathbf{Q}}_y^\top\tilde{\mathbf{Q}}_y\mathbf{I}_{\color{myblue}{\mathcal{X}_{1d}}})\mathbf{Z}_y^\top \\
        (\mathbf{I}_\mathcal{N} - \tilde{\mathbf{Q}}_z^\top\tilde{\mathbf{Q}}_z\mathbf{I}_{\color{myblue}{\mathcal{X}_{1d}}})\mathbf{Z}_z^\top
    \end{bmatrix}    
    \vec{v}_\beta.
\end{alignat*}
Since these operations are performed inside an iterative loop where $\mathcal{U}_M$ is being constructed, the dimensions of $\mathbf{Q}$ are increasing and the same happens for $\tilde{\mathbf{Q}}_x,\tilde{\mathbf{Q}}_y,\tilde{\mathbf{Q}}_z$. Then, we compute  both quantities above incrementally, $i.e.$ we add the contribution for each new set of vectors appearing in $\tilde{\mathbf{Q}}_i \quad i=x,y,z$ at each iteration by updating the result obtained at the previous iteration.
This is possible only because $\tilde{\mathbf{Q}}_x,\tilde{\mathbf{Q}}_y,\tilde{\mathbf{Q}}_z$ store orthonormal vectors.

In conclusion, thanks to a row permutation $\mathbf{P}_\mathrm{ord}$, it is possible to preserve sparsity in matrix-matrix products, reducing the computational burden and the memory allocation at the same time. 

\printglossary[type=abbreviations]  




\end{appendices}


\bibliography{refs}

@BOOK{Rudin1990,
  title     = "Functional Analysis",
  author    = "Rudin, Walter",
  publisher = "McGraw Hill Higher Education",
  series    = "International Series in Pure \& Applied Mathematics",
  edition   =  2,
  month     =  oct,
  year      =  1990,
  address   = "Maidenhead, England",
  language  = "en"
}

@article{Roth2020,
  title = {Global Burden of Cardiovascular Diseases and Risk Factors,  1990–2019},
  volume = {76},
  ISSN = {0735-1097},
  url = {http://dx.doi.org/10.1016/j.jacc.2020.11.010},
  DOI = {10.1016/j.jacc.2020.11.010},
  number = {25},
  journal = {Journal of the American College of Cardiology},
  publisher = {Elsevier BV},
  author = {Roth,  Gregory A. and Mensah,  George A. and others},
  year = {2020},
  month = dec,
  pages = {2982–3021}
}

@article{Dickopf2014,
  title = {Evaluating Local Approximations of the L2-Orthogonal Projection Between Non-Nested Finite Element Spaces},
  volume = {7},
  ISSN = {1004-8979},
  url = {http://dx.doi.org/10.4208/nmtma.2014.1218nm},
  DOI = {10.4208/nmtma.2014.1218nm},
  number = {3},
  journal = {Numerical Mathematics: Theory,  Methods and Applications},
  publisher = {Global Science Press},
  author = {Thomas Dickopf & Rolf Krause},
  year = {2014},
  month = jul,
  pages = {288–316}
}

@article{Strocchi2023,
  title = {Cell to whole organ global sensitivity analysis on a four-chamber heart electromechanics model using Gaussian processes emulators},
  volume = {19},
  ISSN = {1553-7358},
  url = {http://dx.doi.org/10.1371/journal.pcbi.1011257},
  DOI = {10.1371/journal.pcbi.1011257},
  number = {6},
  journal = {PLOS Computational Biology},
  publisher = {Public Library of Science (PLoS)},
  author = {Strocchi,  Marina and Longobardi,  Stefano and Augustin,  Christoph M. and Gsell,  Matthias A. F. and Petras,  Argyrios and Rinaldi,  Christopher A. and Vigmond,  Edward J. and Plank,  Gernot and Oates,  Chris J. and Wilkinson,  Richard D. and Niederer,  Steven A.},
  editor = {Tsaneva-Atanasova,  Krasimira},
  year = {2023},
  month = jun,
  pages = {e1011257}
}

@article{CorralAcero2020,
  title = {The ‘Digital Twin’ to enable the vision of precision cardiology},
  volume = {41},
  ISSN = {1522-9645},
  url = {http://dx.doi.org/10.1093/eurheartj/ehaa159},
  DOI = {10.1093/eurheartj/ehaa159},
  number = {48},
  journal = {European Heart Journal},
  publisher = {Oxford University Press (OUP)},
  author = {Corral-Acero,  Jorge and Margara,  Francesca and Marciniak,  Maciej and Rodero,  Cristobal and Loncaric,  Filip and Feng,  Yingjing and Gilbert,  Andrew and Fernandes,  Joao F and Bukhari,  Hassaan A and Wajdan,  Ali and Martinez,  Manuel Villegas and Santos,  Mariana Sousa and Shamohammdi,  Mehrdad and Luo,  Hongxing and Westphal,  Philip and Leeson,  Paul and DiAchille,  Paolo and Gurev,  Viatcheslav and Mayr,  Manuel and Geris,  Liesbet and Pathmanathan,  Pras and Morrison,  Tina and Cornelussen,  Richard and Prinzen,  Frits and Delhaas,  Tammo and Doltra,  Ada and Sitges,  Marta and Vigmond,  Edward J and Zacur,  Ernesto and Grau,  Vicente and Rodriguez,  Blanca and Remme,  Espen W and Niederer,  Steven and Mortier,  Peter and McLeod,  Kristin and Potse,  Mark and Pueyo,  Esther and Bueno-Orovio,  Alfonso and Lamata,  Pablo},
  year = {2020},
  month = mar,
  pages = {4556–4564}
}

@article{Liu2021, 
title={The impact of myocardial compressibility on organ-level simulations of the normal and infarcted heart}, volume={11}, 
ISSN={2045-2322}, 
DOI={10.1038/s41598-021-92810-y}, number={1}, journal={Scientific Reports}, 
publisher={Springer Science and Business Media LLC}, 
author={Liu, Hao and Soares, JoÃ£o S. and Walmsley, John and Li, David S. and Raut, Samarth and Avazmohammadi, Reza and Iaizzo, Paul and Palmer, Mark and Gorman, Joseph H. and Gorman, Robert C. and Sacks, Michael S.}, 
year={2021}, 
month=jun }

@article{Augustin2016,
  title = {Anatomically accurate high resolution modeling of human whole heart electromechanics: A strongly scalable algebraic multigrid solver method for nonlinear deformation},
  volume = {305},
  ISSN = {0021-9991},
  url = {http://dx.doi.org/10.1016/j.jcp.2015.10.045},
  DOI = {10.1016/j.jcp.2015.10.045},
  journal = {Journal of Computational Physics},
  publisher = {Elsevier BV},
  author = {Augustin,  Christoph M. and Neic,  Aurel and Liebmann,  Manfred and Prassl,  Anton J. and Niederer,  Steven A. and Haase,  Gundolf and Plank,  Gernot},
  year = {2016},
  month = jan,
  pages = {622–646}
}

@article{Nasopoulou2017,
  title = {Improved identifiability of myocardial material parameters by an energy-based cost function},
  volume = {16},
  ISSN = {1617-7940},
  url = {http://dx.doi.org/10.1007/s10237-016-0865-3},
  DOI = {10.1007/s10237-016-0865-3},
  number = {3},
  journal = {Biomechanics and Modeling in Mechanobiology},
  publisher = {Springer Science and Business Media LLC},
  author = {Nasopoulou,  Anastasia and Shetty,  Anoop and Lee,  Jack and Nordsletten,  David and Rinaldi,  C. Aldo and Lamata,  Pablo and Niederer,  Steven},
  year = {2017},
  month = feb,
  pages = {971–988}
}

@article{Xi2013,
  title = {The estimation of patient-specific cardiac diastolic functions from clinical measurements},
  volume = {17},
  ISSN = {1361-8415},
  url = {http://dx.doi.org/10.1016/j.media.2012.08.001},
  DOI = {10.1016/j.media.2012.08.001},
  number = {2},
  journal = {Medical Image Analysis},
  publisher = {Elsevier BV},
  author = {Xi,  Jiahe and Lamata,  Pablo and Niederer,  Steven and Land,  Sander and Shi,  Wenzhe and Zhuang,  Xiahai and Ourselin,  Sebastien and Duckett,  Simon G. and Shetty,  Anoop K. and Rinaldi,  C. Aldo and Rueckert,  Daniel and Razavi,  Reza and Smith,  Nic P.},
  year = {2013},
  month = feb,
  pages = {133–146}
}

@article{Bayer2012,
  title = {A Novel Rule-Based Algorithm for Assigning Myocardial Fiber Orientation to Computational Heart Models},
  volume = {40},
  ISSN = {1573-9686},
  url = {http://dx.doi.org/10.1007/s10439-012-0593-5} ,
  DOI = {10.1007/s10439-012-0593-5},
  number = {10},
  journal = {Annals of Biomedical Engineering},
  publisher = {Springer Science and Business Media LLC},
  author = {Bayer,  J. D. and Blake,  R. C. and Plank,  G. and Trayanova,  N. A.},
  year = {2012},
  month = may,
  pages = {2243–2254}
}

@article{Galarce2023,
  title = {Displacement and Pressure Reconstruction from Magnetic Resonance Elastography Images: Application to an In Silico Brain Model},
  volume = {16},
  ISSN = {1936-4954},
  url = {http://dx.doi.org/10.1137/22M149363X} ,
  DOI = {10.1137/22m149363x},
  number = {2},
  journal = {SIAM Journal on Imaging Sciences},
  publisher = {Society for Industrial & Applied Mathematics (SIAM)},
  author = {Galarce,  Felipe and Tabelow,  Karsten and Polzehl,  J\"{o}rg and Papanikas,  Christos Panagiotis and Vavourakis,  Vasileios and Lilaj,  Ledia and Sack,  Ingolf and Caiazzo,  Alfonso},
  year = {2023},
  month = jun,
  pages = {996–1027}
}

@article{Galarce2020,
  title = {Reconstructing haemodynamics quantities of interest from Doppler ultrasound imaging},
  volume = {37},
  ISSN = {2040-7947},
  url = {http://dx.doi.org/10.1002/cnm.3416} ,
  DOI = {10.1002/cnm.3416},
  number = {2},
  journal = {International Journal for Numerical Methods in Biomedical Engineering},
  publisher = {Wiley},
  author = {Galarce,  Felipe and Lombardi,  Damiano and Mula,  Olga},
  year = {2020},
  month = dec 
}

@article{Barone2020,
  title = {Experimental validation of a variational data assimilation procedure for estimating space-dependent cardiac conductivities},
  volume = {359},
  ISSN = {0045-7825},
  url = {http://dx.doi.org/10.1016/j.cma.2019.112615}       ,
  DOI = {10.1016/j.cma.2019.112615},
  journal = {Computer Methods in Applied Mechanics and Engineering},
  publisher = {Elsevier BV},
  author = {Barone,  Alessandro and Gizzi,  Alessio and Fenton,  Flavio and Filippi,  Simonetta and Veneziani,  Alessandro},
  year = {2020},
  month = jan,
  pages = {112615}
}

@article{Farazmand2024,
  title = {Sparse Discrete Empirical Interpolation Method: State Estimation from Few Sensors},
  volume = {46},
  ISSN = {1095-7197},
  url = {http://dx.doi.org/10.1137/24M1636344} , 
  DOI = {10.1137/24m1636344},
  number = {6},
  journal = {SIAM Journal on Scientific Computing},
  publisher = {Society for Industrial & Applied Mathematics (SIAM)},
  author = {Farazmand,  Mohammad},
  year = {2024},
  month = dec,
  pages = {A3658–A3680}
}

@article{Bidar2024,
  title = {Sensor placement for data assimilation of turbulence models using eigenspace perturbations},
  volume = {36},
  ISSN = {1089-7666},
  url = {http://dx.doi.org/10.1063/5.0182080} ,
  DOI = {10.1063/5.0182080},
  number = {1},
  journal = {Physics of Fluids},
  publisher = {AIP Publishing},
  author = {Bidar,  O. and Anderson,  S. R. and Qin,  N.},
  year = {2024},
  month = jan 
}

@article{Pan2024,
  title = {Unrolled and rapid motion-compensated reconstruction for cardiac CINE MRI},
  volume = {91},
  ISSN = {1361-8415},
  url = {http://dx.doi.org/10.1016/j.media.2023.103017}  ,
  DOI = {10.1016/j.media.2023.103017},
  journal = {Medical Image Analysis},
  publisher = {Elsevier BV},
  author = {Pan,  Jiazhen and Hamdi,  Manal and Huang,  Wenqi and Hammernik,  Kerstin and Kuestner,  Thomas and Rueckert,  Daniel},
  year = {2024},
  month = jan,
  pages = {103017}
}

@article{Gomez2014,
  title = {Accurate High-Resolution Measurements of 3-D Tissue Dynamics With Registration-Enhanced Displacement Encoded MRI},
  volume = {33},
  ISSN = {1558-254X},
  url = {http://dx.doi.org/10.1109/TMI.2014.2311755}  ,
  DOI = {10.1109/tmi.2014.2311755},
  number = {6},
  journal = {IEEE Transactions on Medical Imaging},
  publisher = {Institute of Electrical and Electronics Engineers (IEEE)},
  author = {Gomez,  Arnold D. and Merchant,  Samer S. and Hsu,  Edward W.},
  year = {2014},
  month = jun,
  pages = {1350–1362}
}

@article{Karabelas2022,
  title = {An accurate,  robust,  and efficient finite element framework with applications to anisotropic,  nearly and fully incompressible elasticity},
  volume = {394},
  ISSN = {0045-7825},
  url = {http://dx.doi.org/10.1016/j.cma.2022.114887} ,
  DOI = {10.1016/j.cma.2022.114887},
  journal = {Computer Methods in Applied Mechanics and Engineering},
  publisher = {Elsevier BV},
  author = {Karabelas,  Elias and Gsell,  Matthias A.F. and Haase,  Gundolf and Plank,  Gernot and Augustin,  Christoph M.},
  year = {2022},
  month = may,
  pages = {114887}
}

@article{Wiputra2020,
  title = {Cardiac motion estimation from medical images: a regularisation framework applied on pairwise image registration displacement fields},
  volume = {10},
  ISSN = {2045-2322},
  url = {http://dx.doi.org/10.1038/s41598-020-75525-4}  ,
  DOI = {10.1038/s41598-020-75525-4},
  number = {1},
  journal = {Scientific Reports},
  publisher = {Springer Science and Business Media LLC},
  author = {Wiputra,  Hadi and Chan,  Wei Xuan and Foo,  Yoke Yin and Ho,  Sheldon and Yap,  Choon Hwai},
  year = {2020},
  month = oct 
}

@article{Meng2024,
  title = {DeepMesh: Mesh-Based Cardiac Motion Tracking Using Deep Learning},
  volume = {43},
  ISSN = {1558-254X},
  url = {http://dx.doi.org/10.1109/TMI.2023.3340118}  ,
  DOI = {10.1109/tmi.2023.3340118},
  number = {4},
  journal = {IEEE Transactions on Medical Imaging},
  publisher = {Institute of Electrical and Electronics Engineers (IEEE)},
  author = {Meng,  Qingjie and Bai,  Wenjia and O’Regan,  Declan P. and Rueckert,  Daniel},
  year = {2024},
  month = apr,
  pages = {1489–1500}
}

@article{caforio2021coupling,
	title        = {A coupling strategy for a first 3D-1D model of the cardiovascular system to study the effects of pulse wave propagation on cardiac function},
	author       = {Caforio, Federica and Augustin, Christoph M. and Alastruey, Jordi and Gsell, Matthias A. F. and others},
	year         = 2022,
	month        = {July},
	journal      = {Computational Mechanics},
	publisher    = {Springer Science and Business Media LLC},
	volume       = 70,
	number       = 4,
	pages        = {703â722},
	doi          = {10.1007/s00466-022-02206-6},
	issn         = {1432-0924},
	url          = {http://dx.doi.org/10.1007/s00466-022-02206-6} ,
}

@article{marx2021efficient,
	title        = {Robust and efficient fixed-point algorithm for the inverse elastostatic problem to identify myocardial passive material parameters and the unloaded reference configuration},
	author       = {Marx, Laura and Niestrawska, Justyna A. and Gsell, Matthias A.F. and Caforio, Federica and others},
	year         = 2022,
	month        = {August},
	journal      = {Journal of Computational Physics},
	publisher    = {Elsevier BV},
	volume       = 463,
	pages        = 111266,
	doi          = {10.1016/j.jcp.2022.111266},
	issn         = {0021-9991},
	url          = {http://dx.doi.org/10.1016/j.jcp.2022.111266},
}

@article{Salvador2024,
  title = {Whole-heart electromechanical simulations using Latent Neural Ordinary Differential Equations},
  volume = {7},
  ISSN = {2398-6352},
  url = {http://dx.doi.org/10.1038/s41746-024-01084-x}     ,
  DOI = {10.1038/s41746-024-01084-x},
  number = {1},
  journal = {npj Digital Medicine},
  publisher = {Springer Science and Business Media LLC},
  author = {Salvador,  Matteo and Strocchi,  Marina and Regazzoni,  Francesco and Augustin,  Christoph M. and Dede’,  Luca and Niederer,  Steven A. and Quarteroni,  Alfio},
  year = {2024},
  month = apr 
}

@article{Pfaller2020,
  title = {Using parametric model order reduction for inverse analysis of large nonlinear cardiac simulations},
  volume = {36},
  ISSN = {2040-7947},
  url = {http://dx.doi.org/10.1002/cnm.3320}    ,
  DOI = {10.1002/cnm.3320},
  number = {4},
  journal = {International Journal for Numerical Methods in Biomedical Engineering},
  publisher = {Wiley},
  author = {Pfaller,  M. R. and Cruz Varona,  M. and Lang,  J. and Bertoglio,  C. and Wall,  W. A.},
  year = {2020},
  month = feb 
}

@article{Bonomi2017,
  title = {A matrix DEIM technique for model reduction of nonlinear parametrized problems in cardiac mechanics},
  volume = {324},
  ISSN = {0045-7825},
  url = {http://dx.doi.org/10.1016/j.cma.2017.06.011}    ,
  DOI = {10.1016/j.cma.2017.06.011},
  journal = {Computer Methods in Applied Mechanics and Engineering},
  publisher = {Elsevier BV},
  author = {Bonomi,  Diana and Manzoni,  Andrea and Quarteroni,  Alfio},
  year = {2017},
  month = sep,
  pages = {300–326}
}

@misc{Gong2019,
      title={PBDW method for state estimation: error analysis for noisy data and nonlinear formulation}, 
      author={Helin Gong and Yvon Maday and Olga Mula and Tommaso Taddei},
      year={2019},
      eprint={1906.00810},
      archivePrefix={arXiv},
      primaryClass={math.NA},
      url={https://arxiv.org/abs/1906.00810}, 
}

@article{Balaban2017,
  title = {High‐resolution data assimilation of cardiac mechanics applied to a dyssynchronous ventricle},
  volume = {33},
  ISSN = {2040-7947},
  url = {http://dx.doi.org/10.1002/cnm.2863}       ,
  doi = {10.1002/cnm.2863},
  number = {11},
  journal = {International Journal for Numerical Methods in Biomedical Engineering},
  publisher = {Wiley},
  author = {Balaban,  Gabriel and Finsberg,  Henrik and  Odland,  Hans Henrik and Rognes,  Marie E. and Ross,  Stian and Sundnes,  Joakim and Wall,  Samuel},
  year = {2017},
  month = apr 
}

@article{Balaban2018,
  title = {In vivo estimation of elastic heterogeneity in an infarcted human heart},
  volume = {17},
  ISSN = {1617-7940},
  url = {http://dx.doi.org/10.1007/s10237-018-1028-5}       ,
  DOI = {10.1007/s10237-018-1028-5},
  number = {5},
  journal = {Biomechanics and Modeling in Mechanobiology},
  publisher = {Springer Science and Business Media LLC},
  author = {Balaban,  Gabriel and Finsberg,  Henrik and Funke,  Simon and Håland,  Trine F. and Hopp,  Einar and Sundnes,  Joakim and Wall,  Samuel and Rognes,  Marie E.},
  year = {2018},
  month = may,
  pages = {1317–1329}
}

@article{Chabiniok2016,
  title={Multiphysics and multiscale modelling, data-model fusion and integration of organ physiology in the clinic: ventricular cardiac mechanics},
  author={Chabiniok, R and Wang, VY and Hadjicharalambous, M and others},
  journal={Interface Focus},
  volume={6},
  number={2},
  pages={20150083},
  year={2016}
}

@article{Imperiale2021,
  title={Sequential data assimilation for mechanical systems with complex image data: application to tagged-MRI in cardiac mechanics},
  author={Imperiale, A and Chapelle, D and Moireau, P},
  journal={Advanced Modeling and Simulation in Engineering Sciences},
  volume={8},
  pages={2},
  year={2021}
}

@article{Cicci2024,
  title={Efficient approximation of cardiac mechanics through reduced order modeling with deep learning-based operator approximation},
  author={Cicci, L and Fresca, S and Manzoni, A and Quarteroni, A},
  journal={International Journal for Numerical Methods in Biomedical Engineering},
  volume={40},
  number={1},
  pages={e3783},
  year={2024}
}

@article{Asner2016,
  title={Estimation of passive and active properties in the human heart using 3D tagged MRI},
  author={Asner, L and Hadjicharalambous, M and Chabiniok, R and others},
  journal={Biomechanics and Modeling in Mechanobiology},
  volume={15},
  pages={1121--1139},
  year={2016}
}

@article{Maday2017,
author = {Maday, Yvon and Taddei, Tommaso},
year = {2017},
month = {12},
pages = {},
title = {Adaptive PBDW Approach to State Estimation: Noisy Observations; User-Defined Update Spaces},
volume = {41},
journal = {SIAM Journal on Scientific Computing},
doi = {10.1137/18M116544X}
}

@article{Guccione1991Passive,
	title        = {Passive Material Properties of Intact Ventricular Myocardium Determined From a Cylindrical Model},
	author       = {Guccione, J. M. and McCulloch, A. D. and Waldman, L. K.},
	year         = 1991,
	month        = {February},
	journal      = {Journal of Biomechanical Engineering},
	publisher    = {ASME International},
	volume       = 113,
	number       = 1,
	pages        = {42–55},
	doi          = {10.1115/1.2894084},
	issn         = {1528-8951},
	url          = {http://dx.doi.org/10.1115/1.2894084}    ,
}

@article{Caforio2024,
  title = {Physics-informed neural network estimation of material properties in soft tissue nonlinear biomechanical models},
  volume = {75},
  ISSN = {1432-0924},
  url = {http://dx.doi.org/10.1007/s00466-024-02516-x},
  DOI = {10.1007/s00466-024-02516-x},
  number = {2},
  journal = {Computational Mechanics},
  publisher = {Springer Science and Business Media LLC},
  author = {Caforio,  Federica and Regazzoni,  Francesco and Pagani,  Stefano and Karabelas,  Elias and Augustin,  Christoph and Haase,  Gundolf and Plank,  Gernot and Quarteroni,  Alfio},
  year = {2024},
  month = jul,
  pages = {487–513}
}

@article{Flory1961,
  title = {Thermodynamic relations for high elastic materials},
  volume = {57},
  ISSN = {0014-7672},
  url = {http://dx.doi.org/10.1039/TF9615700829},
  DOI = {10.1039/tf9615700829},
  journal = {Transactions of the Faraday Society},
  publisher = {Royal Society of Chemistry (RSC)},
  author = {Flory,  P. J.},
  year = {1961},
  pages = {829}
}

@article{sarvazyan2011overview,
  title={An overview of elastography-an emerging branch of medical imaging},
  author={Sarvazyan, Armen and J Hall, Timothy and W Urban, Matthew and Fatemi, Mostafa and R Aglyamov, Salavat and S Garra, Brian},
  journal={Current Medical Imaging},
  volume={7},
  number={4},
  pages={255--282},
  year={2011},
  publisher={Bentham Science Publishers},
    doi = {10.2174/157340511798038684},
url = {http://dx.doi.org/10.2174/157340511798038684}  
}

@article{Maday2014,
  doi = {10.1002/nme.4747},
  url = {https://doi.org/10.1002/nme.4747}  ,
  year = {2014},
  month = {8},
  publisher = {Wiley},
  volume = {102},
  number = {5},
  pages = {933--965},
  author = {Yvon Maday and Anthony T. Patera and James D. Penn and Masayuki Yano},
  title = {A parameterized-background data-weak approach to variational data assimilation: formulation,  analysis,  and application to acoustics},
  journal = {International Journal for Numerical Methods in Engineering}
}

@article{Hammond2019,
    title = {PBDW: A non-intrusive Reduced Basis Data Assimilation method and its application to an urban dispersion modeling framework},
    journal = {Applied Mathematical Modelling},
    volume = {76},
    pages = {1-25},
    year = {2019},
    issn = {0307-904X},
    doi = {https://doi.org/10.1016/j.apm.2019.05.012},
    url = {https://www.sciencedirect.com/science/article/pii/S0307904X19302951},
    author = {J.K. Hammond and R. Chakir and F. Bourquin and Y. Maday}
    }

@article{Taddei2017,
  doi = {10.1051/m2an/2017005},
  url = {https://doi.org/10.1051/m2an/2017005},
  year = {2017},
  month = {9},
  publisher = {{EDP} Sciences},
  volume = {51},
  number = {5},
  pages = {1827--1858},
  author = {Tommaso Taddei},
  title = {An Adaptive Parametrized-Background Data-Weak approach to variational data assimilation},
  journal = {{ESAIM}: Mathematical Modelling and Numerical Analysis}
}

@article{Maday2015,
  title = {PBDW State Estimation: Noisy Observations; Configuration-Adaptive Background Spaces; Physical Interpretations},
  volume = {50},
  ISSN = {2267-3059},
  url = {http://dx.doi.org/10.1051/proc/201550008},
  DOI = {10.1051/proc/201550008},
  journal = {ESAIM: Proceedings and Surveys},
  publisher = {EDP Sciences},
  author = {Maday,  Yvon and T,  Anthony and Penn,  James D and Yano,  Masayuki},
  editor = {Boyer,  Franck and Gallouet,  Thierry and Herbin,  Raphaèle and Hubert,  Florence},
  year = {2015},
  month = mar,
  pages = {144–168}
}

@article{Rozza2007,
  doi = {10.1007/bf03024948},
  url = {https://doi.org/10.1007/bf03024948},
  year = {2007},
  month = {9},
  publisher = {Springer Science and Business Media {LLC}},
  volume = {15},
  number = {3},
  pages = {1--47},
  author = {G. Rozza and D. B. P. Huynh and A. T. Patera},
  title = {Reduced basis approximation and a posteriori error estimation for affinely parametrized elliptic coercive partial differential equations},
  journal = {Archives of Computational Methods in Engineering}
}

@phdthesis{Taddei_phdthesis,
  doi = {10.13140/RG.2.2.16001.45928},
  url = {http://rgdoi.net/10.13140/RG.2.2.16001.45928},
  author = {Taddei,  Tommaso},
  title = {Model order reduction methods for data assimilation; state estimation and structural health monitoring},
  language = {en},
  publisher = {Unpublished},
  year = {2016},
  month = {9},
  school = {MIT}
}

@article{Carlberg2008,
  doi = {10.2514/6.2008-5964},
  url = {https://doi.org/10.2514/6.2008-5964},
  year = {2008},
  month = jun,
  publisher = {American Institute of Aeronautics and Astronautics},
  author = {Kevin Carlberg and Charbel Farhat},
  title = {A Compact Proper Orthogonal Decomposition Basis for Optimization-Oriented Reduced-Order Models},
  booktitle = {12th {AIAA}/{ISSMO} Multidisciplinary Analysis and Optimization Conference}
}

@article{Carlberg2010,
  doi = {10.1002/nme.3074},
  url = {https://doi.org/10.1002/nme.3074},
  year = {2010},
  month = dec,
  publisher = {Wiley},
  volume = {86},
  number = {3},
  pages = {381--402},
  author = {Kevin Carlberg and Charbel Farhat},
  title = {A low-cost,  goal-oriented `compact proper orthogonal decomposition' basis for model reduction of static systems},
  journal = {International Journal for Numerical Methods in Engineering}
}

@article{Binev2018,
  doi = {10.1137/17m1157635},
  url = {https://doi.org/10.1137/17m1157635},
  year = {2018},
  month = jan,
  publisher = {Society for Industrial {\&} Applied Mathematics ({SIAM})},
  volume = {6},
  number = {3},
  pages = {1101--1126},
  author = {Peter Binev and Albert Cohen and Olga Mula and James Nichols},
  title = {Greedy Algorithms for Optimal Measurements Selection in State Estimation Using Reduced Models},
  journal = {{SIAM}/{ASA} Journal on Uncertainty Quantification}
}

@article{Cohen2023,
  title = {Nonlinear approximation spaces for inverse problems},
  volume = {21},
  ISSN = {1793-6861},
  url = {http://dx.doi.org/10.1142/S0219530522400140},
  DOI = {10.1142/s0219530522400140},
  number = {1},
  journal = {Analysis and Applications},
  publisher = {World Scientific},
  author = {Cohen, Albert and Dolbeault, Matthieu and Mula, Olga and Somacal, Anthony},
  year = {2023},
  pages = {217--253}
}

@book{Hesthaven2016,
  title = {Certified Reduced Basis Methods for Parametrized Partial Differential Equations},
  ISBN = {9783319224701},
  ISSN = {2191-8201},
  url = {http://dx.doi.org/10.1007/978-3-319-22470-1},
  DOI = {10.1007/978-3-319-22470-1},
  journal = {SpringerBriefs in Mathematics},
  publisher = {Springer International Publishing},
  author = {Hesthaven,  Jan S and Rozza,  Gianluigi and Stamm,  Benjamin},
  year = {2016}
}

@book{Quarteroni2016,
  title = {Reduced Basis Methods for Partial Differential Equations},
  ISBN = {9783319154312},
  ISSN = {2038-5714},
  url = {http://dx.doi.org/10.1007/978-3-319-15431-2},
  DOI = {10.1007/978-3-319-15431-2},
  journal = {UNITEXT},
  publisher = {Springer International Publishing},
  author = {Quarteroni,  Alfio and Manzoni,  Andrea and Negri,  Federico},
  year = {2016}
}

@article{Benaceur2021,
  title = {Reducing sensors for transient heat transfer problems by means of variational data assimilation},
  volume = {7},
  ISSN = {2426-8399},
  url = {http://dx.doi.org/10.5802/smai-jcm.68},
  DOI = {10.5802/smai-jcm.68},
  journal = {The SMAI journal of computational mathematics},
  publisher = {Cellule MathDoc/CEDRAM},
  author = {Benaceur,  Amina},
  year = {2021},
  month = mar,
  pages = {1–25}
}

@article{Gudbjartsson1995,
  title = {The rician distribution of noisy mri data},
  volume = {34},
  ISSN = {1522-2594},
  url = {http://dx.doi.org/10.1002/mrm.1910340618},
  DOI = {10.1002/mrm.1910340618},
  number = {6},
  journal = {Magnetic Resonance in Medicine},
  publisher = {Wiley},
  author = {Gudbjartsson,  HáKon and Patz,  Samuel},
  year = {1995},
  month = dec,
  pages = {910–914}
}

@article{CardenasBlanco2008,
  title = {Noise in magnitude magnetic resonance images},
  volume = {32A},
  ISSN = {1552-5023},
  url = {http://dx.doi.org/10.1002/cmr.a.20124},
  DOI = {10.1002/cmr.a.20124},
  number = {6},
  journal = {Concepts in Magnetic Resonance Part A},
  publisher = {Wiley},
  author = {Cárdenas‐Blanco,  Arturo and Tejos,  Cristian and Irarrazaval,  Pablo and Cameron,  Ian},
  year = {2008},
  month = nov,
  pages = {409–416}
}

@article{Aletras1999,
  title = {DENSE: Displacement Encoding with Stimulated Echoes in Cardiac Functional MRI},
  volume = {137},
  ISSN = {1090-7807},
  url = {http://dx.doi.org/10.1006/jmre.1998.1676},
  DOI = {10.1006/jmre.1998.1676},
  number = {1},
  journal = {Journal of Magnetic Resonance},
  publisher = {Elsevier BV},
  author = {Aletras,  Anthony H. and Ding,  Shujun and Balaban,  Robert S. and Wen,  Han},
  year = {1999},
  month = mar,
  pages = {247–252}
}

@article{Kramer2020,
  title = {Standardized cardiovascular magnetic resonance imaging (CMR) protocols: 2020 update},
  volume = {22},
  ISSN = {1097-6647},
  url = {http://dx.doi.org/10.1186/s12968-020-00607-1},
  DOI = {10.1186/s12968-020-00607-1},
  number = {1},
  journal = {Journal of Cardiovascular Magnetic Resonance},
  publisher = {Elsevier BV},
  author = {Kramer,  Christopher M. and Barkhausen,  J\"{o}rg and Bucciarelli-Ducci,  Chiara and Flamm,  Scott D. and Kim,  Raymond J. and Nagel,  Eike},
  year = {2020},
  month = jan,
  pages = {17}
}

@article{Kramer2013,
  title = {Standardized cardiovascular magnetic resonance (CMR) protocols 2013 update},
  volume = {15},
  ISSN = {1097-6647},
  url = {http://dx.doi.org/10.1186/1532-429x-15-91},
  DOI = {10.1186/1532-429x-15-91},
  number = {1},
  journal = {Journal of Cardiovascular Magnetic Resonance},
  publisher = {Elsevier BV},
  author = {Kramer,  Christopher M and Barkhausen,  J\"{o}rg and Flamm,  Scott D and Kim,  Raymond J and Nagel,  Eike},
  year = {2013},
  month = jan,
  pages = {91}
}

@article{Chan2010,
  title = {ASCI 2010 standardized practice protocol for cardiac magnetic resonance imaging: a report of the Asian society of cardiovascular imaging cardiac computed tomography and cardiac magnetic resonance imaging guideline working group},
  volume = {26},
  ISSN = {1573-0743},
  url = {http://dx.doi.org/10.1007/s10554-010-9708-y},
  DOI = {10.1007/s10554-010-9708-y},
  number = {S2},
  journal = {The International Journal of Cardiovascular Imaging},
  publisher = {Springer Science and Business Media LLC},
  author = {Chan,  Carmen W. S. and Choi,  Byoung Wook and Jinzaki,  Masahiro and Kitagawa,  Kakuya and Tsai,  I-Chen and Yong,  Hwan Seok and Yu,  Wei},
  year = {2010},
  month = oct,
  pages = {187–202}
}

@article{Kocaoglu2020,
  title = {Breath-hold and free-breathing quantitative assessment of biventricular volume and function using compressed SENSE: a clinical validation in children and young adults},
  volume = {22},
  ISSN = {1097-6647},
  url = {http://dx.doi.org/10.1186/s12968-020-00642-y},
  DOI = {10.1186/s12968-020-00642-y},
  number = {1},
  journal = {Journal of Cardiovascular Magnetic Resonance},
  publisher = {Elsevier BV},
  author = {Kocaoglu,  Murat and Pednekar,  Amol S. and Wang,  Hui and Alsaied,  Tarek and Taylor,  Michael D. and Rattan,  Mantosh S.},
  year = {2020},
  month = jan,
  pages = {54}
}

@book{Jung2013,
  title = {Methode der finiten Elemente f\"{u}r Ingenieure: Eine Einf\"{u}hrung in die numerischen Grundlagen und Computersimulation},
  ISBN = {9783658011017},
  DOI = {10.1007/978-3-658-01101-7},
  publisher = {Springer Fachmedien Wiesbaden},
  author = {Jung,  Michael and Langer,  Ulrich},
  year = {2013}
}

@article{Rausch2017,
  title = {An augmented iterative method for identifying a stress-free reference configuration in image-based biomechanical modeling},
  volume = {58},
  ISSN = {0021-9290},
  url = {http://dx.doi.org/10.1016/j.jbiomech.2017.04.021},
  DOI = {10.1016/j.jbiomech.2017.04.021},
  journal = {Journal of Biomechanics},
  publisher = {Elsevier BV},
  author = {Rausch,  Manuel K. and Genet,  Martin and Humphrey,  Jay D.},
  year = {2017},
  month = jun,
  pages = {227–231}
}

@article{Benaceur2024,
  title = {Statistical variational data assimilation},
  volume = {432},
  ISSN = {0045-7825},
  url = {http://dx.doi.org/10.1016/j.cma.2024.117402},
  DOI = {10.1016/j.cma.2024.117402},
  journal = {Computer Methods in Applied Mechanics and Engineering},
  publisher = {Elsevier BV},
  author = {Benaceur,  Amina and Verf\"{u}rth,  Barbara},
  year = {2024},
  month = dec,
  pages = {117402}
}

@article{Romor2023,
  title = {Non-linear Manifold Reduced-Order Models with Convolutional Autoencoders and Reduced Over-Collocation Method},
  volume = {94},
  ISSN = {1573-7691},
  url = {http://dx.doi.org/10.1007/s10915-023-02128-2},
  DOI = {10.1007/s10915-023-02128-2},
  number = {3},
  journal = {Journal of Scientific Computing},
  publisher = {Springer Science and Business Media LLC},
  author = {Romor,  Francesco and Stabile,  Giovanni and Rozza,  Gianluigi},
  year = {2023},
  month = feb 
}

@article{Mula2025,
  title = {Dynamical Approximation and Sensor Placement for Filtering Problems},
  volume = {47},
  ISSN = {1095-7197},
  url = {http://dx.doi.org/10.1137/23M1625548},
  DOI = {10.1137/23m1625548},
  number = {1},
  journal = {SIAM Journal on Scientific Computing},
  publisher = {Society for Industrial & Applied Mathematics (SIAM)},
  author = {Mula,  Olga and Pagliantini,  Cecilia and Vismara,  Federico},
  year = {2025},
  month = feb,
  pages = {A403–A429}
}

@article{Galarce2025,
  title = {A fast food-freezing temperature estimation framework using optimally located sensors},
  volume = {299},
  ISSN = {0020-7403},
  url = {http://dx.doi.org/10.1016/j.ijmecsci.2025.110374},
  DOI = {10.1016/j.ijmecsci.2025.110374},
  journal = {International Journal of Mechanical Sciences},
  publisher = {Elsevier BV},
  author = {Galarce,  Felipe and Rivera,  Diego R. and Pacheco,  Douglas R.Q. and Caiazzo,  Alfonso and Castillo,  Ernesto},
  year = {2025},
  month = aug,
  pages = {110374}
}

@article{Binev2017,
  title = {Data Assimilation in Reduced Modeling},
  volume = {5},
  ISSN = {2166-2525},
  url = {http://dx.doi.org/10.1137/15M1025384},
  DOI = {10.1137/15m1025384},
  number = {1},
  journal = {SIAM/ASA Journal on Uncertainty Quantification},
  publisher = {Society for Industrial & Applied Mathematics (SIAM)},
  author = {Binev,  Peter and Cohen,  Albert and Dahmen,  Wolfgang and DeVore,  Ronald and Petrova,  Guergana and Wojtaszczyk,  Przemyslaw},
  year = {2017},
  month = jan,
  pages = {1–29}
}

@article{Levitt2024,
  title = {Randomized compression of rank-structured matrices accelerated with graph coloring},
  volume = {451},
  ISSN = {0377-0427},
  url = {http://dx.doi.org/10.1016/j.cam.2024.116044},
  DOI = {10.1016/j.cam.2024.116044},
  journal = {Journal of Computational and Applied Mathematics},
  publisher = {Elsevier BV},
  author = {Levitt,  James and Martinsson,  Per-Gunnar},
  year = {2024},
  month = dec,
  pages = {116044}
}

@article{Romor2025,
  title = {Friedrichs’ systems discretized with the DGM: domain decomposable model order reduction and Graph Neural Networks approximating vanishing viscosity solutions},
  volume = {531},
  ISSN = {0021-9991},
  url = {http://dx.doi.org/10.1016/j.jcp.2025.113915},
  DOI = {10.1016/j.jcp.2025.113915},
  journal = {Journal of Computational Physics},
  publisher = {Elsevier BV},
  author = {Romor,  Francesco and Torlo,  Davide and Rozza,  Gianluigi},
  year = {2025},
  month = jun,
  pages = {113915}
}

@article{Corrochano2023,
  title = {Higher order dynamic mode decomposition to model reacting flows},
  volume = {249},
  ISSN = {0020-7403},
  url = {http://dx.doi.org/10.1016/j.ijmecsci.2023.108219},
  DOI = {10.1016/j.ijmecsci.2023.108219},
  journal = {International Journal of Mechanical Sciences},
  publisher = {Elsevier BV},
  author = {Corrochano,  Adrián and D’Alessio,  Giuseppe and Parente,  Alessandro and Le Clainche,  Soledad},
  year = {2023},
  month = jul,
  pages = {108219}
}

@article{Mazzilli2022,
  title = {Reduced-order modelling based on non-linear modes},
  volume = {214},
  ISSN = {0020-7403},
  url = {http://dx.doi.org/10.1016/j.ijmecsci.2021.106915},
  DOI = {10.1016/j.ijmecsci.2021.106915},
  journal = {International Journal of Mechanical Sciences},
  publisher = {Elsevier BV},
  author = {Mazzilli,  Carlos E.N. and Gon\c{c}alves,  Paulo B. and Franzini,  Guilherme R.},
  year = {2022},
  month = jan,
  pages = {106915}
}

@article{Do2025,
  title = {Sparse Wasserstein Barycenters and Application to Reduced Order Modeling},
  volume = {102},
  ISSN = {1573-7691},
  url = {http://dx.doi.org/10.1007/s10915-024-02766-0},
  DOI = {10.1007/s10915-024-02766-0},
  number = {3},
  journal = {Journal of Scientific Computing},
  publisher = {Springer Science and Business Media LLC},
  author = {Do,  Minh-Hieu and Feydy,  Jean and Mula,  Olga},
  year = {2025},
  month = jan 
}

@book{Benner2017,
author = {Benner, Peter and Ohlberger, Mario and Cohen, Albert and Willcox, Karen},editor = {Peter Benner and Mario Ohlberger and Albert Cohen and Karen Willcox},
title = {Model Reduction and Approximation},
publisher = {Society for Industrial and Applied Mathematics},
year = {2017},
doi = {10.1137/1.9781611974829},
address = {Philadelphia, PA},
edition   = {},
URL = {https://epubs.siam.org/doi/abs/10.1137/1.9781611974829},
eprint = {https://epubs.siam.org/doi/pdf/10.1137/1.9781611974829}
}

@article{Bui2023,
  title = {A component-based data assimilation strategy with applications to vascular flows},
  volume = {73},
  ISSN = {2267-3059},
  url = {http://dx.doi.org/10.1051/proc/202373089},
  DOI = {10.1051/proc/202373089},
  journal = {ESAIM: Proceedings and Surveys},
  publisher = {EDP Sciences},
  author = {Bui,  Duc-Quang and Mollo,  Pierre and Nobile,  Fabio and Taddei,  Tommaso},
  editor = {Ehrlacher,  Virginie and Lombardi,  Damiano and Mula,  Olga and Nobile,  Fabio and Taddei,  Tommaso},
  year = {2023},
  pages = {89–106}
}

\end{document}